\documentclass[twocolumn,showpacs,
preprintnumbers,amsmath,amssymb]{revtex4}
\usepackage{amssymb,latexsym}
\usepackage{graphicx}

\renewcommand{\theequation}{\arabic{section}.\arabic{equation}}

\newcommand{\dis}{\displaystyle}
\newcommand{\id}{\equiv}
\newcommand{\bequ}{\begin{equation}}
\newcommand{\eequ}{\end{equation}}
\newcommand{\barr}{\begin{array}}
\newcommand{\earr}{\end{array}}
\newcommand{\bea}{\begin {eqnarray}}
\newcommand{\eea}{\end {eqnarray}}
\newcommand{\lb}{\label}

\begin{document}
\def \G {{\cal G}}
\def \L {{\ell_1\ell_2\ell_3\ell_4}}
\def \S {{s_1s_2s_3s_4}}
\def \ul {{\underline{\ell}}}
\def \Gp {{\cal G}'}
\let\la=\lambda
\def \Z {\mathbb{Z}}
\def \Zt {\mathbb{Z}_o^4}
\def \R {\mathbb{R}}
\def \C {\mathbb{C}}
\def \La {\Lambda}
\def \ka {\kappa}
\def \vphi {\varphi}
\def \x0 {x^0}
\def \dx {\left( x_1;x_2\right)}
\def \qx {\left( x_1,x_2;x_3,x_4\right)}
\def \tp {\left( p,q,k\right)}
\def \rc {\left( \xi,\eta,\tau\right)}
\def \vx {\vec{x}}
\def \vy {\vec{y}}
\def \vp {\vec{p}}
\def \vq {\vec{q}}
\def \K {K\qx }
\def \k {k\dx }
\def \P {{\cal P}(\varphi({x}))}
\def \P' {{\cal P}'(\varphi({x}))}
\def \P'' {{\cal P}''(\varphi({x}))}
\def \Cdot {\dot{\cal C}}
\def \Zd {\Z ^d}
\def \Sc {S^c(x_1,\dots ,x_n)}
\def \be {\begin {equation}}
\def \ee {\end {equation}}
\def \s {\tilde{S}^{(2)}_\la}
\def \SS {\tilde{S}^{(4)}_\la}
\def \KK { K^{(2)}}
\def \DD {D_L}
\title{Baryon-Baryon Bound States From First Principles}
\author{Paulo A. Faria da Veiga} \email{veiga@icmc.usp.br}
\author{Michael O'Carroll}
\affiliation{Departamento de Matem\'atica Aplicada e
Estat\'{\i}stica, ICMC-USP\\ C.P. 668, 13560-970 S\~ao Carlos SP,
Brazil.} \pacs{11.15.Ha, 02.30.Tb, 11.10.St, 24.85.+p.\\Keywords:
Excitation Spectrum, Lattice QCD, Bethe-Salpeter Equation, Spectral
Analysis}
\date{October 10, 2006\vspace{.3cm}}


\begin{abstract}
We determine baryon-baryon bound states in $3+1$ dimensional ${\rm
SU}(3)$ lattice QCD with two flavors, $4\times 4$ spin matrices, and
in an imaginary time formulation. For small hopping parameter,
$\kappa>0$, and large glueball mass (strong coupling), we show the
existence of three-quark isospin $1/2$ particles (proton and
neutron) and isospin $3/2$ baryons (delta particles), with
asymptotic masses $-3\ln\kappa$ and isolated dispersion curves. We
only consider the existence of bound states of total isospin
$I=0,3$. Using a ladder approximation to a lattice Bethe-Salpeter
equation, baryon-baryon bound states are found in these two sectors,
with asymptotic masses $-6\ln\kappa$ and binding energies of order
$\kappa^2$. The dominant baryon-baryon interaction is an
energy-independent spatial range-one potential with an ${\cal
O}(\kappa^2)$ strength.  There is also attraction arising from gauge
field correlations associated with six overlapping bonds, but it is
counterbalanced by Pauli repulsion to give a vanishing zero-range
potential. The overall range-one potential results from a quark,
antiquark exchange with {\it no} meson exchange interpretation; the
repulsive or attractive nature of the interaction does depend on the
isospin and spin of the two-baryon state.
\end{abstract}

\maketitle

\section{Introduction and results}
Our aim in this paper is to obtain the low-lying energy-momentum
(EM) spectrum such as particles and their bound states in ${\rm
SU}(3)$ lattice quantum chromodynamics (QCD)
\cite{Wil,Creu2,Sei,MM}. We treat a lattice QCD model in $3+1$
dimensions, with $4\times 4$ Dirac spin matrices, and with
two-flavor quarks ({\em up} and {\em down}). This model has a global
${\rm SU}(2)$ flavor (isospin) symmetry, and accommodates twenty,
three-quark, one-particle states (baryons), and their antiparticles,
associated with the proton ($p$), the neutron ($n$) and the delta
($\Delta$) particles.

This work is an important part of a recent series of papers given in
Refs. \cite{QCD,CMP,JMP,2baryon,2meson,2flavor2baryon,2flavor2meson}
where, treating less complex QCD models, we started a research
program aiming at developing a building up process and techniques to
understand, from first principles, the hadronic particles and their
bound states, bridging the gap between QCD and nuclear physics. The
existence of baryons and mesons is a manifestation of confinement.
Our main goal is to understand the mechanisms responsible for
attraction and repulsion between the hadronic particles, when and
how bound states occur and how their binding is related to the
effective Yukawa meson-exchange theory.

Here, as in previous papers (see Refs.
\cite{QCD,CMP,JMP,2baryon,2meson,2flavor2baryon,2flavor2meson}), the
analysis is performed in an imaginary-time formulation of lattice
QCD, and is restricted to the strong coupling regime (small hopping
parameter $0<\kappa\ll 1$, and large glueball mass $\beta\equiv
1/g_0^2\ll \kappa$). Even though the strong coupling regime is far
from the scaling limit (continuum quantum field theory), these
models do exhibit particles and we hope, as for confinement, that
the main features of attraction and repulsion of the continuum model
are present in the strong coupling regime.

Of course, there is the pointwise repulsive interaction originating
from Pauli exclusion. As expected, there is also a nonlocal
interaction arising from a quark-antiquark (\emph{not} a meson
particle!) exchange mechanism like in the effective Yukawa theory.
However, the attractive or repulsive nature of this interaction does
depend on the spin and isospin of the two-baryon state. How this
dependence occurs is to be understood. Not less important, our
analysis displays another essential mechanism of attraction arising
from gauge field correlation effects associated with integration
over (six) overlapping gauge bonds: if these gauge correlations are
not present the bound states cease to exist.

Our general approach to detect particles and bound states in the
low-lying part of the spectrum is to derive spectral representations
for the two- and four-baryon functions which allow us to relate
their complex momentum singularities to the EM spectrum via a
Feynman-Kac (F-K) formula. For other approaches to the determination
of the spectrum both at the theoretical and numerical levels, see
Refs.
\cite{Wil,Banks,OS,Sei,FK,Mon,MW,Schreiber,Creu,Mach,Mach2,Savage}
and Refs. \cite{num1,num2,num3,num4} dealing with numerical
simulations.

It must be pointed out that the detection of particle (and their
bound states) masses through the exponential decay rate of suitable
correlations, {\em without} a spectral representation, is
meaningless, and the resulting values may be far from the correct
ones, especially in cases where degeneracies are broken with small
spectral separations.

To determine the bound state spectrum, in a first step, we use a
ladder approximation to a lattice version of the Bethe-Salpeter
(B-S) equation for suitable four-baryon correlations. However, our
work does incorporate the main ingredients to go beyond this ladder
approximation and analytically validate our results to the full
model in the strong coupling regime, similar to what is done in some
classical and stochastic spin systems (see Refs. \cite{SO,CMP2}).

In our previous papers, the simplest case in which a two-baryon
bound state appeared is in the two-flavor total isospin $I=0,1$
sectors with $2\times 2$ spin matrices in $2+1$ dimensions (see Ref.
\cite{2flavor2baryon}); there are {\em no} bound states for $I=2,3$.
However, this model is not complex enough to accommodate states like
protons and neutrons in the one-particle spectrum.

Here, considering a more realistic lattice QCD model, it is shown
that there are twenty one-baryon states with asymptotic masses given
by $-3\ln \kappa-3\kappa^3/4+{\cal O}(\kappa^6)$ and isolated
dispersion curves (the upper gap property). The upper gap property
is unknown in the Hamiltonian formulation of the model.

Concerning the two-particle spectrum, we determine the two-baryon
bound states for the total isospin $I=0$ and $I=3$ sectors and below
the two-baryon threshold, which is given by twice the smallest of
the baryon mass. We show that there are diverse two-baryon bound
states. Their asymptotic masses are $-6\ln \kappa$ and their binding
energies are of order $\kappa^2$. In the $I=0$ sector, the most
strongly bound, two-baryon bound states correspond to a
superposition of $p-n$ and $\Delta - \Delta$ total spin $S=1$
states, and also $\Delta-\Delta$, $S=3$ states. The more weakly
bound, bound states are associated with a superposition of $p-n$ and
$\Delta - \Delta$ total spin $S=0$ states and, in addition, $\Delta
- \Delta$, $S=2$ states. In contrast to the $I=0$ states, we find
that for the maximum isospin $I=3$ sector there are also strongly
bound and weakly bound bound states in the lowest spin sectors
$S=0,1$ and {\em no} bound states if $S=2,3$. These results are in
agreement with our previous results of Ref. \cite{2flavor2baryon}
that the attraction between the two particles decreases with
increasing $I$.

It is important to note that although our results are obtained using
fairly complicated machinery, in the end, a simple picture emerges
for the formation of baryon-baryon bound states. There is a
correspondence between the lattice B-S equation, in relative
coordinates, and a lattice resolvent operator Schr\"odinger
equation. The two-baryon system in relative coordinates behaves
approximately like that of a non-relativistic one-particle lattice
Hamiltonian $T+V$ with lattice kinetic energy
$-\kappa^3\Delta_\ell$, where $\Delta_\ell$ is the spatial lattice
Laplacian and the potential energy $V$ is $\kappa^2 V^\prime$, the
quasi-meson exchange space range-one potential which dominates the
kinetic energy for small $\kappa$. The attractive or repulsive
nature of the interaction depends on the isospin, spin spectral
structure of $V$ at a single site of space-range one. Because of the
$\kappa^3$ dependence of the kinetic energy $T$ and the $\kappa^2$
dependence of the potential energy, there is no minimal critical
value of the interaction strength needed for the presence of a bound
state. This is in contrast to the case of the Birman-Schwinger (see
Ref. \cite{RS1}) bound for a $-\Delta_\ell+\lambda V^\prime$. If
$V^\prime$ is everywhere positive, there is a critical negative
$\lambda$ needed for a bound state to occur in three space
dimensions.

A brief report on the results given in this paper appears in Ref.
\cite{physlettb}. The present article is meant to be much more
self-contained, describing in a relatively complete way the details
about our methods and techniques. It is organized as follows. In
Section \ref{sec2}, we mathematically define the model, describe the
standard construction of the underlying physical Hilbert state
${\cal H}$, and introduce the fields which naturally emerge as those
quantities associated with the physical baryon particles. To do
this, we use a hyperplane decoupling expansion as done for quantum
field theories in the continuum (see \cite{Sp}), but adapted to the
lattice. In Section \ref{sec3}, the one-particle analysis is done,
and the baryon-baryon bound states are determined below the
two-baryon threshold in Section \ref{sec4}. In order to keep the
text as fluid as possible, trying to separate the main ideas of our
methods from unavoidable sometimes heavy technical details, Sections
\ref{sec3} and \ref{sec4} are subdivided according to their
principal logical steps and details about specific points are given
in the Appendices A-F. The similarity between the lattice B-S, in
relative coordinates, and the lattice resolvent operator
Schr\"odinger equation is exploited in Section \ref{sec5}. Finally,
in Section \ref{sec6}, we end the paper with conclusions and
prospective ideas, and a list of important open problems.

To close the Introduction, it is useful to make some comments on the
seemingly awkward form of the two- and four- baryon correlation
functions that we use in the sequel and which play an important role
in our method. We explain the guiding principle for obtaining the
form of correlation functions, which also holds in the continuum
spacetime. Consider the physical Hilbert space matrix elements
$(\chi, e^{-H|t|}\;\psi)$, where $H$ is the energy operator,
$e^{-H|t|}$ is the imaginary-time evolution operator (semi-group)
and $\chi$ and $\psi$ are both in the subspace of one- or two-baryon
states. We have a F-K formula at our disposal and applying it to the
two time orderings in $t$ gives the form of the correlation
functions. It is these correlation functions, that have appropriate
imaginary-time decay properties, which allows our analysis to go
through.
\section{The model, the physical Hilbert space and baryon fields}
\lb{sec2} We now introduce our ${\rm SU}(3)$, $d+1$ dimensional,
imaginary-time lattice QCD model, where $d=3$ is the space
dimension. The partition function is given formally by $ \lb{part}
  Z
=\int\,e^{-S(\psi,\bar
 \psi,g)}\,d\psi\,d\bar\psi\,d\mu(g)\,,$ and
for a function $F(\bar\psi, \psi, g)$, the normalized correlations
are denoted by \bequ\lb{avf}\langle F(\bar\psi, \psi,
g)\rangle\,=\,\frac1{Z}\int\,F(\bar\psi, \psi, g)\,e^{-S(\psi,\bar
 \psi,g)}\,d\psi\,d\bar\psi\,d\mu(g)\,.\eequ
The model
 action $S\equiv S(\psi,\bar\psi,g)$ is Wilson's action \cite{Wil} \bequ \barr{lll}
S&=&\displaystyle\frac{\kappa}{2}\sum\,\bar\psi_{a,\alpha,f}(u)\Gamma^{\sigma
e^\mu}_{\alpha\beta}(g_{u,u+\sigma e^\mu})_{ab}
\psi_{b,\beta,f}(u+\sigma e^\mu)\vspace{.1cm}\\&&+
\dis\sum_{u\in\Zt} \bar\psi_{a,\alpha,f}(u)M_{\alpha\beta}
\psi_{a,\beta,f}(u)-\frac {1}{g_0^2}\dis\sum_{p}\chi(g_p),\earr
\lb{action}\eequ where, besides the sum over repeated indices
$\alpha,\:\beta=1,2,3,4$ (spin), $a=1,2,3$ (color) and
$f=+1/2,-1/2\equiv +,-$ (isospin), the first sum runs over
$u=(u^0,\vec u)=(u^0,u^1,u^2,u^3)\in \Zt\equiv\left\{\pm 1/2,\pm
3/2,\pm 5/2...\right\}\times \mathbb{Z}^3$, $\sigma=\pm1$ and
$\mu=0,1,2,3$. Here, we are adopting the label $0$ for the time
direction and $e^\mu$, $\mu=0,1,2,3$, denotes the unit lattice
vector for the $\mu$-direction. The choice of the shifted lattice
for the time direction, avoiding the zero-time coordinate, is
 so that, in the continuum limit, two-sided equal time
 limits of quark Fermi fields correlations can be accommodated.
 At each site $u\in\Zt$, there are fermionic Grassmann fields
 $\psi_{a\alpha f}(u)$, associated with a quark, and
fields $\bar\psi_{a\alpha f}(u)$, associated with an anti-quark,
which carry a Dirac spin index $\alpha$, an ${\rm SU}(3)$ color
index $a$ and isospin $f$. We refer to $\alpha=1,2$ as {\em
 upper} spin indices and $\alpha=3,4$ or $+$ or $-$ respectively, as {\em lower} ones.
For each nearest neighbor oriented lattice bond $< u,u\pm e^\mu>$
there is an ${\rm SU}(3)$ matrix $U(g_{u,u\pm e^\mu})$ parametrized
by the gauge group element $g_{u,u\pm e^\mu}$ and satisfying
$U(g_{u,u+e^\mu})^{-1}=U(g_{u+e^\mu,u})$. Associated with each
lattice oriented plaquette $p$ there is a plaquette variable
$\chi(U(g_p))$ where $U(g_p)$ is the orientation-ordered product of
matrices of ${\rm SU}(3)$ of the plaquette oriented bonds, and
$\chi$ is the trace. For notational simplicity, we sometimes drop
$U$ from $U(g)$. Concerning the parameters, we take the quark-gauge
coupling or hopping parameter $\kappa>0$. Also, $g_0>0$ describes
the pure gauge strength and $M\equiv M(m,\kappa) =(m+2\kappa)I_4$,
$I_n$ being the $n\times n$ identity matrix. Given $\kappa$, for
simplicity and without loss of generality, $m>0$ is chosen such that
$M_{\alpha\beta}= \delta_{\alpha\beta}$, meaning that $m+2\kappa=1$.
Also, we take $\Gamma^{\pm e^\mu}=-I_4\pm\gamma^\mu$, where
$\gamma_\mu$, satisfying
$\{\gamma_\mu,\gamma_\nu\}=2\delta_{\mu\nu}I_4$, are the $4\times 4$
hermitian traceless anti-commuting Dirac matrices $\gamma^0=
\left(\barr{cc} I_2&0
\\0&-I_2\earr\right)$ and  $\gamma^j= \left(\barr{cc} 0&i{\sigma}^j
\\-i{\sigma}^j&0\earr\right)$; and $\sigma^j$, $j=1,2,3$, denotes
the hermitian traceless anti-commuting $2\times 2$ Pauli spin
matrices. The measure $d\mu(g)$ is the product measure over
non-oriented bonds of normalized ${\rm SU}(3)$ Haar measures (see
\cite{Si2}). There is only one integration variable per bond, so
that $g_{uv}$ and $g^{-1}_{vu}$ are not treated as distinct
integration variables. The integrals over Grassmann fields are
defined according to \cite{Ber}. For a polynomial in the Grassmann
variables with coefficients depending on the gauge variables, the
fermionic integral is defined as the coefficient of the monomial of
maximum degree, i.e. of
$\prod_{u,\ell}\bar\psi_{\ell}(u)\psi_{\ell}(u)$, $\ell\equiv
(\alpha,a,f)$. In Eq. (\ref{avf}), $d\psi\,d\bar\psi$ means
$\prod_{u,\ell}d\psi_{\ell}(u)\,d\bar\psi_{\ell}(u)$ such that, with
a normalization ${\cal N}_1=\langle 1\rangle$, we have $\langle
\psi_{\ell_1}(x)\,\bar\psi_{\ell_2}(y)\rangle\,=\,$ $(1/{\cal N}_1
)\;\int\,\psi_{\ell_1}(x)\,\bar\psi_{\ell_2}(y)\,e^{-\sum_{u,\ell_3,\ell_4}\,
\bar\psi_{\ell_3}(u)O_{\ell_3\ell_4}
\psi_{\ell_4}(u)}\:d\psi\,d\bar\psi$ $=O^{-1}_{\alpha_1,\alpha_2}
\delta_{a_1a_2}\delta_{f_1f_2}\delta(x-y)$, with a Kronecker delta
for space-time coordinates, and where ${\cal O}$ is diagonal in the
color and isospin indices.

Throughout the paper, we work in the strong coupling regime i.e. we
take $0< g_0^{-1}\ll\kappa\ll 1$, $m=1-2\kappa\lesssim 1$. With this
restriction on the parameters, and since we have chosen $\Gamma^{\pm
e^\mu}$ matrices within the two-parameter family described in Ref.
\cite{Sei}, positivity is preserved and there is a quantum
mechanical Hilbert space ${\cal H}$ of physical states (see below).
Furthermore, the condition $m>0$ guarantees that the one-particle
free Fermion dispersion curve increases in each positive momentum
component and is convex for small momenta.

The action of Eq. (\ref{action}) is invariant by the gauge
transformations given by, for $x\in\Zt$ and $h(x)\in{\rm SU}(3)$,
\bequ \barr{c}
\psi(x) \,\mapsto\,h(x)\,\psi(x)\\
\bar\psi(x) \,\mapsto\,\bar\psi(x)\,[h(x)]^{-1}\\U(g_{x+e^\mu,x})
\,\mapsto\,[h(x+e^\mu)]^{-1}\,U(g_{x+e^\mu,x})\,h(x) \,.
\earr\lb{gauge}\eequ Concerning the global ${\rm SU}(2)$ isospin
symmetry of the action, we follow the treatment given in Ref.
\cite{2flavor2meson}. We treat isospin symmetry at the level of
correlation functions (rather than as operators in the physical
Hilbert space) in Appendix A. Other symmetries of the action in Eq.
(\ref{action}), such as time reversal, charge conjugation, parity,
coordinate reflections and spatial rotations, which can be
implemented by unitary (anti-unitary for time reversal) operators on
the physical Hilbert space ${\cal H}$, were treated in Ref.
\cite{CMP} and do {\em not} affect the isospin indices. For
$\kappa=0$, we recover the spin ${\rm SU}(2)$ symmetry, separately
in the lower and the upper spin indices. Furthermore, in Appendix B,
a new time {\em reflection} symmetry, to be distinguished from time
reversal, of the correlation functions is shown to be useful for
obtaining additional relations between correlations.

By polymer expansion methods (see Refs. \cite{Sei,Si}), the
thermodynamic limit of correlations exists and truncated
correlations have exponential tree decay. The limiting correlation
functions are lattice translational invariant. Furthermore, the
correlation functions extend to analytic functions in the coupling
parameters $\kappa$ and $\beta=1/g_0^2$. For the formal hopping
parameter expansion, see Refs. \cite{Creu2,MM}.

The underlying quantum mechanical physical Hilbert space ${\cal H}$
and the EM operators $H$ and $P^j$, $j=1,2,3$ are defined as in
Refs. \cite{CMP,JMP}. We start from gauge invariant correlations,
with support restricted to $u^0=1/2$ and we let $T_0^{x^0}$,
$T_i^{x^i}$, $i=1,2,3$, denote translation of the functions of
Grassmann and gauge variables by $x^0\geq0$, $\vec
x=(x^1,x^2,x^3)\in\Z^{3}$. For $F$ and $G$ only depending on
coordinates with $u^0=1/2$, we have the F-K formula
\bequ\lb{FeyKa}(G,\check \check T_0^{x^0}\check T_1^{x^1}\check
T_2^{x^2}\check T_3^{x^3}F)_{{\cal H}} =\langle [T_0^{x^0}\vec
T^{\vec x}F]\Theta G\rangle\,\,,\eequ where $T^{\vec
x}=T_1^{x^1}T_2^{x^2} T_3^{x^3}$ and $\Theta$ is an anti-linear
operator which involves time reflection. Following Ref. \cite{Sei},
the action of $\Theta$ on single fields is given by
$$\begin{array}{l} \Theta\bar
\psi_{a\alpha}(u)=(\gamma_0)_{\alpha\beta}\psi_{a\beta}(tu)\,
,\vspace{.1cm}\\\Theta\psi_{a\alpha}(u)=\bar
\psi_{a\beta}(tu)(\gamma_0)_{\beta\alpha}\,;\end{array}$$where
$t(u^0,\vec u)=(-u^0,\vec u)$, for $A$ and $B$ monomials,
$\Theta(AB)=\Theta(B)\Theta(A)$; and for a function of the gauge
fields $\Theta f(\{g_{uv}\})=f^*(\{g_{(tu)(tv)}\})$,
$u,v\in\Z_o^{d+1}$, where $*$ means complex conjugate. $\Theta$
extends anti-linearly to the algebra. For simplicity, we do not
distinguish between Grassmann, gauge variables and their associated
Hilbert space vectors in our notation. As linear operators in ${\cal
H}$, $\check T_\mu$, $\mu=0,1,2,3$, are mutually commuting; $\check
T_0$ is self-adjoint, with $-1\leq \check T_0\leq1$, and $\check
T_{j=1,2,3}$ are unitary. So, we write $\check T_j=e^{iP^j}$ and
$\vec P=(P^1,P^2,P^3)$ is the self-adjoint momentum operator. Its
spectral points are $\vec p\in{\bf T}^3\equiv (-\pi,\pi]^3$. Since
$\check T_0^2\geq0$, the energy operator $H\geq0$ can be defined by
$\check T_0^2=e^{-2H}$. We call a point in the EM spectrum
associated with spatial momentum $\vec p=\vec 0$ as a mass and, to
be used below, we let ${\cal E}(\la^0,\vec \la)$ be the product of
the spectral families for the operators $\check T_0$, $\check T_1$,
$\check T_2$ and $\check T_3$. By the spectral theorem (see Ref.
\cite{RS1}, we have
$$
\check T_0=\int_{-1}^1\lambda^0 dE_0(\lambda^0) \quad,\quad \check
T_{j=1,2,3}=\int_{-\pi}^\pi\lambda^0 dF_j(\lambda^j)\,,
$$
so that ${\cal E}(\la^0,\vec \la)=E_0(\lambda^0)\prod_1^3\,
F_j(\lambda^j)$. The positivity condition $\langle F\Theta
F\rangle\geq 0$ is established in \cite{Sei}, but there may be
nonzero $F$'s such that $\langle F\Theta F\rangle= 0$. If the
collection of such $F$'s is denoted by ${\cal N}$, a pre-Hilbert
space ${\cal H}'$ can be constructed from the inner product $\langle
G\Theta F\rangle$ and the physical Hilbert space ${\cal H}$ is the
completion of the quotient space ${\cal H}'/{\cal N}$,  including
also the cartesian product of the inner space sectors, the color
space $\mathbb{C}^3$, the spin space $\mathbb{C}^4$ and the isospin
space $\mathbb{C}^2$.

Here, we analyze the one and two-baryon sectors of ${\cal H}$.
Points in the EM spectrum are detected as singularities in momentum
space spectral representations of suitable two- and four-baryon
field correlations given below.
\section{One-particle spectrum}\lb{sec3}
In this section, we restrict our attention to the subspace ${\cal
H}_o\subset {\cal H}$ generated by an odd number of $\tilde\psi$'s
where the baryons lie (the tilde means the presence or absence of a
bar). The mesons lie in the even subspace ${\cal H}_e$. We treat the
one-baryon states adapting the methods of Refs.
\cite{CMP,2flavor2baryon}.
\subsection{Baryon fields and particle identifications}\lb{sec31}
The decoupling of hyperplane method (see Refs. \cite{CMP,JMP})
reveals that the gauge invariant barred fields listed below create
baryon particles. We point out that the same method applies to QCD
models with any number of flavors. In particular, for the
${\mathrm{SU}}(3)$ invariant three flavor case, the method would
give the eightfold way. Here, the baryon particles are labeled by
one-baryon total isospin ($1^{st}$ index), the one baryon
$z$-component isospin ($2^{nd}$ index) and $z$-component of total
spin  ($3^{rd}$ index).

Let a hat $\hat\:$ mean the presence or absence of a bar
simultaneously for {\em all} $\psi$'s. The decoupling of hyperplane
expansion shows that the baryon fields are given by (with all fields
evaluated at the same point)
\begin{eqnarray}\lb{baryon1}&\!\!\!\!\!\!\!\!&\left\{
\barr{l}\hat{B}_{{\small \frac12}{\small \frac 1 2}{\small
\frac{\pm1}2}}=\frac{\large \epsilon_{abc}}{3\sqrt{2}}\,
(\hat\psi_{a\pm +}\hat\psi_{b\mp -}-\hat\psi_{a\pm-}
\hat\psi_{b\mp +})\hat\psi_{c\pm+},\vspace{.14cm}\\
\hat{B}_{{\small \frac12}{\small \frac {-1} 2}{\small
\frac{\pm1}2}}=\frac{\large \epsilon_{abc}}{3\sqrt{2}}\,
(\hat\psi_{a\pm +}\hat\psi_{b\mp -}-\hat\psi_{a\pm -}
\hat\psi_{b\mp +})\hat\psi_{c\pm +},\earr\right.\vspace{.9cm}\\
 \lb{baryon2}&\!\!\!\!\!\!\!\!&\left\{ \barr{l}\hat{B}_{{\small
\frac32}{\small \frac {1} 2}{\small \frac {1} 2}}=\!\frac{\large
\epsilon_{abc}}{6}
(\hat\psi_{a++}\hat\psi_{b--}\!\!+\!\!2\hat\psi_{a+
-}\hat\psi_{b- +})\hat\psi_{c++} ,\vspace{.14cm}\\
\hat{B}_{{\small \frac32}{\small \frac {1} 2}{\small \frac {-1}
2}}=\!\frac{\large \epsilon_{abc}}{6}
(2\hat\psi_{a++}\hat\psi_{b--}\!\!+\!\!\hat\psi_{a+
-}\hat\psi_{b- +})\hat\psi_{c-+} ,\vspace{.14cm}\\
 \hat{B}_{{\small \frac32}{\small \frac
{1} 2}{\small \frac{\pm3}2}}=\frac{\large
\epsilon_{abc}}{2\sqrt{3}}\, \hat\psi_{a\pm +}\hat\psi_{b\pm
+}\hat\psi_{c\pm -},\earr\right.\vspace{.9cm}\\
 \lb{baryon3}&\!\!\!\!\!\!\!\!&\left\{\barr{l}\hat{B}_{{\small
\frac32}{\small \frac {-1} 2}{\small \frac {1} 2}}=\!\frac{\large
\epsilon_{abc}}{6}
(2\hat\psi_{a++}\hat\psi_{b--}\!\!+\!\!\hat\psi_{a+
-}\hat\psi_{b- +})\hat\psi_{c+-} ,\vspace{.14cm}\\
\hat{B}_{{\small \frac32}{\small \frac {-1} 2}{\small \frac {-1}
2}}=\!\frac{\large \epsilon_{abc}}{6}
(\hat\psi_{a++}\hat\psi_{b--}\!\!+\!\!2\hat\psi_{a+
-}\hat\psi_{b- +})\hat\psi_{c--} ,\vspace{.14cm}\\
 \hat{B}_{{\small \frac32}{\small \frac
{-1} 2}{\small \frac{\pm3}2}}=\frac{\large
\epsilon_{abc}}{2\sqrt{3}}\, \hat\psi_{a\pm -}\hat\psi_{b\pm
-}\hat\psi_{c\pm +},\earr\right.\vspace{.9cm}\\
\lb{baryon4}&\!\!\!\!\!\!\!\!&\left\{ \barr{l}\hat{B}_{{\small
\frac32}{\small \frac {-3} 2}{\small \frac {\pm1}2}}=\frac{\large
\epsilon_{abc}}{2\sqrt{3}}\,
\hat\psi_{a\pm-}\hat\psi_{b\pm-}\hat\psi_{c\mp-}\,,\vspace{.14cm}\\
\hat{B}_{{\small \frac32}{\small \frac {-3} 2}{\small \frac{
\pm3}2}}=\frac{\large \epsilon_{abc}}{6}\,
\hat\psi_{a\pm-}\hat\psi_{b\pm-}\hat\psi_{c\pm-}\,,\earr\right.\vspace{.9cm}\\
 \lb{baryon5}&\!\!\!\!\!\!\!\!&\left\{ \barr{l}\hat{B}_{{\small
\frac32}{\small \frac {3} 2}{\small \frac {\pm1}2}}=\frac{\large
\epsilon_{abc}}{2\sqrt{3}}\,
\hat\psi_{a\pm+}\hat\psi_{b\pm+}\hat\psi_{c\mp+}\,,\vspace{.14cm}\\
\hat{B}_{{\small \frac32}{\small \frac {3} 2}{\small \frac{
\pm3}2}}=\frac{\large \epsilon_{abc}}{6}\,
\hat\psi_{a\pm+}\hat\psi_{b\pm+}\hat\psi_{c\pm+}\,, \earr\right.
\end{eqnarray}
where the index $\alpha$ in $\hat\psi_{a,\alpha,f}$ is a {\em lower}
spin component $\alpha=3,4\equiv+,-$.

The baryon fields in Eqs. (\ref{baryon1})-(\ref{baryon5}) obey the
normalization $\langle B_i \bar B_j\rangle^{(0)}=-\delta_{ij}$,
where $i,j$ are collective indices and the superscript $^{(0)}$
denotes $\kappa=0$ in the hopping term in the action $S$ given in
Eq. (\ref{action}). Below, the superscript $^{(n)}$ means the {\em
coefficient} of $\kappa^n$, $n=0,1,...$ We recall that, at
$\kappa=0$, our model also has a global ${\rm SU}(2)$ spin
invariance in the space of lower and upper components.

A brief description of the hyperplane decoupling expansion method,
and an important product structure for special third order $\kappa$
derivatives, at $\kappa=0$, for the two-baryon correlation function
defined below is given in Appendix B [see Eq. (\ref{product3})],
together with the isospin identification for the above baryon fields
with physical particles. The method reveals the fields and
associated particles so that no guesswork or ansatz has to be used.

In analogy with the two-flavor ${\rm SU}(3)$ gauge QCD in the
continuum with only the up (electric charge $+2/3$, $z$-component
isospin $+1/2$) and the down (electric charge $-1/3$, $z$-component
isospin $-1/2$) quarks, we identify the particles associated with
the barred fields in Eqs. (\ref{baryon1})-(\ref{baryon5}). The
proton ($p_\pm$) and the neutron ($n_\pm$) are, respectively,
described in Eq. (\ref{baryon1}). $\Delta^+$, $\Delta^0$ and
$\Delta^-$ are identified with the fields in Eqs.
(\ref{baryon2})-(\ref{baryon4}). Finally, $\Delta^{++}$ corresponds
to the baryon fields in Eq. (\ref{baryon5}). The corresponding
$z$-component isospin conjugated particles are described by the
unbarred fields.

We remark that, although not obvious, the fields in Eqs.
(\ref{baryon1})-(\ref{baryon5}) are the same as those obtained by
the construction of gauge-invariant (colorless) wave functions as
products of spin symmetric, flavor symmetric wave functions, except
for the proton and neutron. For the proton and neutron, wave
functions of Eq. (\ref{baryon1}) are the same as the sum of products
of wave functions that are antisymmetric in both spin and flavor in
the pairs $12$, $13$ and $23$.
\subsection{Two-baryon correlations and the one-particle analysis}\lb{sec32}
The baryon-baryon correlation function is defined by ($\chi$ here
denotes the characteristic function)
\bequ\lb{GG}\barr{lll}\!\!G_{\ell_1\ell_2}(u,v)&\!\!\!=\!&\langle
B_{\ell_1}(u)\bar B_{\ell_2}(v) \rangle\,\chi_{u^0\leq
v^0}\vspace{.14cm}\\&\!\!\!\!&-\langle \bar B_{\ell_1}(u)
B_{\ell_2}(v) \rangle^*\,\chi_{u^0> v^0}\equiv
G_{\ell_1\ell_2}(u-v),\earr\eequ where $\ell=(I,I_z,s)$ is a
collective index. By isospin symmetry $G_{\ell_1\ell_2}(x=u-v)$ is
diagonal in $I,I_z$ and, for $I$ fixed, independent of $I_z$. The
consequences of isospin symmetry for the two-point functions, as
well as for the four-point functions given below, are discussed in
 Appendix A, where important orthogonality relations and
correlation function identities are derived. Also, the $z$-component
of spin $s$ is restricted to $\pm 1/2$, for $I=1/2$ (for $p$ and
$n$) or $\pm 3/2,\pm 1/2$, for $I=3/2$ (the $\Delta$ particles).

Setting $x^0= u^0-v^0\not=0$,  we have the following spectral
representation for $G$, with $\bar B_{\ell}\equiv \bar
B_{\ell}(1/2,\vec 0)$, \bequ \lb{FK1}\barr{lcl}
 G_{\ell_1\ell_2}(x)&\,=\,&-\dis\int_{-1}^1\,\dis
 \int_{{\mathbb T}^3}\,(\la^0)^{|x^0|-1}
 e^{i\vec \la.\vec x}\vspace{.15cm}\\& & \times\:
  d_{\la}(\bar B_{\ell_1},{\cal
E}(\la^0,\vec \la)\bar B_{\ell_2})_{{\cal H}}\,; \earr\eequ for
$x\in\Z^{d+1}$, $x^0\not= 0$, and is an even function of $\vec x$.

Following Proposition 1 of Ref. \cite{CMP}, we verify how Eq.
(\ref{FK1}) is obtained. First, we derive a spectral representation
for $u^0<v^0$. With this time ordering, we have
$$\barr{lll}\!\!\!G_{\ell_1\ell_2}\!(u,v)&\!\!\!=\!\!\!& \langle B_{\ell_1}(u)\bar B_{\ell_2}(v)
\rangle \vspace{.15cm}\\&\!\!\!=\!\!\!&- \langle
[T_0^{v^0-u^0-1}\vec T^{\vec v-\vec u} \bar B_{\ell_2}\!(1/2,\vec
0)] B_{\ell_1}(-1/2,\vec 0) \rangle.\earr$$But $B_{\ell}(-1/2,\vec
0)= \Theta \bar B_{\ell}(1/2,\vec 0)$, so that
$$G_{\ell_1\ell_2}\!(u,v)\!=\!-\langle [T_0^{v^0-u^0-1}\vec T^{\vec v-\vec
u} \bar B_{\ell_2}\!(1/2,\vec 0)] \Theta \bar B_{\ell_1}(1/2,\vec 0)
\rangle.$$ By parity symmetry, $\bar B_\ell(w^0,\vec 0)$ has parity
$-1$ so that $\vec v-\vec u$ can be replaced by $\vec u-\vec v$ here
and below. The F-K formula of Eq. (\ref{FeyKa}) gives the result.
For $u^0>v^0$, we write $G_{\ell_1\ell_2}\!(u,v)=- \langle
[T_0^{u^0-v^0-1}\vec T^{\vec u-\vec v} \bar B_{\ell_1}\!(-1/2,\vec
0)] \Theta \bar B_{\ell_2}(1/2,\vec 0) \rangle^*$ and use the F-K
formula. For $u^0=v^0$ there is no spectral representation. By going
backwards, the correlation of Eq. (\ref{GG}) is obtained from the
Hilbert space inner product.

With this, upon separating the time zero contribution, the Fourier
transform $\tilde
G_{\ell_1\ell_2}(p)=\sum_{x\in\mathbb{Z}^4}\,G_{\ell_1\ell_2}(x)e^{-ip.x}$
of $G_{\ell_1\ell_2}(x)$ admits the spectral representation
\bequ\lb{specrep} \barr{lcl}\tilde
G_{\ell_1\ell_2}(p)&\!\!=\!\!&\tilde G_{\ell_1\ell_2}(\vec p)\!
-\!(2\pi)^3
\!\int_{-1}^{1} f(p^0,\la^0)d_{\la^0}\alpha_{\vec
p,\ell_1\ell_2}(\la^0)\,,\earr\eequ where
\bequ\lb{dalpha}d_{\la^0}\alpha_{\vec
p,\ell_1\ell_2}(\la^0)\!=\!\int_{\mathbb{T}^3}\,\delta(\vec p-\vec
\la) d_{\la^0}d_{\vec\la}(\bar B_{\ell_1},{\cal E}(\la^0,\vec
\la)\bar B_{\ell_2})_{\cal H}\,,\eequ with $f(x,y)\equiv
(e^{ix}-y)^{-1}+(e^{-ix}-y)^{-1}$, and we set $\tilde G (\vec
p)=\sum_{\vec x}\,e^{-i\vec p.\vec x}G(x^0=0,\vec x)$.

The spectral representation given in Eq. (\ref{specrep}) allows us
to identify complex $p$ singularities of $\tilde G(p)$ with points
in the EM spectrum.

 Using symmetries, $\tilde G_{\ell_1\ell_2}(\vec
p)$ is real diagonal and independent of isospin and color indices,
diagonal and independent of the spin, at least up to order
$\kappa^6$.

From isospin symmetry, we see that the $20\times 20$ two-point
function matrix decomposes into blocks. There is one $2\times 2$
block for the proton $p$ and the same for the neutron $n$, and there
are four $4\times 4$ blocks associated to the $\Delta$'s. Turning to
the determination of the dispersion curves, as $G_{\ell_1\ell_2}$ is
diagonal in $I$ and $I_z$, and for fixed $I$, independent of $I_z$,
we consider $I$ and $I_z$ fixed and denote the spin dependent
two-point function by $G_{s_1s_2}$. The dispersion curves $w(\vec
p)\equiv w(\vec p,\kappa)$ are given by the zeroes of the
determinant of $\tilde \Gamma_{s_1s_2}(p)$, for $p^0$ in the
positive imaginary axis, i.e.
$${\mathrm{det}}\, \tilde \Gamma_{s_1s_2}(p^0=iw(\vec p),\vec p)=0\,.$$

In Appendix B, we show that the $20\times 20$ two-point function
matrix $\tilde \Gamma_{s_1s_2}(p)$ is {\sl not} diagonal at $\vec
p\not=\vec 0$ (it is diagonal for $\vec p=\vec 0$!). Thus, if we
want to obtain the dispersion curves, this property prevents us to
follow closely the analysis given in Ref. \cite{CMP} for the mass
spectrum. Using a different method, we show that there are $2$ ($4$)
dispersion curves, {\em not} necessarily distinct, for $I=1/2$
($3/2$), denoted by $w_j(\vec p)$, $j=1,2$ for $I=1/2$ and
$j=1,2,3,4$ for $I=3/2$; the masses $m_j$ are given by $m_j=w_j(\vec
p=\vec 0)$. The twenty approximate one-particle dispersion curves
are identical up to and including ${\cal O}(\kappa^5)$ and, with
\bequ\lb{pl}p_\ell^2\equiv 2\sum_{i=1}^{3}(1-\cos p^i)\,,\eequ are
given by \bequ \lb{disp}w_j(\vec p)\equiv w_j(\vec p,\kappa)=\left[
-3\ln {\kappa}-3\kappa^3/4+ p_\ell^2\kappa^3/8\right]+r_j(\vec
p)\,,\eequ where $r_j(\vec p)$ is of  ${\cal O}(\kappa^6)$, and its
$r_j(\vec p=\vec 0)$ at ${\cal O}(\kappa^6$ is zero. Thus, if there
is a mass splitting, it is at least of ${\cal O}(\kappa^7)$. As also
shown in Appendix B, for the case of $2\times 2$ blocks associated
with the proton and neutron, since the $2\times 2$ matrices are
diagonal by a parity-charge conjugation-time reflection symmetry,
 we can apply the more powerful method of Ref.
\cite{CMP}, which includes an application of the analytic implicit
function theorem, to show that the corresponding (identical)
$r_j(\vec p)$ are analytic in $\kappa$ and in each component
$p^{i=1,2,3}$ of the momentum $\vec p$.

The spectral measure $d_{\la^0}\alpha_{\vec p,s_1s_2}(\la^0)$ of Eq.
(\ref{dalpha}) has the decomposition
$$\barr{ll}d_{\la^0}\alpha_{\vec p,s_1s_2}(\la^0)=&\sum_j\,[
Z_{j,s_1s_2}(\vec p)\delta(\la^0-e^{-w_j(\vec
p)})]d\la^0\vspace{.14cm}\\&+d_{\la^0}\nu_{s_1s_2}(\la^0,\vec
p)\,,\earr$$ where the first term corresponds to the one-particle
contributions, and $$Z_{j,s_1s_2}(\vec p)^{-1}\!=\!-(2\pi)^3
e^{w_{j}(\vec p)}\frac{\partial\tilde\Gamma_{s_1s_2}}{\partial\chi}
(p^0=i\chi,\vec p)|_{\chi=w_j(\vec p)}\,.$$ Let $
\Gamma_{s_1s_2}(u,v)$ denote the convolution inverse of the
two-point function $G_{s_1s_2}(u,v)$. We can show that $
\Gamma_{s_1s_2}(u,v)$ decays faster than $G_{s_1s_2}(u,v)$, and
hence the Fourier transform $\tilde\Gamma_{s_1s_2}(p)= [\tilde
G]_{s_1s_2}^{-1}(p)$ of
$\Gamma_{s_1s_2}(x=u-v)\equiv\Gamma_{s_1s_2}(u,v)$ has a larger
region of analyticity in $p^0$. Thus, the cofactor $[{\rm cof}\,
\tilde\Gamma]_{s_1s_2}(p)/{\rm det}[\tilde\Gamma(p)]\equiv \tilde
\Gamma_{s_1s_2}^{-1}(p)$ provides a meromorphic extension of $\tilde
G_{s_1s_2}(p)$ up to near the two-particle threshold (precisely, the
meson-baryon threshold, which is $\approx -5\ln \kappa$), and the
one-particle EM spectrum occurs, for each $\vec p$, as points given
by the $p^0$ imaginary axis zeroes of ${\rm det}[\tilde\Gamma(p)]$.
 The two-point function convolution inverse
$\Gamma(x)$ is defined by
\bequ\Gamma=(1+G_d^{-1}G_n)^{-1}\,G_d^{-1}\,,\lb{Neumann}\eequ using
a Neumann series, where $G_d$ is the diagonal part of $G$ given by
\begin{eqnarray}
G_{d,s_1s_2}(u,v)&\,=\,&G_{s_1s_2}(u,u)\delta_{s_1s_2}
\delta_{uv}\,,\lb{Gd}\vspace{.14cm}
\end{eqnarray}
and $G_n$ is the remainder \bequ \lb{Gn}
G_{n,s_1s_2}(u,v)=G_{s_1s_2}(u,v)-G_{d,s_1s_2}(u,v)\,.\eequ By the
decoupling of hyperplane method, we show the decay of $G$ and
$\Gamma$; $G_d^{-1}$, $G_n$ and $\Gamma$ are bounded as matrix
operators on $\ell_2({\mathbb C}^{20}\times\Zt)$; for small
$\kappa$, $G_n$ is small in norm so that the Neumann series in Eq.
(\ref{Neumann}) converges. To obtain the results on the dispersion
curves $w_{j}(\vec p)$, up to order $\kappa^6$, we need detailed
information on the small distance behavior of $\Gamma_{s_1s_2}$,
which are obtained by showing explicit cancelations in the Neumann
series and go beyond the hyperplane decoupling method (see Ref.
\cite{CMP} and Appendix B for more details).

We now give an intuitive argument for obtaining the mass and
dispersion curves. The precise value requires a more refined
analysis adapting the method used in Ref. \cite{CMP}, and depends on
the short distance behavior of $\Gamma_{s_1s_2}(x)$, which in turn
depends on that of $G_{s_1s_2}(x)$, via the Neumann series. By
explicit calculation, $G_{s_1s_2}(0)=-1$, $G_{s_1s_2}(\pm
e^0)=-\kappa^3$, $G_{s_1s_2}(\pm e^j)=-\kappa^3/8$, $j=1,2,3$,
(correcting the misprinted values in Ref. \cite{CMP}) such that
$$\tilde G_{s_1s_2}(p)=[-1-2\cos p^0 \kappa^3-\dfrac{\kappa^3}4
\sum_{j=1}^3\;\cos p^j ]\delta_{s_1s_2}+{\cal O}(\kappa^6)\;,$$ and
then
$$\tilde \Gamma_{s_1s_2}(p)=[-1+2\cos p^0 \kappa^3+\dfrac{\kappa^3}4
\sum_{j=1}^3\;\cos p^j ]\delta_{s_1s_2}+{\cal O}(\kappa^6)
\;.$$Setting $\tilde \Gamma_{s_1s_2}(p)=0$ for $p^0=iw_{j}(\vec p)$,
we obtain the dispersion curves of Eq. (\ref{disp}), which are shown
to be the correct ones up to ${\cal O}(\kappa^5)$ by adapting the
methods of Ref. \cite{CMP}.

Finally, with this, the $\la^0$ support of the measure
$d\nu_{s_1s_2}(\la^0,\vec \la)$ is contained in $|\la^0|\leq
|\kappa|^{5-\epsilon}$.

The decomposition of $\tilde G_{s_1s_2}(p)$ is given by
\bequ\lb{specrep2} \barr{lcl}\tilde G_{s_1s_2}(p)&\!\!=\!\!&\tilde
G_{s_1s_2}(\vec p)\! -\sum_j\,[ Z_{j,s_1s_2}(\vec p)\left[\dfrac
1{e^{ip^0}-e^{w_j(\vec p)}}\vspace{.15cm}\right.\\&&\left.+\dfrac
1{e^{-ip^0}-e^{w_j(\vec p)}}\right]-(2\pi)^3 \!\int_{-1}^{1}
f(p^0,\la^0)\vspace{.15cm}\\&&\times\;d_{\la^0}\nu_{\vec
p,s_1s_2}(\la^0)\,,\earr\eequ
\section{Two-baryon spectrum}\lb{sec4}
In this section, we restrict our attention to the subspace ${\cal
H}_e$ generated by an even number of $\tilde\psi$'s. We treat the
two-baryon states adapting the methods of Ref. \cite{2flavor2baryon}
(see also \cite{physlettb}).
\subsection{Two-baryon states and four-baryon correlations
in the individual and total spin basis}\lb{sec41} To determine
baryon-baryon bound states, we consider the subspace of the subspace
${\cal H}_e\subset{\cal H}$ generated by the product of the twenty
one-baryon fields given by $\bar B_{\ell_1}(\vec x_1) \bar
B_{\ell_2}(\vec x_2)$. If we ignore possible linear dependencies,
this space has dimension $400$, which -upon using isospin symmetry-
decomposes into two-baryon total isospin $I=0,1,2,3$ sectors, of
dimensions 20, 108, 160 and 112, respectively. In this subsection,
to reduce the complexity of the analysis, here we first restrict our
attention to the $I=0$ sector, where we expect to find a deuteron,
$p-n$ bound state. The $I=3$ states will be treated in Section 4.4.
The more algebraically complex $I=1,2$ sectors are presently being
analyzed \cite{I124D2flavor2baryon}. The isospin orthogonality
relations and correlation identities used here are derived in
Appendix A.

The $I=0$ states are given by the isospin Clebsch-Gordan (C-G)
linear combinations
$$\bar \tau^{1/2}_{s_1s_2}(x_1,x_2)=\sum_{I_1+I_2=0}\, c^{ \frac 12\,\frac12}_{I_1I_2}
\bar B_{\frac12,I_1,s_1}(x_1) \bar B_{\frac1 2,I_2,s_2}(x_2)\,,
$$
for the coupling of two isospin $1/2$ baryons, and similarly, with
$(1/2\rightarrow 3/2,s_i\rightarrow t_i)$,
for the coupling of two $3/2$ isospins. We use $s_j$ ($t_j$) for
isospin $1/2$ ($3/2$) baryons and recall that the $1/2$, $3/2$
coupling does {\em not} give $I=0$; and we refer to this description
of the two-particle states as the individual spin basis.

It turns out that the dominant interaction between baryons admits a
simpler description in terms of total spin. The total spin states
are obtained by taking C-G combinations of the $\tau$'s. Namely,
\bequ\lb{1/2b} \bar {\cal
T}^{1/2}_{SS_Z}(x_1,x_2)=\sum_{s_1+s_2=S_z}\, a^{\frac
12\,\frac12}_{s_1s_2} \bar \tau^{1/2}_{s_1,s_2}(x_1,x_2)\,, \eequ
and similarly, with $(1/2\rightarrow 3/2,S\rightarrow
T,s_i\rightarrow t_i)$,
where the $a$'s are the spin C-G coefficients. 
Using their properties, we find that \bequ \lb{par}\barr{lll} \bar
{\cal T}^{1/2}_{SS_Z}(x_1,x_2)&=& (-1)^{S+1}\bar {\cal
T}^{1/2}_{SS_Z}(x_2,x_1)\,,
\vspace{.15cm}\\
\bar {\cal T}^{3/2}_{TT_Z}(x_1,x_2)&=&(-1)^{T+1}\bar {\cal
T}^{3/2}_{TT_Z}(x_2,x_1) \,,\earr\eequ
 so that, for coincident points, the states vanish for $S=0$, and
 for $T=0,2$, a kinematic statistical effect. The states are in the space
symmetric (anti-symmetric) for $S=1$, $T=1,3$ ($S = 0$, $T=0,2$).

We order the individual spin pairs as $(s_1,s_2)=(\frac 12, \frac
12)$, $(\frac 12, -\frac 12)$, $(-\frac 12, \frac 12)$, $(-\frac
12,- \frac 12)$; $(t_1,t_2)=(\frac 32, \frac 32)$, $(\frac 32, \frac
12)$, $(\frac 32, -\frac 12)$, $(\frac 32,- \frac 32)$, $(\frac 12,
\frac 32)$, ..., $(\frac 12, -\frac 32)$, $(-\frac 12,\frac 32)$,
... $(-\frac 12, -\frac 32)$, $(-\frac 32, \frac 32)$, ..., $(-\frac
32,- \frac 32)$, and the total spin pairs as $(S,S_z)=(1,1)$,
$(T,T_z)=(1,1)$, $(S,S_z)=(1,0)$, $(T,T_z)=(1,0)$, $(S,S_z)=(1,-1)$,
$(T,T_z)=(1,-1)$, $(T,T_z)=$ $(3,3)$, $(3,2)$, ..., $(3,-3)$,
$(S,S_z)=(0,0)$, $(T,T_z)=$ $(0,0)$, $(T,T_z)=(2,2)$, $(2,1)$, ...,
$(2,-2)$. From Eq. (\ref{par}) and in the total spin basis we see
that the two baryon states are space symmetric (anti-symmetric) in
the $1$st through the $13$th (last $7$) components.

The individual and the total spin basis, in the $I=0$ sector, are
related by a real orthogonal transformation whose entries are C-G
coefficients. Let ${\cal A}=\left[ {\cal A}
\right]_{i,j=1,\ldots,20}$ be the $20\times 20$ transformation
matrix from the individual to the total spin basis. The nonzero
elements of ${\cal A}$ are: ${\cal A}_{11}=1= {\cal A}_{45}= {\cal
A}_{75}={\cal A}_{13\,20}$; ${\cal A}_{27}=\sqrt{3/10}= {\cal
A}_{2\,13}={\cal A}_{6\,12}={\cal A}_{6\,18}$; ${\cal
A}_{2\,10}=-\sqrt{2/5}={\cal A}_{6\,15}$;
 ${\cal A}_{32}=\sqrt{1/2}={\cal
A}_{33}={\cal A}_{86}={\cal A}_{89}={\cal A}_{12\,16}={\cal
A}_{12\,19}={\cal A}_{14\,2}={\cal A}_{16\,6}={\cal A}_{17\,7}=
{\cal A}_{19\,12}={\cal A}_{20\,16}$; ${\cal
A}_{48}=\sqrt{9/20}={\cal A}_{4\,17}={\cal A}_{10\,11}={\cal
A}_{10\,14}$; ${\cal A}_{4\,11}=-\sqrt{1/20}={\cal A}_{4\,14}$;
${\cal A}_{97}=\sqrt{1/5}={\cal A}_{9\,13}={\cal A}_{11\,12}= {\cal
A}_{11\,18}$; ${\cal A}_{9\,10}=\sqrt{3/5}={\cal A}_{11\,15}$;
${\cal A}_{10\,8}=\sqrt{1/20}={\cal A}_{10\,17}$;
 ${\cal A}_{14\,3}=-\sqrt{1/2}={\cal
A}_{16\,9}={\cal A}_{17\,13}={\cal A}_{19\,18}={\cal A}_{20\,19}$;
${\cal A}_{15\,8}=1/2={\cal A}_{15\,14}={\cal A}_{18\,8}= {\cal
A}_{18\,11}$ and, finally, ${\cal A}_{15\,11}=-1/2={\cal
A}_{15\,17}={\cal A}_{18\,14}= {\cal A}_{18\,17}$.
\subsection{Four-baryon correlations and two-baryon spectrum}\lb{sec42}
As we do not know a priori which linear combination of states
corresponds to a bound state, we consider the $20\times 20$ matrix
of (truncated) four-baryon functions
$$M(x_1,x_2;x_3,x_4)=\left(
\barr{ll} M_{11}&M_{12}\\M_{21}&M_{22} \earr
\right)(x_1,x_2;x_3,x_4)\,,$$ where, in the in\-di\-vi\-dual spin
ba\-sis,
and for $x_1^0=x_2^0$, $x_3^0=x_4^0$, $x_1^0\leq x_3^0$, we have
$M_{11;s_1s_2s_3s_4}(x_1,x_2;x_3,x_4)=\langle
\tau^{1/2}_{s_1s_2}(x_1,x_2)\bar \tau^{1/2}_{s_3s_4}(x_3,x_4)\rangle
$, $M_{12;s_1s_2t_3t_4}(x_1,x_2;x_3,x_4)=\langle
\tau^{1/2}_{s_1s_2}(x_1,x_2) \bar
\tau^{3/2}_{t_3t_4}(x_3,x_4)\rangle$, $
M_{21;t_1t_2s_3s_4}(x_1,x_2;x_3,x_4)=\langle
\tau^{3/2}_{t_1t_2}(x_1,x_2) \bar
\tau^{1/2}_{s_3s_4}(x_3,x_4)\rangle$ and $
M_{22;t_1t_2t_3t_4}(x_1,x_2;x_3,x_4)=\langle
\tau^{3/2}_{t_1t_2}(x_1,x_2) $ $\times\, \bar
\tau^{3/2}_{t_3t_4}(x_3,x_4)\rangle$; and similarly for the total
spin basis. For $x_1^0>x^0_3$, $M_{ij}$ is defined by interchanging
the barred and unbarred fields and taking the complex conjugate. As
in Ref. \cite{2flavor2baryon}, $x_1$, ..., $x_4$ are now taken in
$\mathbb{Z}^4$.

$M_{ij}$ has spectral representations which we use to detect bound
states below the two-baryon threshold. We now obtain a spectral
representation for $M_{ij;s_1s_2s_3s_4}$. To simplify and make the
analysis more transparent, we derive a spectral representation in
terms of the basic baryon fields $B_\ell$. By simply writing the
suitable linear combinations of these representations, we obtain the
spectral representations for the four-point functions of the
$\tau_\ell$ fields.

The starting point employs the F-K formula which furnishes a 4-point
correlation function in terms of $B_\ell$'s. For $x^0\not= 0$, with
$x=(x^0=v^0-u^0,\vec x)\in\mathbb{Z}^4$,
$$\barr{l}(\bar B_{\ell_1}(1/2,\vec u_1) \bar B_{\ell_2}(1/2,\vec u_2),
(\check T_0)^{|x^0|-1}\vspace{.14cm}\\\hspace{.9cm}\times\,\vec
T^{\vec x}\bar B_{\ell_3}(1/2,\vec u_3)\, \bar B_{\ell_4}(1/2,\vec
u_4))_{{\cal H}}=-{\cal G}_{\ell_1\ell_2\ell_3\ell_4}(x),\earr$$
where ${\cal G}_{\ell_1\ell_2\ell_3\ell_4}(x)\equiv {\cal
G}_{\ell_1\ell_2\ell_3\ell_4}(u_1,u_2,u_3+\vec x,u_4+\vec x)$ and,
for $u^0_1=u^0_2 =u^0$ and $u^0_3=u^0_4=v^0$,$$\barr{l} {\cal
G}_{\ell_1\ell_2\ell_3\ell_4}(u_1,u_2,u_3,u_4)=\vspace{.15cm}\\
\qquad\qquad\langle B_{\ell_1}(u_1)B_{\ell_2}(u_2)\bar
B_{\ell_3}(u_3)\bar B_{\ell_4}(u_4) \rangle\,\chi_{u^0\leq
v^0}\vspace{.15cm}\\\qquad\qquad+\;\langle \bar B_{\ell_1}(u_1)\bar
B_{\ell_2}(u_2)B_{\ell_3}(u_3) B_{\ell_4}(u_4) \rangle^*\,\chi_{u^0>
v^0}\;.\earr$$

We obtain a spectral representation for   ${\cal
G}_{\ell_1\ell_2\ell_3\ell_4}(x)$ and its Fourier transform
$\tilde{\cal G}_{\ell_1\ell_2\ell_3\ell_4}(k)$ which allows us to
relate the complex $k$ singularities of $\tilde{\cal
G}_{\ell_1\ell_2\ell_3\ell_4}(k)$ to the EM spectrum. Inserting the
spectral representations for $\check T_0$ and $\check T_{i=1,2,3}$,
and taking the Fourier transform gives
$$\barr{lll}\hspace{-.06cm}\tilde
\G_{\ell_1\ell_2\ell_3\ell_4}(k)\!\!\!&=\!\!&
 \tilde \G_{\ell_1\ell_2\ell_3\ell_4} (\vec k)
-(2\pi)^3\int_{-1}^1\int_{{\bf
T}^3}\,f(k^0,\la^0)\vspace{.14cm}\\&&\times\delta (\vec k-\vec
\la)\,\!d_{\la^0} d_{\vec \la} ( \bar B_{\ell_1}(1/2,\vec u_1)\bar
B_{\ell_2}(1/2,\vec u_2),\vspace{.14cm}\\&&\times{\cal E}(\la^0,\vec
\la)\bar B_{\ell_3}(1/2,\vec u_3)\bar B_{\ell_4}(1/2,\vec
u_4))_{{\cal H}},\earr$$where $\tilde
\G_{\ell_1\ell_2\ell_3\ell_4}(\vec k)\!\!=\!\!\sum_{\vec x\in{\bf
T}^3}\! e^{-i\vec k.\vec x }\G_{\ell_1\ell_2\ell_3\ell_4}(x^0=0,\vec
x)$. The singularities of $\tilde \G_{\ell_1\ell_2\ell_3\ell_4}(k)$,
for $k=(k^0=i\chi,\vec k =0)$ and $e^{\pm \chi}\leq 1$, are points
in the mass spectrum.

To analyze $\tilde \G_{\ell_1\ell_2\ell_3\ell_4}(k)$, we follow the
method of analysis for spin models as in Ref. \cite{SO}. To use a
notation closer to that of Ref. \cite{SO}, we relabel the time
direction coordinates in $\G_\L(x)$ by integer labels, with
$u_i^0-1/2=x_i^0$, $\vec u_i=\vec x_i$, $i=1,\ldots, 4$, and denote
by $D_\L$  the shifted correlation functions. Next, we write
$D_\L(x_1,x_2,x_3+\vec x,x_4+\vec x)$, $x_1^0=x_2^0$ and
$x_3^0=x_4^0$, $x^0=x_3^0-x_2^0$,
 where $x_i$ and $x$ are points on the $\Z^4$ lattice.
 Now, by translation invariance, we pass to difference
 coordinates and then to lattice relative coordinates
 $\xi=x_2-x_1,\quad \eta=x_4-x_3\quad {\rm and}\quad
 \tau=x_3-x_2$ (see Figure 1 below). These lattice relative coordinates are a substitute for
 the usual center of mass and relative coordinates customarily used in the continuum.
After doing this, we obtain
$$\barr{ll}D&\!\!\!\!_\L(x_1,x_2,x_3+\vec x,x_4+\vec
x)\vspace{.15cm}\\&=D_\L(0,x_2-x_1,x_3-x_1+\vec x,x_4-x_1+\vec x)
\vspace{.15cm}\\& \equiv D_\L(\vec \xi,\vec \eta,\tau+\vec
x)\,,\earr$$ and $$\tilde \G_\L (k)=e^{i\vec k.\vec \tau}\hat
D_\L(\vec \xi,\vec \eta,k)\,,$$ where $\hat D_\L(\vec \xi,\vec
\eta,k)$ $=$ $\sum_{\tau\in\Z^4}\,D_\L(\vec \xi, \vec
\eta,\tau)e^{-ik.\tau}$. Explicitly, we have \bequ\lb{DD}\barr{ll}
\!\!D_\L\!\!\!\!&(x_1,x_2,x_3,x_4)\!=\!\langle
B_{\ell_1}(x_1^0+1/2,\vec
x_1)\vspace{.15cm}\\&\times\;B_{\ell_2}(x_2^0+1/2, \vec x_2) \bar
B_{\ell_3}(x_3^0+1/2, \vec x_3) \vspace{.15cm}\\&\times\;\bar
B_{\ell_4}(x_4^0+1/2, \vec x_4) \rangle\,\chi_{x_2^0\leq
x_3^0}\vspace{.15cm}\\&+\;\langle \bar B_{\ell_1}(x_1^0+1/2,\vec
x_1)\bar B_{\ell_2}(x_2^0+1/2, \vec x_2)\vspace{.15cm}\\&\times\;
B_{\ell_3}(x_3^0+1/2, \vec x_3) B_{\ell_4}(x_4^0+1/2, \vec x_4)
\rangle^*\,\chi_{x_2^0> x_3^0}\;.\earr \eequ

The point of the considerations above is that the singularities of
$\tilde \G_\L (k)$ are the same as those of $\hat D_\L(\vec \xi,\vec
\eta,k)$.

\begin{center}\vspace{.6cm}

\setlength{\unitlength}{.6cm}\hspace{1cm}
\begin{picture}(4,4)
\thicklines \put(-3.6,3.3){$\vec\eta=x_4-x_3$}
\put(-3.6,2.5){$\tau=x_3-x_2$} \put(-3.6,1.7){$\vec\xi=x_2-x_1$}
\put(1.3,1){\circle*{0.147}} \put(1.2,1.3){$x_1$}
\put(1.3,1){\line(3,0){3.7}} \put(3.,.84){$>$} \put(4.9,1.3){$x_2$}
\put(5,1){\line(-1,1){3}} \put(5,1){\circle*{0.147}}
\put(1.9,4.33){$x_3$} \put(2,4){\line(3,0){3.7}}
\put(2,4){\circle*{0.147}} \put(5.8,4){\circle*{0.147}}
\put(5.7,4.33){$x_4$} \put(3.,.2){$\vec\xi$} \put(3.58,2.6){$\tau$}
\put(3.5,2.5){\line(3,-1){.3}} \put(3.5,2.5){\line(1,-2){.15}}
\put(3.9,4.33){$\vec\eta$} \put(3.9,3.83){$>$} \thinlines
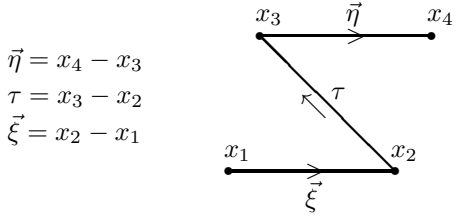
\put(2.9,2.3){$\nwarrow$} \put(-3.5,-.53){Figure 1: Lattice relative
coordinates.}
\end{picture}
\vspace{.6cm}\end{center}

Analogously, reinstating the linear combinations, for $M_\L$, we
obtain the spectral representation
$$\barr{lll}\hspace{-.06cm}\tilde
M_{\ell_1s_2s_3s_4}(k)\!\!\!&\!=\!\!&
 \tilde M_{s_1s_2s_3s_4} (\vec k)
-(2\pi)^3\int_{-1}^1\int_{{\bf
T}^3}\,f(k^0,\la^0)\vspace{.14cm}\\&&\!\times\,\delta (\vec k-\vec
\la)\,d_{\la^0} d_{\vec \la} ( \bar \tau_{s_1}(1/2,\vec u_1)\bar
\tau_{s_2}(1/2,\vec u_2),\vspace{.14cm}\\\!&&\times\,{\cal
E}(\la^0,\vec \la)\bar \tau_{s_3}(1/2,\vec u_3)\bar
\tau_{s_4}(1/2,\vec u_4))_{{\cal H}}\,,\earr$$where, for $x^0\not=
0$, and with $x=(x^0=v^0-u^0,\vec x)\in\mathbb{Z}^4$,
 $$\tilde
M_{s_1s_2s_3s_4}(\vec k)=\sum_{\vec x\in{\mathbb {Z}}^3} e^{-i\vec
k.\vec x }M_{s_1s_2s_3s_4}(x^0=0,\vec x)\,.$$

 The singularities of $\tilde
M_{s_1s_2s_3s_4}$, for $k=(k^0=i\chi, \vec k=\vec 0)$ and
$e^{\pm\chi}\leq 1$ are points in the mass spectrum. To determine
$\tilde M_{s_1s_2s_3s_4}$ we employ a lattice version of the B-S
equation, as discussed below. To detect baryon-baryon bound states
below the lowest two-baryon threshold, we write $k^0=i(2\bar
m-\epsilon)$, where $\bar m$ is the smallest baryon mass and
determine the value of $\epsilon >0$ that gives the singularity. We
call this special value of $\epsilon$ the binding energy of the
corresponding bound state.
\subsection{The B-S equation, the ladder approximation and $I=0$ bound states}\lb{sec43}
Next, we write a B-S equation for $M$ as,
\bequ\lb{BS}M=M_0+M_0KM\,,\eequ and in contrast to the continuum
case, we use what we call an equal time representation.

In terms of kernels, with $x^0_1=x^0_2$ and $x^0_3=x^0_4$, we
have$$\barr{lll}\!
\!M(x_1,x_2,x_3,x_4)\!\!\!&=&\!\!\!M_0(x_1,x_2,x_3,x_4)\!\!+\!\!\int\!\!
M_0(x_1,x_2,y_1,y_2)\vspace{.14cm}\\&&\!\!\!\times\,K(y_1,y_2,y_3,y_4)
M(y_3,y_4,x_3,x_4)\vspace{.14cm}\\&&\!\!\!\times\,
\delta(y^0_1-y^0_2) \delta(y^0_3-y^0_4) dy_1 dy_2 dy_3 dy_4,\earr$$
where we use a continuum notation for sums over lattice points.
$K(y_1,y_2,y_3,y_4)$ is called the B-S kernel. We remark that in the
imaginary time continuum quantum field theory formulation with
Euclidean invariance all the B-S equation coordinates are taken to
run over all spacetime.

Formally, this B-S equation defines $K=M_0^{-1}-M^{-1}$. Our
strategy is to compute $K$ in the leading $\kappa$ order, which we
call a ladder approximation $L$ to $K$, and then solve the B-S
equation for $M$ using $L$. In the B-S equation, $M_0$ is obtained
from $M$ by erroneously applying Wick's theorem to the one-baryon
composite fields $\tilde B$ in $M$ such as,
$$\barr{l}-\langle B_{\ell_1}(x_1)\bar
B_{\ell_3}(x_3)\rangle\,\langle B_{\ell_2}(x_2)\bar
B_{\ell_4}(x_4)\rangle\;\vspace{.14cm}\\+\langle B_{\ell_1}(x_1)\bar
B_{\ell_4}(x_4)\rangle\;\langle B_{\ell_2}(x_2)\bar
B_{\ell_3}(x_3)\rangle\,,\earr$$for the term $\langle
B_{\ell_1}(x_1) B_{\ell_2}(x_2)\bar B_{\ell_3}(x_3) \bar
B_{\ell_4}(x_4)\rangle$ in $M_\L(x_1,x_2,x_3,x_4)$. With this, we
obtain \bequ\lb{m0}\barr{ll}
F_{0,\L}(x_1,x_2,x_3,x_4)\!\equiv&\!\!-G_{\ell_1\ell_3}(x_1,x_3)G_{\ell_2\ell_4}(
x_2,x_4)\vspace{.15cm}\\&\!\!+G_{\ell_1\ell_4}(x_1,x_4)
G_{\ell_2\ell_3}(x_2,x_3)\,.\earr\eequ By referring to the
definition of the two-baryon function $G_{\ell_1\ell_4}$ of Eq.
(\ref{GG}), we see that Eq. (\ref{m0}) holds for all temporal
orderings.

In terms of the $(\vec \xi,\vec \eta,\tau)$ lattice relative
coordinates and taking the Fourier transform in $\tau$ only, the B-S
equation becomes (see \cite{SO}) $$\barr{ll}\hat M(\vec \xi,\vec
\eta,k)=&\hat M_0(\vec \xi,\vec \eta,k)+\int \,\hat M_0(\vec
\xi,\vec \xi\hspace{.045cm}',k) \vspace{.15cm}\\&\times\:\hat
K(-\vec \xi\hspace{.045cm}',-\vec \eta\hspace{.045cm}',k)\hat M(\vec
\eta\hspace{.045cm}',\vec \eta,k)\, d\vec\xi\hspace{.045cm}'d\vec
\eta\hspace{.045cm}'\,.\earr$$

The above B-S equation corresponds, roughly speaking, to an equation
for the kernel of a one-particle lattice Schr\"odinger operator
resolvent equation. The B-S kernel $\hat K(-\vec
\xi\hspace{.045cm}',-\vec \eta\hspace{.045cm}',k)$, in general, acts
as an energy-dependent, non-local potential (see Refs.
\cite{2flavor2baryon,2flavor2meson}). $\hat M_0$ is like the
resolvent of the dressed noninteracting particles and $\hat M$ is
the interacting resolvent. To facilitate the analysis, it is
desirable to have $\hat K$ with the shortest range as possible. For
example, in weakly coupled scalar models, with local interactions
and in the weak coupling regime, as well as in other spin and
stochastic systems, the dominant contribution to $\hat K$ is local,
and the ladder approximation dominates in the full model (see Refs.
\cite{DE,CMP2,SO}). In general, the hyperplane decoupling method is
used to obtain decay properties of $K$, which allows us to control
the solution to the B-S equation.

To make sense of the B-S equation and the kernel $K$, we now define
the space where $M$ and $M_0$ act as matrix operators, which is
given by $\ell^\prime_2=\ell^s_2\oplus\ell^a_2$, such that
$f=f_1+f_2\in\ell^\prime_2$, $f_1$ $(f_2)$ is associated with the
first $13$
 (last $7$) components; $f_1\in\ell^s_2(A_{\sl s})$,
$f_2\in\ell^s_2(A_{\sl a})$, where ${\sl s}$ (${\sl a}$) means the
symmetric (antisymmetric) subspace and $$A_{\sl
s}\!=\!\{(x_1,x_2;j)\in
\mathbb{Z}^4\times\mathbb{Z}^4\times(1,2,...,13)|
x^0_1\!=\!x^0_2\}\,,$$ and $$A_{\sl a}\!=\!\!\{(x_1,x_2;j)\in
\mathbb{Z}^4\!\times\mathbb{Z}^4\!\times(14,...,20)|
x^0_1\!=\!x^0_2, \vec x_1\!\not=\!\vec x_2\}.$$

Using the orthogonality properties of the two-point functions,
in the total spin basis, we have
$$\barr{ll}
M^{(0)}_{0;S,S_z,S^\prime,S_z^\prime}(x_1,x_2;x_3,x_4) =&
\delta_{SS^\prime}\delta_{S_zS_z^\prime}[-\delta(x_1-x_3)\;\times
\\&\delta(x_2-x_4)
+(-1)^S\;\times \\&\delta(x_1-x_4)\delta(x_2-x_3)]\;,\earr$$and the
same for $M^{(0)}_{0;T,T_z,T^\prime,T_z^\prime}$, $T=0,1,2,3$.
Hence, $M_0^{(0)}$ acts as $-21\!\!1\not= 0$ on $\ell_2^\prime$, and
$M_0$ is invertible. To define the B-S operator $K=M_0^{-1}-M^{-1}$,
we also need $M^{-1}$. For $M(x_1,x_2,x_3,x_4)$, setting
$$M(0)\equiv M(x,x,x,x)=M(0,0,0,0)\,,$$ we find that the $\kappa^0$
order is given by
$$\barr{ll}M^{(0)}(x_1,x_2,x_3,x_4)\!=&\!M^{(0)}(0)\delta(x_1-x_2)
\delta(x_1-x_3)\vspace{.14cm}\\&\times\,
\delta(x_1-x_4)+M^{(0)}_0[1-\vspace{.14cm}\\&\!
\delta(x_1-x_2)\delta(x_1-x_3)\delta(x_1-x_4)]\,,\earr$$ such that
$M^{(0)}$ agrees with $M^{(0)}_0$, except at coincident points.
Hence, to show $M$ is invertible we need the invertibility of the
matrix $M^{(0)}(0)$ which we analyze next.

By a lengthy computation given in Appendix C, we find that, in the
total spin basis, \bequ\lb{Mdecompo}M^{(0)}(0)=M^{(0)}_1(0)\oplus
M^{(0)}_1(0)\oplus M^{(0)}_1(0)\oplus M^{(0)}_b(0)\,,\eequ where
\bequ\lb{Mdec2}M^{(0)}_1(0)=-\dfrac19{\small \left( \barr{cc} 20&-8
\sqrt{5}\vspace{.1cm}\\-8 \sqrt{5}&16\earr \right)}\,,\eequ and
$M^{(0)}_b(0)$ acts as $-41\!\!1$ on the components with indices
from 7 to 13. The eigenvalues of $M^{(0)}_1(0)$ are $-4$ and $0$.
The null space prohibits the invertibility of $M^{(0)}(0)$. The zero
eigenvector is $\frac 13\left(\barr{c}2\\\sqrt{5}\earr\right)$ and
the first line of the eigenvalue equation is $$\langle {\cal
T}_{S=1,S_z}[{\cal T}_{S=1,S_z} +\frac{\sqrt{5}}2 {\cal
T}_{T=1,T_z}] \rangle^{(0)}=0\,.$$ The vanishing of this average
suggests that there is a pointwise linear dependence relation at
coincident points. Indeed, this turns out to be the case, and is
proven in Appendix C that, at coincident points, \bequ\lb{LD}{\cal
T}_{S=1,S_z} +\frac{\sqrt{5}}2 {\cal T}_{T=1,T_z}=0\,,\eequ for
$S_z=T_z=-1,0,1$.

The truncated four-point correlation $M$ still has a null space in
the symmetric subspace of $\ell_2^\prime$. For this reason, we are
forced to reduce the symmetric subspace for coincident points only,
and we take the restriction of $M^{(0)}(0)$ to the orthogonal
complement of the null space in the symmetric subspace. So, for
coincident points we have a $10$ dimensional space. For
non-coincident points, $M^{(0)}$ agrees with $M^{(0)}_0\not= 0$, so
that $K$ is well-defined.

With this definition of $K$, and using its decay properties shown in
Appendix D, there is a local potential, and energy-dependent space
range-zero potential, as well as an energy-independent range-one
potential. As in Ref. \cite{2flavor2baryon}, we modify $M$ at
coincident points by a multiplicative constant $1/2$. Precisely, we
replace $M(x_1,x_2,x_3,x_4)$ by
$$M^\prime(x_1,x_2,x_3,x_4)=h(x_1,x_2) M(x_1,x_2,x_3,x_4)
h(x_3,x_4)\,,$$ where $$h(x,y)=[1 - \delta(x -y)]  + \frac
1{\sqrt{2}} \delta(x-y)\,,$$ and the B-S equation becomes
\bequ\lb{modifBS}M^\prime=M_0+M_0K^\prime M^\prime\,.\eequ The
$\tau$ variable Fourier transforms $\hat M$ and $\hat M^\prime$
clearly have the same singularities. For simplicity of notation, we
drop the prime and keep using $M$, $\hat M$, $K$ and $\hat K$.

Next, to define our ladder approximation, we write a Neumann
expansion representation for $K=M_0^{-1}-M^{-1}$. With $$\delta M
=M-M^{(0)}\quad ,\quad\delta M_0=M_0-M_0^{(0)}\,,$$ we have
\bequ\barr{ll}K=&\sum_{n=0}^\infty(-1)^n [ ( M^{(0)-1}_0\delta
M_0)^n M^{(0)-1}_0\vspace{.14cm}\\&- (M^{(0)-1} \delta M)^n
M^{(0)-1} ]\,.\earr\lb{n}\eequ For small $\kappa$, the dominant
contributions to $K$ come from coincident points at $\kappa=0$ and a
space range-one potential of order $\kappa^2$. As discussed in
Appendix E, it turns out that, because of cancelations
$K^{(0)}(0)=(M^{(0)}_0)^{-1}(0)-(M^{(0)})^{-1}(0)=0$. Surprisingly,
there is an attractive gauge field correlation effect which is
responsible for this result. More explicitly, as a contribution to
the definition of $h(x,y)$, we must consider the effect of a gauge
integral over six overlapping bonds with the same orientation, which
is given by ${\cal I}_6$ in Eq. (\ref{I6}). Moreover, it is worth
mentioning that if these six gauge fields are uncorrelated three by
three, meaning that ${\cal I}_6$ is replaced by $({\cal I}_3)^2$,
${\cal I}_3$ being the integral over three overlapping bonds with
the same orientation [see Eq. (\ref{I3})], the associated attractive
contribution to $K^{(0)}(0)$ becomes repulsive, leading to an
overall repulsive $K^{(0)}(0)$.

Going further in the $\kappa$ expansion, since $\delta M$ is ${\cal
O}(\kappa^2)$ and $\delta M_0$ is ${\cal O}(\kappa^3)$, we find that
the dominant contribution to $K$ is given by the $n=1$ term in the
Neumann expansion of Eq. (\ref{n}). This defines our ladder
approximation to $K$ and is given by \bequ \lb{L1}L\equiv
K^{(2)}\kappa^2=(M^{(0)})^{-1} (\delta M)^{(2)}
(M^{(0)})^{-1}\kappa^2\,.\eequ

To obtain an explicit formula for the kernel of $L$ of Eq.
(\ref{L1}), first we consider the contribution to $L$ coming from
the non-oriented geometric bond linking the sites $0$ and $e^1$. It
is given by
$$\barr{ll}\!\!K^{(2)}(0,e^1;0,e^1)&\!\!\!\!+
K^{(2)}(0,e^1;e^1,0)\!=\!(M^{(0)})^{-1}\!(0,e^1;0,e^1)
\vspace{.15cm} \\&\!\!\!\!\times
\;M^{(2)}(0,e^1;0,e^1)(M^{(0)})^{-1}(0,e^1;0,e^1)
\vspace{.15cm}\\&\!\!\!\!+(M^{(0)})^{-1}(0,e^1;e^1,0)
M^{(2)}(0,e^1;e^1,0)\vspace{.15cm}\\&\!\!\!\!\times
\;(M^{(0)})^{-1}(0,e^1;e^1,0)\,. \earr $$ Since
$M^{(0)}=M^{(0)}_0=-2\;1\!\!1$, for non-coincident points, each
$(M^{(0)})^{-1}$ factor can be replaced by $-1/2$. In the total spin
basis, taking into account the space coordinate exchange symmetry
(anti-symmetry) of ${\cal T}$ for $S=1, T=1,3$ ($S=0, T=0,2$) given
in Eq. (\ref{par}), the second term is the same as the first one.

The contribution $M^{(2)}(0,e^1;0,e^1)\,\kappa^2$ is obtained by
expanding the hopping term exponential in the action appearing in
the numerator, which brings down a pair of oppositely oriented
overlapping gauge field bounds. The gauge field integral of Eq.
(\ref{I2}) is performed using the Peter-Weyl theorem (see Ref.
\cite{Creu2} for other more complicated and useful gauge integrals
and Appendix D), and gives a factor $1/3$. In this way, the $11$
element $M^{(2)}(0,e^1;0,e^1)$ of the matrix $M^{(2)}$ is
$$M^{(2)}_{11,\S}(0,e^1;0,e^1)
=\dfrac 1{12} v^{(2)}_{\S}\,,$$ where, in the individual spin basis,
\bequ\lb{v}\barr{ll}\!\!
v^{(2)}_{s_1,s_2,s_3,s_4}\!=\!\!&\!\!\displaystyle\sum\,^\prime
c^{\small \frac 12\,\frac12}_{I_1I_2} c^{\small \frac
12\,\frac12}_{I_3I_4} \,\times\vspace{.14cm}\\&\!\!\langle
B_{{\small \frac12}I_1s_1}\bar B_{{\small \frac12}I_3s_3}
\bar\psi_{a_1\alpha_1f_1}
\psi_{a_1\beta_2f_2}\rangle^{(0)}\,\times\vspace{.14cm}\\&\!\!\langle
B_{{\small \frac12}I_2s_2}\bar B_{{\small \frac12}I_4s_4}
\bar\psi_{a_2\alpha_2f_2} \psi_{a_2\beta_1f_1}\rangle^{(0)}
\Gamma^{e^1}_{\alpha_1\beta_1}
\Gamma^{-e^1}_{\alpha_2\beta_2},\earr\eequ with all fields at a same
site and where the prime here means the sum over $I_1, ...,I_4$,
with $I_1+I_2=0=I_3+I_4$. A similar expression holds for the
components $12$ ($s_1s_2t_3t_4$), with $(1/2,I_3,s_3)$ and
$(1/2,I_4,s_4)$, respectively, replaced by $(3/2,J_3,t_3)$ and
$(3/2,J_4,t_4)$; with $(1/2,I_1,s_2)$ and $(1/2,I_2,s_2)$,
respectively, replaced by $(3/2,J_1,t_1)$ and $(3/2,J_2,t_2)$ for
the component $21$ ($t_1t_2s_3s_4$); and with $(1/2,I_1,s_1)$,
$(1/2,I_2,s_2)$, $(1/2,I_3,s_3)$ and $(1/2,I_4,s_4)$, respectively,
replaced by $(3/2,J_1,t_1)$, $(3/2,J_2,t_2)$, $(3/2,J_3,t_3)$ and
$(3/2,J_4,t_4)$ for the component $22$ ($t_1t_2t_3t_4$).

Taking into account the other directions and using translation and
rotation invariance, we obtain for the kernel of $K^{(2)}$, in the
total spin basis
\bequ\barr{ll}K^{(2)}(x_1,x_2;x_3,x_4)=&\frac{1}{24} v^{(2)}
\sum_{\sigma,j}\delta(x_2-x_1-\sigma
e^j)\;\times\vspace{.14cm}\\&\delta(x_1-x_3)\delta(x_2-x_4)\delta(x^0_2-x^0_3)
\,,\lb{K22}\earr\eequ where here $v^{(2)}$ is the total spin basis
matrix.

By a lengthy computation (see Appendix E),  in the individual spin
basis, the symmetric matrix $v^{(2)}=[v_{ij}]_{i,j=1,\ldots , 20}$
has the following nonzero elements: $v_{11}=-1/3=v_{44}$;
$v_{17}=-4/\sqrt{6}=v_{1\,13}=v_{4\,12}=v_{4\,18}$;
$v_{1\,10}=4\sqrt{2}/3=v_{4\,15}$; $v_{22}=10/3=v_{33}$;
$v_{23}=-11/3$; $v_{28}=2\sqrt{2}=v_{3\,17}$; $v_{2\,
11}=-4\sqrt{2}/3=v_{3\,14}$; $v_{3\, 11}=2\sqrt{2}/3= v_{2\,14}$;
$v_{55}=-3=v_{20\,20}$; $v_{66}=-2=v_{99}=v_{16\,16}=v_{19\,19}$;
$v_{69}=-1=v_{77}=v_{8\,11}=v_{12\,12}=v_{13\,13}=v_{14\,17}=
v_{16\,19}=v_{18\,18}$;
$v_{7\,10}=-2\sqrt{3}=v_{10\,13}=v_{12\,15}=v_{15\,18}$;
$v_{10\,10}=-5/3=v_{15\,15}$; $v_{11\,
11}=-4/3=v_{11\,14}=v_{14\,14}$.

It turns out that $v^{(2)}$ has an especially simple, almost
diagonal form, in the total spin basis. Using the basis
transformation ${\cal A}$ given in Section 4.1, it is written as
\bequ\lb{vv2}v^{(2)}=W\oplus W\oplus W\oplus W_{1d}\oplus W_2\oplus
W_{2d}\,,\eequ with $W=\dfrac13{\small
\left(\barr{cc}-1&-\sqrt{80}\\-\sqrt{80}&1\earr \right)}$ (with
eigenvalues $\pm 3$), $W_2={\small \left(\barr{cc}7&4\\4&1\earr
\right)}$ (with eigenvalues $-1$ and $9$), $W_{1d}=-3P_{13}$,
$W_{2d}=-P_{22}$, where $P_{13}$ is the orthogonal projection on the
$7^{th}$ through $13^{th}$ components and $P_{22}$ on the $16^{th}$
through $20^{th}$.

The interaction present in Eq. (\ref{v}) arises from a
quark-antiquark exchange but, by inspection of the meson particle
fields (see Refs. \cite{JMP,2flavor2meson}), this is {\em not} a
meson particle exchange. To see this, consider the field averages in
Eq. (\ref{v}). The spin indices of the $B$ and $\bar B$ fields are
both lower. This forces the spin indices of the $\bar \psi \psi$
fields, in each of the two averages, to be either both lower or both
upper. A true meson particle has one upper and one lower spin index.
We refer to this interaction as a {\em quasi-meson exchange}.

Using the total spin basis representation for $v^{(2)}$, we return
to the determination of bound states and solve the B-S equation
using $L$ (see Refs. \cite{2flavor2baryon,2flavor2meson} for
details). In terms of relative coordinates, we have the space
range-one energy-independent potential \bequ\lb{lhat}\hat
L(\xi,\eta,k^{0})=\kappa^2 v^{(2)}\sum_{j,\sigma}\delta(\vec
\eta-\vec \xi)\delta(\sigma e^j-\vec \xi)\,,\eequ and the B-S
equation, in the ladder approximation, has the solution,
\bequ\lb{solucao}\barr{lll}\hat M(\vec \xi,\vec \eta, k^0)&=&\hat
M_0(\vec \xi,\vec \eta) +\sum_{\sigma_1,\sigma_2,j_1,j_2} \hat
M_0(\vec \xi,\sigma_1\vec
e^{j_1})\times\vspace{.15cm}\\&&\kappa^2v^{(2)}\left(1-\kappa^2v^{(2)}\breve{M}_0\right)^{-1}\!\!(\sigma_1\vec
e^{j_1},\sigma_2\vec e^{j_2})\times\vspace{.15cm}\\&& \hat
M_0(\sigma_2\vec e^{j_2},\eta)\,,\earr\eequ where the $\breve{M}_0$
lattice indices are restricted to the spatial nearest neighbors of
zero, and we suppress the $k^0$ dependence and spin indices on the
r.h.s. Reinstating the spin indices, the resulting matrix is
$120\times 120$, where we have implicit matrix multiplication in the
spin indices $\breve{M}_0$ means $\hat{M}_0$ with indices restricted
by the multiplication of $v^{(2)}$ and we suppress $k^0$ dependence
on the r.h.s.

To go further, we need properties of $\hat M_0$. In Appendix F,
using the spectral representation for the two-baryon functions in
Eq. (\ref{m0}), after passing to lattice relative coordinates, we
derive a general spectral representation for $\hat M_0$. Assuming
that the two-baryon function is diagonal in the spin indices in the
individual spin basis [which holds up to and including ${\cal
O}(\kappa^5)$, see Ref. \cite{CMP}, we show that the dominant
contribution is given by, in the total spin basis,
\bequ\lb{m00}\barr{l} \!\hat M_{0,SS_zS^\prime S_z^\prime}\!(\vec
\xi,\vec \eta,k^0)\!=\!\!\left[ -2(2\pi)^{-3}\!
\int_{{\mathbb{T}}^3} \tilde G(\vec p) \tilde G(\vec p)d\vec
p\right.\vspace{.1cm}\\\left.\qquad-2(2\pi)^{3}\int_{-1}^1\int_{-1}^1
\int_{{\mathbb{T}}^3} f(k^0,\la^0\la^{\prime 0})
d_{\la^0}\alpha_{\vec
p}(\la^0)\right.\vspace{.1cm}\\\qquad\left.\times\,d_{\la^{\prime
0}}\alpha_{\vec p}(\la^{\prime 0})\,\Xi(\vec p,\vec\xi,\vec
\eta)d\vec p\;\right]\,\delta_{S S^\prime}\,\delta_{S_z
S_z^\prime}\,,\earr \eequ where $$\Xi(\vec p,\vec\xi,\vec
\eta)\!=\!\delta_{S1}  \cos \vec p.\vec \xi\cos \vec p.\vec
\eta+\delta_{S0}\sin \vec p.\vec \xi\sin \vec p.\vec\eta\;;$$ and
the same for $T=1,3$ ($T=0,2$) replacing $S=1$ ($S=0$).

To continue, we set $k^0=i(2\bar m-\epsilon)$, where we recall that
$\bar m$ denotes the minimum of the baryon masses and $\epsilon >0$
is the bound state binding energy. Recalling the definition of
$p_\ell^2$ given in Eq. (\ref{pl}), we further make the following
approximations: a) we retain only the product of one-particle
contributions; b) $w(\vec p)-m\approx \kappa^3p_{\ell}^2/8$; c)
$\bar m\approx -3\ln \kappa$ and d) $Z(\vec p)\approx
-(2\pi)^{-3}e^{-w(\vec p)}\approx -(2\pi)^{-3}\kappa^3$. The
approximate $\hat M_0$ becomes \bequ\lb{mmm}\barr{lll}\hat
M_{0,SS_zS^\prime S_z^\prime}&=&-\dfrac
2{(2\pi)^3}\int_{\mathbb{T}^3} \,\left[
1+\dfrac1{e^{\kappa^3p_\ell^2/4+\epsilon}-1}
\right]\vspace{.15cm}\\&&\times\; \Xi(\vec p,\vec\xi,\vec
\eta)\,\!d\vec p\,\,\delta_{S S^\prime}\,\delta_{S_z
S_z^\prime}\,.\earr\eequ

With all these approximations, the singularities of $\hat M$, below
the two-baryon threshold $2\bar m$, occur as zeroes of the
determinant of the $120\times 120$ matrix $1-\kappa^2v^{(2)}\hat
M_0(\sigma_1\vec e^{j_1},\sigma_2\vec e^{j_2})$. Anticipating that
the binding energy $\epsilon$ is of order $\kappa^2$, we also
approximate the denominator in $\hat M_0$ to get $$\hat
M_{0,SS_zS^\prime S_z^\prime}(\vec \xi,\vec \eta,k^0)\!\approx\!
-\frac2{(2\pi)^{3}}(1-e^{-\epsilon})^{-1}\!\int_{{\mathbb{T}}^3}\!\Xi(\vec
p,\vec\xi,\vec \eta)\,\!d\vec p,$$ and the same for $T$'s instead of
$S$'s. These approximations are consistent with finding a bound
state solution with binding energy greater than ${\cal
O}(\kappa^3)$. Note that $\hat M_0(\sigma_1 e^{j_1},\sigma_2
e^{j_2})$ is diagonal in $j_1$ and $j_2$, and that the integral
takes the values $\pm (2\pi)^{3}/2$. Thus, the bound state condition
becomes \bequ\det
[1+4(1-e^{-\epsilon})^{-1}v^{(2)}\kappa^2]=0\,,\lb{bsc}\eequ for a
${20\times 20}$ matrix. Next, using the spectral representation for
$v^{(2)}$, if $\lambda$ is a negative eigenvalue, then the bound
state condition is
$$1+4\lambda(1-e^{-\epsilon})^{-1}\kappa^2=0\,,$$ with approximate
solution $\epsilon=4|\lambda|\kappa^2$, which is the bound state
binding energy. Hence, we have a bound state for each negative
eigenvalue of $v^{(2)}$, with multiplicity given by the space
dimension $3$. For these bound states we have attractive potential
wells at $\pm e^j$, $j=1, ...,d=3$, which would give us a
multiplicity $2d$. However, we only have symmetric or antisymmetric
wave functions, which reduces the multiplicity to $d=3$.

Referring to the spectral decomposition of $v^{(2)}$ in the total
spin basis, and recalling the ordering of the total spin basis, we
see that the most strongly bound, bound states ($\lambda=-3$) are
associated with a superposition of $p-n$ and $\Delta-\Delta$ total
spin $1$ states; and also the $\Delta-\Delta$ total spin $3$ states.
The more weakly bound, bound states ($\lambda=-1$) are associated
with the superposition of $p-n$ and $\Delta-\Delta$ total spin $0$
states; as well as the $\Delta-\Delta$ total spin $2$ states.

We now explain the superposition of two-baryon states occurring in
the bound state associated with the superposition of $p$-$n$ and
$\Delta-\Delta$. Similar considerations apply to the other bound
states. Going back to the decomposition of $v^{(2)}$ given in Eq.
(\ref{vv2}), this state is associated with the negative eigenvalue
$-3$ of one of the three $2\times 2$ $W$'s. Consider the first $W$
and label the basis elements $(S,S_z)=(1,1)$, $(T,T_z)=(1,1)$ by
$\sigma=1,2$. The corresponding eigenvector is
$w_{-3}=(\sqrt{5}/3,2/3)$. If we start from the B-S equation for
$(w_{-3},Mw_{-3})$, it reduces to a single equation with $v^{(2)}$
substituted by $-3$. Here, we have used the fact that $\hat M_0$
commutes with $v^{(2)}$ in our approximation. But $(w_{-3},Mw_{-3})$
can be written, for $x_1^{(0)}=x_2^{(0)}\leq x_3^{(0)}=x_4^{(0)}$,
as the average $$\begin{array}{l} \langle \, [(\sqrt{5}/3) {\cal
T}^{1/2}_{S=1,S_z=1}(x_1,x_2)+(2/3){\cal
T}^{3/2}_{T=1,T_z=1}(x_1,x_2)]\vspace{.15cm}\\ \;\times\:
[(\sqrt{5}/3) \bar{\cal T}^{1/2}_{S=1,S_z=1}(x_3,x_4)+(2/3)\bar
{\cal T}^{3/2}_{T=1,T_z=1}(x_3,x_4) ]\,\rangle,
\end{array}$$so that the superposition of two-baryon states given by
the field $[(\sqrt{5}/3) \bar{\cal
T}^{1/2}_{S=1,S_z=1}(x_1,x_2)+(2/3)\bar {\cal
T}^{3/2}_{T=1,T_z=1}(x_1,x_2) ]$ more closely describes the field
which creates the bound state.
\subsection{Bound states for the $I=3$ sector}\lb{sec44}
In this subsection, we show that the application of our method to
the maximum total isospin $I=3$ sector (freezing all isospins to
$+$) leads to {\em no} baryon-baryon bound states if $S=2,3$, but
bound states do appear if $S=0,1$. The analysis here parallels that
of the $I=0$ case, but it is algebraically much simpler, and we will
be brief.

The one-baryon states are associated with the fields $\bar
B_t\equiv\bar B_{{\small \frac32}{\small\frac32}t}$,
$t=\pm3/2,\pm1/2$ of Eq. (\ref{baryon5}). The two-baryon states that
we consider here, in the total $I=3$, $I_z=3$ sector, are those
generated by the product of the above one-baryon fields
$\bar\tau_{t_1t_2}(x_1,x_2)$, $x^0_1=x^0_2$. The bound states are
formed with a quasi-meson exchange $\kappa^2$ space range-one
potential which dominates and, in our ladder approximation, the
space range-zero potential is zero.

After interchanging spin and isospin, the calculation of this
potential is the {\em same} as for the lattice QCD model in $2+1$
dimensions and  with $2\times 2$ spin matrices and two flavors (see
Ref. \cite{2flavor2baryon}), and the bound state spectral results
are the same, apart from the additional space dimension.

In the total spin basis, the two-baryon states are given by
$${\cal T}_{TT_z}(x_1,x_2)=\sum_{t_1+t_2=T_z}\,c_{t_1t_2}^{TT_z}
\tau_{t_1t_2}(x_1,x_2)\,,$$ where $c_{t_1t_2}^{TT_z}$ are the C-G
coefficients for the coupling of two spin $3/2$ states to give total
spin $T=3,2,1,0$, and $z$-component $T_z$ of total spin. They obey
the symmetry property
$$
{\cal T}_{TT_z}(x_1,x_2)=(-1)^T\,{\cal T}_{TT_z}(x_2,x_1)\,,$$ for
$T=3,1$; for $T=0,2$ ($1,3$) they are space symmetric
(antisymmetric). So, at coincident points, they vanish for $T=3,1$;
for $T=0,2$ ($1,3)$ they are symmetric (anti-symmetric). Also, from
the pointwise relations $\tilde B_{\pm\frac32}\tilde B_{\pm{\small
\frac32}}=\tilde B_{\pm{\small \frac32}}\tilde B_{\pm{\small
\frac12}}=\tilde B_{\pm{\small \frac32}}\tilde B_{\mp{\small
\frac12}}=\tilde B_{\pm{\small \frac12}}\tilde B_{\pm{\small
\frac12}}=0$, we find that ${\cal T}_{T=2\,T_z}$ is also zero at
coincident points. A computation gives the coincident point linear
dependency relation $\tilde B_{{\small\frac32}}\tilde B_{-{\small
\frac32}}=-\tilde B_{{\small\frac12}}\tilde B_{-{\small \frac12}}$,
and we find that $\tilde {\cal T}_{00}=2\tilde
B_{{\small\frac32}}\tilde B_{-{\small\frac32}}$.

The four-point correlation in the unmodified, shifted form, and in
the individual spin basis, is taken as, with $x^0_1=x^0_2$,
$x^0_3=x^0_4$, and  for $x^0_1\leq x^0_3$,
$$M_{t_1t_2t_3t_4}\!(x_1,x_2,x_3,x_4)\!=\!\langle B_{t_1}(x_1\!)
B_{t_2}(x_2\!) \bar B_{t_3}(x_3\!)\bar B_{t_4}(x_4\!)\rangle\,\!,$$
and with the usual interchange of barred and unbarred fields and
taking the complex conjugate of the average, for $x^0_1> x^0_3$. The
associated uncorrelated or {\em Wickified} four baryon correlation
is, for $x^0_1\leq x^0_3$,
$$\barr{ll}M_{0,t_1t_2t_3t_4}\!\!\!\!&(x_1,x_2,x_3,x_4)=
\vspace{.15cm}\\&-\,\langle B_{t_1}(x_1)
 \bar B_{t_3}(x_3) \rangle\;\langle B_{t_2}(x_2)\bar B_{t_4}(x_4)\rangle
 \vspace{.15cm}\\&+\,\langle B_{t_1}(x_1)
 \bar B_{t_4}(x_4)\rangle\;\langle B_{t_2}(x_2)\bar
 B_{t_3}(x_3)\rangle \,.\earr$$

The four-baryon correlations in the total spin basis are obtained by
taking the corresponding C-G linear combinations. For
$T=T^\prime=T_z=T_z^\prime=0$, we obtain $M^{(0)}_{TT_zT^\prime
T_z^\prime}(0)=-4$ and $M^{(0)}_{0,TT_zT^\prime T_z^\prime}(0)=-2$.

As in Section 4.3, an $h$-modified B-S equation and operator
kernels are introduced. 
Then, the energy-independent space range-zero potential
$K^{(0)}=[M_0^{(0)}]^{-1}-[M^{(0)}]^{-1}=0$. Analogous to Eq.
(\ref{L1}), the dominant interaction is a $\kappa^2$ quasi-meson
space range-one potential which we take as our ladder approximation
$L=K^{(2)}\kappa^2$ to $K$. Here, with $K^{(2)}$ and $v^{(2)}$ in
the total spin basis, as in Eq. (\ref{K22})
$$\barr{ll}K^{(2)}(x_1,x_2;x_3,x_4)=&\frac{1}{24} v^{(2)}
\sum_{\sigma,j}\delta(x_2-x_1-\sigma
e^j)\;\times\vspace{.14cm}\\&\delta(x_1-x_3)\delta(x_2-x_4)\delta(x^0_2-x^0_3)
\,,\earr $$ where, in the individual spin basis,
\bequ\lb{vvv}\barr{ll}
v^{(2)}_{t_1,t_2,t_3,t_4}=&\Gamma^{e^1}_{\alpha_1\beta_1}
\Gamma^{-e^1}_{\alpha_2\beta_2}\,\langle B_{t_1}\bar B_{t_3}
\bar\psi_{a_1\alpha_1f_1}
\psi_{a_1\beta_2f_2}\rangle^{(0)}\vspace{.15cm}\\&\times\,\langle
B_{t_2}\bar B_{t_4} \bar\psi_{a_2\alpha_2f_2}
\psi_{a_2\beta_1f_1}\rangle^{(0)} \,.\earr\eequ

Next, using similar arguments to those in Appendix E, we can write
$$\barr{ll}v^{(2)}_{t_1,t_2,t_3,t_4}=&\langle B_{t_1}\bar B_{t_3}
:\bar\psi_{a_1\alpha f_1} \psi_{a_1\beta
f_2}:\rangle^{(0)}\vspace{.15cm}\\&\times\:\langle B_{t_2}\bar
B_{t_4} :\bar\psi_{a_2\beta f_2} \psi_{a_2\alpha
f_1}:\rangle^{(0)}\,,\earr$$ where in the expansion of the
$\kappa=0$ average using Wick's theorem the Wick ordering $:\::$
forbids the contraction between the {\em explicit internal} $\bar
\psi$ and $\psi$ fields. This restriction implies that $f_1=f_2=+$.
In order to calculate $v^{(2)}_{t_1,t_2,t_3,t_4}$, we also take into
account spin constraints and use conjugation and the $\kappa=0$ spin
flip symmetry.

Doing this and suppressing the $+$ isospin indices of the $\tilde
\psi$, it is enough to calculate $$\barr{lll}\langle B_{{\small
\frac 12}}\bar B_{{\small \frac 12}} :\bar\psi_{a+}
\psi_{a+}:\rangle^{(0)}\!\!&=&\!\!-2=2 \langle B_{{\small \frac
12}}\bar B_{{\small \frac 12}} :\bar\psi_{a-}
\psi_{a-}:\rangle^{(0)}\,, \vspace{.15cm}\\\langle B_{{\small \frac
32}}\bar B_{{\small \frac 12}} :\bar\psi_{a+}
\psi_{a-}:\rangle^{(0)}\!\!&=&\!\!-\sqrt{3}\,,\vspace{.15cm}\\
\langle B_{{\small \frac 32}}\bar B_{{\small \frac 32}}
:\bar\psi_{a+} \psi_{a+}:\rangle^{(0)}\!\!&=&\!\!-3\,,\earr$$ and
$\langle B_{{\small \frac 12}}\bar B_{{\small \frac {-1}2}}
:\bar\psi_{a+} \psi_{a-}:\rangle^{(0)}=-2$.

In this way, in the individual spin basis, using the same ordering
as in the zero isospin case, for the $16\times 16$ submatrix, the
symmetric matrix $v^{(2)}=[v_{ij}]_{i,j=1,\ldots , 16}$ has the
following nonzero elements: $v_{11}=9=v_{16\,16}$;
$v_{22}=6=v_{55}=v_{12\,12}=v_{15\,15}$;
$v_{25}=3=v_{33}=v_{47}=v_{88}=v_{99}=v_{10\,13}=v_{12\,15}=v_{14\,14}$;
$v_{36}=2\sqrt{3}=v_{69}=v_{8\, 11}=v_{11\,14}$;
$v_{66}=5=v_{11\,11}$; $v_{77}=4=v_{7\,10}=v_{10\,10}$. The matrix
is easily diagonalized in the following reordered basis: $3$, $6$,
$9$, $4$, $7$, $10$, $13$, $8$, $11$, $14$, $2$, $5$, $12$, $15$,
$1$, $16$. Letting ${\cal V}$ denote this matrix, we have ${\cal
V}={\cal V}_1\oplus{\cal V}_2\oplus{\cal V}_1\oplus{\cal
V}_3\oplus{\cal V}_3\oplus{\cal V}_4\oplus{\cal V}_4$, with ${\cal
V}_{1,2,3,4}$ symmetric, and where ${\cal V}_1$ is $3\times 3$ with
nonvanishing elements ${\cal V}_{1;11}=3={\cal V}_{1;33}$, ${\cal
V}_{1;22}=5$, ${\cal V}_{1;12}=2\sqrt{3}={\cal V}_{1;23}$; ${\cal
V}_2$ is $4\times 4$ with nonvanishing elements ${\cal
V}_{2;22}=4={\cal V}_{2;33}={\cal V}_{2;23}$, ${\cal
V}_{2;12}=3={\cal V}_{2;34}$; ${\cal V}_3$ is $2\times 2$ with
nonvanishing elements ${\cal V}_{3;11}=6={\cal V}_{3;22}$, ${\cal
V}_{3;12}=3$; ${\cal V}_4$ is $1\times 1$ with nonvanishing elements
${\cal V}_{4;11}=9$.

Concerning their eigenvalues and eigenvectors $(T,T_z)$ (the total
and $z$-component of the total spin), we have: $9$ $(3,1)$, $3$
$(2,1)$ and $-1$ $(1,1)$ for ${\cal V}_1$ in the first spot; $9$
$(3,0)$, $3$ $(2,0)$, $-1$ $(1,0)$ and $-3$ $(0,0)$ for ${\cal
V}_2$; $9$ $(3,-1)$, $3$ $(2,-1)$ and $-1$ $(1,-1)$ for ${\cal V}_1$
in the second spot; $9$ $(3,2)$ and $3$ $(2,2)$ for ${\cal V}_3$
first spot; $9$ $(3,-2)$ and $3$ $(2,-2)$ for ${\cal V}_3$ in the
second spot; $9$ $(3,3)$ for ${\cal V}_4$ in the first spot; $9$
$(3,-3)$ for ${\cal V}_4$ in the second spot.

As we see, in the total spin basis, $v^{(2)}$ is diagonal.
Alternatively, it can be also diagonalized by the orthogonal
transformation relating the total and individual spin basis. Similar
to what occurs for the $I=0$ sector, we shall see that the negative
eigenvalues are associated with bound states. Referring to the
above, the negative eigenvalue $-3$ ($-1$) is associated with
$T,T_z=(0,0)$ ($T,T_z=(1,1),(1,0), (1-1)$).

Proceeding as in the $I=0$ case, in terms of lattice relative
coordinates, the space range-one  energy-independent potential in
the ladder approximation is, as in Eq. ({\ref{lhat}), given by
$$\hat L(\vec \xi,\vec\eta,k^{0})=\kappa^2 v^{(2)}\sum_{j,\sigma}\delta(\vec
\eta-\vec \xi)\delta(\sigma e^j-\vec \xi)\,,$$ with $v^{(2)}$ in the
total spin basis and the B-S equation has the solution given by Eq.
(\ref{solucao}). Here $\hat M_0(\vec \xi,\vec \eta,k^0)$ is derived
as in Appendix F and in the total spin representation is given as in
Eq. (\ref{m00}) with $S$, $S_z$ replaced by $T$, $T_z$, as well as
their primes and with now $$\barr{ll}\Xi(\vec p,\vec\xi,\vec
\eta)=&(\delta_{T2}+\delta_{T0})  \cos \vec p.\vec \xi\cos \vec
p.\vec \eta\vspace{.15cm}\\&+(\delta_{T3}+\delta_{T1})\sin \vec
p.\vec \xi\sin \vec p.\vec\eta\,.\earr$$

Finally, approximating $\hat M_0$ as before, for each negative
eigenvalue $\la$ of $v^{(2)}$ we have the bound state condition
$1+4\la(1-e^{-\epsilon})^{-1}\kappa^2=0$, with the approximate
solution of binding energy $\epsilon=4|\la|\kappa^2$. Hence, we only
have bound states of spin one and zero, with multiplicities $3$ and
$1$, respectively.
\section{Connection between the B-S equation and a lattice Schr\"odinger equation}\lb{sec5}
In this section, to understand our spectral results in a simple way,
we exploit the similarity between our lattice B-S equation for $\hat
M(\vec \xi, \vec\eta, k^0)$ and a one-particle Schr\"odinger
resolvent operator equation for the lattice Hamiltonian
$$H=H_0+V\,,$$ with a kinetic energy operator $H_0$ and a
potential energy operator $V$.

For $\kappa$ and $\epsilon$ small, we further approximate $\hat
M_0(\vec \xi, \vec\eta, k^0)$ of Eq. (\ref{mmm}) by $$\barr{lll}\hat
M_{0}&=&-\dfrac 2{(2\pi)^3}\dis\int_{\mathbb{T}^3}
\,\dfrac1{\kappa^3p_\ell^2/4+\epsilon}\, \Xi(\vec p,\vec\xi,\vec
\eta)\,\!d\vec p\,,\earr$$for $p_\ell$ given in Eq. (\ref{pl}) and
$\Xi$ depends on the total spin.

Consider the Schr\"odinger resolvent equation,
$$(H-z)^{-1} = (H_0-z)^{-1}-(H_0-z)^{-1}\,V\,(H-z)^{-1}\,,$$ for
$z$ not in the spectrum of $H_0$ and $H$.

Upon comparing with minus the B-S equation in our ladder
approximation, i.e. $-\hat M=-\hat M_0-\hat M_0\hat L\hat M$, we can
make the identifications
$$H_0=\frac{{\kappa}^3}8\,(-\Delta_\ell)\quad,\quad
\hat L=V\quad,\quad z=-\frac\epsilon 2\;\; (z<0)\,,$$ where
$\Delta_\ell$ is the lattice Laplacian with Fourier transform
$-p_\ell^2$. The Hamiltonian $H=H_0+V$ acts on the space
$\ell_{20}^2(\mathbb{Z}^3)$, which decomposes into a direct sum of
space-symmetric and antisymmetric subspaces depending on the total
spin indices (omitted here for simplicity of notation!). For $f\in
\ell_{20}^2(\mathbb{Z}^3)$, for each $x$, $f_R(x)$ is the $R$th
component where $R$ runs over the $20$ total spin basis indices. The
subspace is symmetric (antisymmetric) for $\Xi(\vec p,\vec\xi,\vec
\eta)=\cos \vec p.\vec\xi\,\cos \vec p.\vec\eta$ ($\sin \vec
p.\vec\xi\,\sin \vec p.\vec\eta$), corresponding to total spin
$S=1$, $T=1,3$ ($S=0$, $T=0,2$).

The kernel of $V$ is $$V(\vec\xi,\vec
\eta)=\kappa^2v^{(2)}\sum_{\sigma,j}\delta(\vec \xi-\sigma
e^j)\delta (\vec\xi-\vec \eta)\,,$$ and, for small $\kappa$, $V$
dominates $H_0$.

In this case, the negative energy bound state eigenfunctions of $H$
with eigenvalue $\lambda=z<0$ are those of $V$ which are easily seen
to have components $$f_R^{(k)}(x)=\alpha_R[\delta(x-e^k)\pm
\delta(x+e^k)]\,,$$where $\alpha$ is an eigenvector of the
eigenvalue equation
$$\kappa^2\,v^{(2)}\alpha=\lambda \alpha\,.$$ The plus (minus) sign
occurs for the symmetric (antisymmetric) subspace of
$\ell_{20}^2(\mathbb{Z}^3)$. The eigenvectors are localized at $\pm
e^k$, respectively. When the free Hamiltonian $H_0$ is not
neglected, the negative eigenvectors of $H$ get spread out.

We now reinstate $H_0$ and obtain the spectral representation of $H$
where we write $V(\vec \xi,\vec \eta)=2\kappa^2\mu\sum_j\delta(\vec
\xi-e^j)\delta(\vec\xi-\vec \eta)$, where $\mu$ is a negative
eigenvalue of $v^{(2)}$, and we are in the space symmetric or
antisymmetric subspace. As $H$ acts in the space of symmetric or
antisymmetric subspace of $\ell_{20}^2(\mathbb{Z}^3)$ the
$\delta(\vec\xi+e^j)$ term  in the potential can be dropped and the
term $\delta(\vec\xi-e^j)$ gets multiplied by $2$. The space
symmetry or antisymmetry is incorporated automatically in the
definition of $(H_0-z)^{-1}$ through the function $\Xi(\vec p,
\vec\xi, \vec \eta)$.

The spectral results below hold for all $\kappa>0$ while the
correspondence with the B-S equation is valid only for $\kappa\ll
1$. With the above observations, the resolvent equation becomes
$$\barr{ll}(H-z)^{-1}(\vec \xi,\vec \eta) =&
(H_0-z)^{-1}(\vec \xi,\vec \eta)-(H_0-z)^{-1}(\vec
\xi,e^j)\vspace{.14cm}\\&\times\;2\kappa^2\mu\,(H-z)^{-1}(e^j,\vec
\eta)\,,\earr$$ with solution
$$\barr{ll}(H-z)^{-1}(\vec \xi,\vec \eta) =&(H_0-z)^{-1}(\vec \xi,\vec \eta)-(H_0-z)^{-1}(\vec
\xi,e^j)\vspace{.14cm}\\&\times\;2\kappa^2\mu\,B_{jk}(H_0-z)^{-1}(e^k,\vec
\eta)\,,\earr$$where $B\equiv A^{-1}$ and
$A=[1+(H_0-z)^{-1}\,2\kappa^2\mu]$ restricted to
$\ell_2(\{e^1,e^2,e^3\})$ so it is a $3\times 3$ matrix. The matrix
$A$ has the
structure$$A=\left(\barr{ccc}a&b&b\\b&a&b\\b&b&a\earr\right)\,,$$
where  $a=1+2\kappa^2\mu (H_0-z)^{-1}(\vec r,\vec t)$, $(\vec r,\vec
t)=$ $(e^1,e^1)$, $(e^2,e^2)$ or $(e^3,e^3)$, and $b=2\kappa^2\mu
(H_0-z)^{-1}(\vec r,\vec t)$, $(\vec r,\vec t)=$ $(e^1,e^2)$,
$(e^1,e^3)$ or $(e^2,e^3)$. Furthermore, ${\rm
det}\,A=(a-b)^2(a+2b)$, and the eigenvalues of $A$ are
$\lambda_1=\lambda_2=(a-b)$, $\lambda_3=a+2b$, with corresponding
normalized eigenvectors $f_1=(r_1,r_2,r_3)$, $f_2=(t_1,t_2,t_3)$,
$f_3=\frac1{\sqrt{3}}(1,1,1)$, where $\sum_ir_i=\sum_it_i=0$ and
$f_1\perp f_2$.

The matrix $B$ has the same structure as $A$ and is given by
$$B=\dfrac1{(a-b)(a+2b)}\left(\barr{ccc}a+b&-b&-b\\-b&a+b&-b\\-b&-b&a+b\earr\right)\,.$$
The negative energy bound states occur as zeroes of the denominator
of $B$, and are given by the zeroes of $(a-b)$ and $(a+2b)$.

Denoting minus the binding energies by $z=-\epsilon$, for
$\epsilon_1=\epsilon_2$, $\epsilon_3$, we have the conditions, for
$z=-\epsilon_1=-\epsilon_2$
$$
0=1+\!2\kappa^2\mu[(H_0-z)^{-1}(e^1,e^1)-(H_0-z)^{-1}(e^1,e^2)]\,,$$and,
for $ z=-\epsilon_3$
$$
0=1+\!2\kappa^2\mu[(H_0-z)^{-1}(e^1,e^1)+2(H_0-z)^{-1}(e^1,e^2)]\,.
$$

The corresponding numerators of the resolvents $(H_0-z)^{-1}$, in
Fourier space, are $(\cos p^1)^2-\cos p^1\cos p^2$ and $(\cos
p^1)^2+2\cos p^1\cos p^2$, in addition to the factor $\Xi(\vec
p,\vec \xi, \vec\eta)$. Expanding the denominator $(p_\ell^2-z)$
shows that $\epsilon_3>\epsilon_1$.

Recalling that the resolvent contains all spectral information, we
now obtain the bound state eigenfunctions. To this end, from the
spectral theorem
$$\barr{ll}\dis\lim_{z\rightarrow -\epsilon_i}&\!\!\!(-\epsilon_i-z)(H-z)^{-1}(\vec \xi,\vec
\eta)= \sum_{j=1}^{m_i} \psi_{ij}(\vec \xi)\bar\psi_{ij}(\vec
\eta)\vspace{.14cm}\\&=-(H_0+\epsilon_i)^{-1}(\vec
\xi,e^j)F_{ijk}(H_0+\epsilon_i)^{-1}(e^k, \vec\eta)\,,\earr$$where
$F_{ijk}=\lim_{z\rightarrow -\epsilon_i}2\kappa^2\mu B_{jk}$ and
$\psi_{ij}(\vec\xi)$ are eigenfunctions of $H$ with eigenvalue
$-\epsilon_i$; $j=1,2,\ldots,m_i$ labels the $m_i$ degenerate
states.

For $i=3$, $a=-2b$ and
$$F_3=\dfrac1{3c_3}\left(\barr{ccc}1&1&1\\1&1&1\\1&1&1\earr\right)\,,$$
where, upon expanding $(a+2b)$ in the denominator of $B$ in powers
of $(z+\epsilon_3)$, we have
$c_3=2[(H_0-z)^{-2}(e^1,e^1)+2(H_0-z)^{-1}(e^1,e^2)]|_{z=-\epsilon_3}$.
The eigenvalues of the matrix are $0$ and $3$, with multiplicity two
and one, respectively.  The corresponding eigenvectors are $h_1$,
$h_2$ and $h_3=(1,1,1)/\sqrt{3}$; in $h_1$ and $h_2$ the sum of the
components is zero and $h_1$ is orthogonal to $h_2$. Noting that
$h_{3j}h_{3k}$ is the $jk$ matrix element of the orthogonal
projection on the multiplicity one eigenspace, we have
$$
\psi_3(\vec\xi)\psi_3(\vec\eta)\!=\!\!\frac1{c_3}(H_0+\epsilon_1)^{-1}\!(\vec\xi,e^j)h_{3j}h_{3k}(H_0+\epsilon_1)^{-1}\!(e^k,
\vec\eta),
$$
so that the normalized eigenfunction is
$$
\psi_3(\vec\xi)=\dfrac1{\sqrt{c_3}}(H_0+\epsilon_1)^{-1}(\vec\xi,e^j)h_{3j}\,.
$$
Similarly, for the doubly degenerate eigenvalue
$\epsilon_1=\epsilon_2$, we have
$$F_1=\dfrac1{3c_1}\left(\barr{rrr}2&-1&-1\\-1&2&-1\\-1&-1&2\earr\right)\,,$$
and the matrix has eigenvalues $0$ (multiplicity one) and $3$
(multiplicity two) and eigenvectors $h_3=(1,1,1)/\sqrt{3}$; and
$h_1$, $h_2$ where $h_1$ is orthogonal to $h_2$ and the sum of
components is zero. Hence,
$$\barr{ll}
\sum_{m=1,2}\psi_{1m}(\vec\xi)\psi_{1m}(\vec\eta)=&\dfrac1{c_1}(H_0+\epsilon_1)^{-1}(\vec\xi,e^j)[h_{1j}\,h_{1k}
\vspace{.15cm}\\&+h_{2j}\,h_{2k}](H_0+\epsilon_1)^{-1}(e^k,
\vec\eta)\,,\earr
$$
so that the normalized eigenfunctions $\psi_{11}$ and $\psi_{12}$
are given by
$$
\psi_{11}(\vec \xi)=\dfrac1{\sqrt{c_1}} (H_0+\epsilon_1)^{-1}(\vec
\xi,e^j)\,h_{1j}\,,
$$and $$
\psi_{12}(\vec \xi)=\dfrac1{\sqrt{c_1}} (H_0+\epsilon_1)^{-1}(\vec
\xi,e^j)\,h_{2j}\,,
$$
which are orthogonal to each other since
$$
\sum_{\vec\xi}\psi_{11}(\vec\xi)\psi_{12}(\vec\xi)=\frac1{c_1}h_{1j}(H_0+\epsilon_1)^{-2}(e^j,e^k)\,h_{2k}\,,
$$
and from the structure of $(H_0+\epsilon_1)^{-1}$, $h_1$ and $h_2$
are also its orthogonal eigenvectors.

The eigenfunctions $\psi_{11,12,3}(\vec\xi)$ have exponential decay
by the Paley-Wiener theorem (see Ref. \cite{RS2}), and in fact the
integrals of the form $(H_0+\epsilon)^{-1}(\vec\xi, \vec \eta)$
occurring in the wave functions as well as in the bound state
condition can be evaluated explicitly in terms of elliptic functions
(see Refs. \cite{Mont,Lif,BR2,Kos,katso,joy} and also Refs.
\cite{2p,ift}). We remark that another approach to the
eigenfunctions is group theoretical noting that the potential is
invariant under the symmetric group ${\rm S}_3$ of permutations of
the coordinate components. We see that the eigenfunction $\psi_3$ is
a basis for the one-dimensional identity representation of ${\rm
S}_3$; $\psi_{11}$, $\psi_{12}$ form a basis for the two-dimensional
representation of ${\rm S}_3$.
\section{Conclusions and final remarks}\lb{sec6}
We point out that we can determine the baryon-baryon bound state
spectrum using the more traditional unmodified B-S equation
(\ref{BS}). The solution of the B-S equation is more complicated as
it deals with a three-part energy-dependent potential, as explained
in Appendix D. Using this approach, we do obtain the same bound
state results.

To conclude, if we consider contributions to the B-S kernel
comprised of linear chains of quark, anti-quark pairs they result in
an exponentially decreasing potential with decay rate $-2\ln
\kappa$, as for the Yukawa theory. Surely, it would be interesting
to look for the expected {\em distance}$^{-1}$ Ornstein-Zernicke
like correction to this potential and to determine the spin and
isospin dependencies of the binding energy, and the dependence on
the number of flavors. Especially, the analysis of the ${\rm SU}(3)$
flavor case can shed some light in understanding bound states when
strangeness and, consequently, $\Lambda$ particles are present.
Also, there is the problem of determining bound states of more than
two baryons. However, lattice effects are expected to be relevant
and unrealistic in determining the resulting geometric spatial
configuration of possible bound states.

We make some remarks on pitfalls that can occur when trying to
determine bound states by simulation methods. For example, consider
the total isospin $I=3$, $I_z=3$ sector treated in Section 4.4. We
have found bound states with binding energies proportional to
$3\kappa^2$ and $\kappa^2$ which are associated with the total spin
$(T,T_z)=(0,0)$ and $(T,T_z)=(1,(-1,0,1))$, respectively. If we look
at the correlation $\langle {\cal T}_{TT_z} (x_1,x_2)\bar{\cal
T}_{TT_z} (x_3,x_4)\rangle$ with $x_1^0<x_3^0$, $\vec x_1=\vec x_2$
and $\vec x_3=\vec x_4$, it is zero for $(T,T_z)=(1,(-1,0,1))$, as
${\cal T}_{TT_z} (x_1,x_2)$ is space antisymmetric. Thus, these
bound states would be missed! Even if we take $\vec x_1\not=\vec
x_2$ and $\vec x_3\not=\vec x_4$, the decay rate in $x_3^0-x_1^0$
would not be $2\bar m-\alpha\kappa^2$, $\alpha>0$, but the smaller
one of $2\bar m-3\alpha\kappa^2$ since we expect there is overlap
with the $(T,T_z)=(0,0)$ state. This overlap occurs as there is {\em
no} spin invariance and associated decomposition into disjoint spin
sectors, at $\kappa\not=0$. Thus it seems to be very difficult to
see all bound states with small spectral separations.

Finally, and more importantly, we would like to know how the baryon
spectrum and the bound state binding energies behave near the
scaling limit.

This work was supported by CNPq and FAPESP. Algebraic computations
were checked with Maple and Matlab, for which we thank Louis
O'Carroll. We also thank P.H.R. dos Anjos for help in unveiling the
time reflection symmetry of Appendix B.\vspace{.4cm}
\appendix{{ \small {\bf APPENDIX A: Isospin orthogonality relations and correlation identities}}}
\lb{appb}\setcounter{equation}{0}
\renewcommand{\theequation}{A.\arabic{equation}}\vspace{.2cm}\\Here
we obtain isospin orthogonality relations for the two and
four-baryon correlations used throughout this paper, and which hold
for all $\kappa$. These relations also hold for spin at $\kappa=0$,
since the action has a ${\rm SU}(2)$ symmetry in the lower
components of the spin index, and spin is also conserved under this
condition.

We consider, typically, the two-point correlation $\langle
\phi_{I_1m_1}(x_1)\bar\phi_{I_2m_2}(x_2)\rangle$ and the four-point
correlation $\langle
\phi_{I_1m_1}(x_1)\phi_{I_2m_2}(x_2)\bar\phi_{I_3m_3}(x_3)
\bar\phi_{I_4m_4}(x_4)\rangle$, where we suppress the spin index and
display only the $I$ and $I_z$ labels.

Considering first the two-point correlation and suppressing the
gauge and spin indices and site arguments, the two-point correlation
is given by linear combinations in the gauge and spin indices, and
linear C-G combinations in the isospin indices of the basic
correlation
$$ \langle
\psi_{i_1}\psi_{i_2}\psi_{i_3}\bar\psi_{i_4}\bar\psi_{i_5}
\bar\psi_{i_6}\rangle\id T_{123;456}\,.$$

We start from the identity, with arbitrary $U\in{\rm U}(2)$,
$$\barr{l}\langle \psi_{i_1}\psi_{i_2}\psi_{i_3}(\bar\psi\,
U)_{i_4}(\bar\psi\, U)_{i_5}(\bar\psi\,
U)_{i_6}\rangle\vspace{.15cm}\\\qquad\qquad = \langle
(U\,\psi)_{i_1}(U\,\psi)_{i_2}(U\,\psi)_{i_3}\bar\psi _{i_4}\bar\psi
_{i_5}\bar\psi _{i_6}\rangle\,,\earr$$ where the r.h.s. is obtained
from the l.h.s. using the two-flavor isospin symmetry. We
schematically rewrite this relation as\bequ\lb{TT}
(TU_3)_{123;456}=(U_3T)_{123;456}\,,\eequ where we set $U_3\id
U\otimes U\otimes U$.

We obtain orthogonality relations for the two-baryon correlation by
relating this identity to the usual quantum mechanical sum of
angular momentum relations. We write $U=e^{i\theta\sigma/2}$,
$I=\sigma/2$, where
$\theta\sigma=\theta_x\sigma_x,\,\theta_y\sigma_y,\,\theta_z\sigma_z$,
or $\theta_{x,y,z}\in\mathbb{R}$. Expanding in $\theta$, we have
$U=1+i\theta I-\frac{\theta^2}2+\ldots$ such that
\bequ\lb{u3}\barr{ll}U_3=&1\otimes 1\otimes 1+i\theta(I\otimes
1\otimes 1+1\otimes I\otimes 1+1\otimes 1\otimes
I)\vspace{.15cm}\\&-\frac{\theta^2}2\,(I^2\otimes 1\otimes 1+
1\otimes I^2\otimes 1+1\otimes 1\otimes 1\otimes
I^2\vspace{.15cm}\\&+ 2\, I\otimes I\otimes 1+2\,I\otimes 1\otimes
I+2\,1\otimes I\otimes I)\ldots\earr\eequ

Multiplying Eq. (\ref{TT}) by $\bar w_{123}$ on the left and
$v_{123}$ on the right, and equating the coefficients of $i\theta_z$
and $-(\theta_x^2+\theta_y^2+\theta_z^2)/2$, we obtain the
identities \bequ\lb{iz} \bar w_{123}(TI_z)_{123;456}v_{456}=\bar
w_{123}(I_zT)_{123;456}v_{456}\,, \eequ and
$$
\bar w_{123}(TI^2)_{123;456}v_{456}=\bar
w_{123}(I^2T)_{123;456}v_{456}\,,
$$ where, with an abuse of notation, $I_z=I_z\otimes 1\otimes 1
+ 1\otimes I_z\otimes 1+1\otimes 1\otimes I_z$ and $I^2=(\vec
I_1+\vec I_2+\vec I_3)^2$, and $\vec I_j$, as usual, acts on the
$j$th spot of the tensor product.

We will see that $I_z$ ($I^2$) has the interpretation of the
$z-$component (square) of total isospin. We can use the usual
complex Hilbert space notation $(\;,\;)$ (as $\sigma_{x,y,z}$ are
self-adjoint)  and write the above as \bequ\lb{amz}
(w,TI_zv)=(I_zw,Tv)\:\;,\:\;  (w,TI^2v)=(I^2w,Tv)\,.\eequ Thus, if
$v$ and $w$ correspond to eigenfunctions with distinct eigenvalues
we have orthogonality. To be more explicit, using the usual quantum
mechanics notation for $\alpha$ ($\beta$) for spin up (down) state,
and suppressing the tensor product notation, an isospin state of
total isospin $\frac 32$  and $z-$ component $\frac32 $ ($\frac 12$)
are represented as the C-G linear combination $v_{\frac 32\frac
32}=\alpha \alpha \alpha$ and $v_{\frac 32\frac 12}=\frac
1{\sqrt{3}}(\alpha \alpha\beta +\alpha\beta\alpha+
\beta\alpha\alpha)$, respectively.

Taking these states into Eqs. (\ref{amz}), and using the usual
quantum mechanical relations gives the orthogonality relations.
Furthermore, taking the appropriate gauge and spin linear
combinations to obtain $\tilde \phi_{I_\ell m_\ell}$ gives the
orthogonality relations for the two-baryon function.

Similar to the $I_z$ identity of Eq. (\ref{iz}), we also have, with
$I_\pm\equiv I_x\pm iI_y$, \bequ\lb{ams2}(w,TI_\pm v)=(I_\mp
w,Tv)\,.\eequ

Next, we take $v=\chi_{Im}$ and $v=\chi_{I(m-1)}$, where
$\chi_{I\ell}$ is the usual normalized eigenfunction of total
isospin $I$ and $z-$ component $\ell$. Hence, $\chi_{I\ell}$
satisfies $I_\pm \chi_{I\ell}=c_\ell^\pm\,\chi_{I(\ell\pm 1)}$,
where $$c_\ell^\pm= [I(I+1)-\ell(\ell\pm 1)]^{1/2}\,.$$ Substituting
in Eq. (\ref{ams2}), with the lower sign, and noting that
$c^-_\ell=c_{\ell-1}^+$, gives $$(\chi_{I(\ell-1)},
T\chi_{I(\ell-1)})=(\chi_{I\ell},T\chi_{I\ell})\,,$$which implies
the identity \bequ\lb{ams3} \langle\phi_{I(\ell-1)}(x)
\bar\phi_{I(\ell-1)}(y)\rangle=\langle\phi_{I\ell}(x)
\bar\phi_{I\ell}(y)\rangle\,.\eequ

Of course, Eq. (\ref{ams3}) implies that all associated one-baryon
spectral properties are the same, for $I$ fixed, and all $I_z$.

Let us now turn to the four-baryon functions. The basic four-point
function is
$$\langle \psi_{i_1}\ldots \psi_{i_6}\bar\psi_{i_7}\ldots
\bar\psi_{i_{12}}\rangle\equiv T_{1\ldots 6;7\ldots 12}\,.$$

As before, we have the identity \bequ\lb{ttt}T{\rm U}_6={\rm
U}_6T\,,\eequ where ${\rm U}_6\equiv \bigotimes_{j=1}^6\,{U}$.
Decomposing ${\rm U}_6$ as ${\rm U}_3\otimes {\rm U}_3$ and using
similar considerations as above, we obtain orthogonality relations
for the $z-$component and square of the total isospin by the
appropriate $C-G$ linear combinations of the product of the
one-particle isospin states $\phi_{I\ell}$.

Concerning the {\em Wickified} four baryon function $M_0$ of Section
4.3, the basic correlation is
$$\barr{ll}\hspace{-.1cm}T_{0,1\ldots 6;7\ldots 12}\!\equiv\!\! &
-\langle\psi_{1}\ldots \psi_3\bar\psi_{7}\ldots
\bar\psi_{9}\rangle\langle\psi_{4}\ldots \psi_{6}\bar\psi_{10}\ldots
\bar\psi_{12}\rangle\vspace{.15cm}\\&
+\langle\psi_{1}\ldots \psi_{3}\bar\psi_{10}\ldots
\bar\psi_{12}\rangle\langle\psi_{4}\ldots \psi_{6}\bar\psi_{7}\ldots
\bar\psi_{9}\rangle .\earr$$

We have the identity
$$\barr{l}-\langle\psi_{1}\ldots \psi_3(\bar\psi U)_{7}\ldots
(\bar\psi U)_{9}\rangle\,\langle\psi_{4}\ldots
\psi_{6}\vspace{.15cm}\\\times\,(\bar\psi U)_{10}\ldots \!(\bar\psi
U)_{12}\rangle\! +\langle\psi_{1}\ldots\! \psi_{3}(\bar\psi
U)_{10}\ldots (\bar\psi
U)_{12}\rangle\vspace{.15cm}\\\times\,\langle\psi_{4}\ldots
\psi_{6}(\bar\psi U)_{7}\ldots (\bar\psi U)_{9}\rangle
=-\langle(U\psi)_{1}\ldots
(U\psi)_3\vspace{.15cm}\\\times\,\bar\psi_{7}\ldots \bar\psi
_{9}\rangle\,\langle (U\psi)_{4}\ldots
(U\psi)_{6}\bar\psi_{10}\ldots \bar\psi_{12}\rangle
\vspace{.15cm}\\+\langle (U\psi)_{1}\ldots
(U\psi)_{3}\bar\psi_{10}\ldots \bar\psi
_{12}\rangle\,\langle(U\psi)_{4}\ldots\!
(U\psi)_{6}\vspace{.15cm}\\\times\,\bar\psi _{7}\ldots
\bar\psi_{9}\rangle \,,\earr$$which can be compactly written as
\bequ\lb{t0}T_0\,{\rm U}_6={\rm U}_6\,T_0\,.\eequ

We conclude by remarking that the transformation properties for $T$
and $T_0$ of Eqs (\ref{ttt}) and (\ref{t0}) are the same.
Consequently, we have the same orthogonality relations for the
$z-$component and square of total isospin of two-baryon states.
\appendix{{ \small {\bf APPENDIX B: The hyperplane decoupling
method}}}\lb{appa}
\setcounter{equation}{0}\renewcommand{\theequation}{B.\arabic{equation}}
\vspace{.2cm}

In this Appendix, we use the hyperplane decoupling method to see how
the twenty baryon particles arise in the energy-momentum spectrum
for our $\rm{SU}(3)$, two-flavor lattice QCD model. It is understood
here that we first work in a finite-volume model and then with
bounds on the normalized correlations which are uniform in the
volume, obtained via a polymer expansion, the arguments can be
extended to the infinite volume theory in a standard way (see Ref.
\cite{Si,Sei,GJ}). This step is {\em not} treated here, and we work
directly in the infinite volume. Also, we explicitly discuss the
hyperplane decoupling method applied to the time direction, the
treatment for the three space directions being similar.

In order to give a brief description of the method, we recall that
the starting point for the time hyperplane decoupling method is to
consider the model with interpolating parameters replacing the
hopping terms giving couplings in the temporal direction. In terms
of the model action, for the hopping terms connecting the
hyperplanes $x^0=p$ and $x^0=p+1$, the hopping parameter $\kappa$ is
replaced by a real parameter $\kappa_p$. (Note that setting
$\kappa_p=0$ erases all connections in the time direction between
$p$ and $p+1\:$!) (see Ref. \cite{Complex}). Doing so, as in
Sections 3 and 4, we define $\kappa_p$-dependent correlations e.g.
the $\kappa_p$ dependent zero, two and four-baryon correlations,
which provide a continuation to complex $\kappa_p$ afterwards.

The main point here is that decay properties of these correlations
in direct lattice spacetime are related to the vanishing of the
first $n$ derivatives with respect to $\kappa_p$, at $\kappa_p=0$.
In our case, the first two temporal hyperplane derivatives of $G$
vanish, which leads to a $-3\ln\kappa$ decay rate. The third
derivative has an important product structure which implies that the
first four derivatives of the $G$ convolution inverse $\Gamma$ are
zero. This in turn shows that $\Gamma$ has a faster decay rate of at
least $-4\ln\kappa$. Associated to these properties, by Fourier
transform, there are analyticity properties in energy-momentum
Fourier dual variables. The Fourier transform $\tilde\Gamma(p)$ is
analytic in a larger $p^0$ strip than $\tilde G(p)$ so that
$\tilde\Gamma^{-1}(p)={\rm cof}^t\,[\tilde\Gamma](p)/{\rm
det}\tilde\Gamma(p)$ provides a meromorphic extension of $\tilde
G(p)$ to the larger strip. As seen in Section \ref{sec3}, the
dispersion curves are determined as the zeroes of ${\rm det}
\tilde\Gamma(p^0=iw(\vec p),\vec p)$, so that they are isolated.
This is a brief account of how, using the
 spectral representations for the Fourier transformed correlations,
 the method leads to spectral properties.

Next, we show how the $20$ baryon particles arise in the two-flavor
case, using the hyperplane decoupling method. Also, we show how to
classify these states according to total isospin $I$, $z$-component
isospin $I_z$ and spin $s_z$. Let $\epsilon\tilde\psi^3_{\vec
\alpha\vec f}\equiv \epsilon_{abc}\tilde\psi_{a\alpha_1f_1}
\tilde\psi_{b\alpha_2f_2} \tilde\psi_{c\alpha_3f_3}$, which is
invariant under the exchange $\alpha_jf_j\leftrightarrow
f_k\alpha_k$, which we call the {\em totally symmetric property}. As
seen below, we can restrict our analysis to the lower spin
components, $\alpha_j=3,4$.

For $x^0<y^0$, we consider a generic two-baryon correlation
$$
G_{\vec \alpha_1\vec f_1\vec \alpha_2\vec f_2}(x,y)\equiv\langle
\epsilon \psi^3_{\vec \alpha_1\vec f_1}(x)\,\epsilon
\bar\psi^3_{\vec \alpha_2\vec f_2}(y)\rangle\,.
$$

We now consider the derivatives with respect to the hyperplane
parameters $\kappa_p$, at $\kappa_p=0$. It is easy to see that
$G^{(0),(1),(2)}_{\vec \alpha_1\vec f_1\vec \alpha_2\vec f_2}=0$
because of imbalance of  fermion fields $\psi$.
For the third derivative, after performing the gauge integral using
[see Ref. \cite{Creu2}] \bequ \barr{lll}{\cal I}_3&\equiv & \int
U(g)_{a_1b_1}U(g)_{a_2b_2}U(g)_{a_3b_3}d\mu(g)\vspace{.15cm}\\&\equiv&
\int g_{a_1b_1}g_{a_2b_2}g_{a_3b_3}d\mu(g)\vspace{.15cm}\\&=&
\frac16\,\epsilon_{a_1a_2a_3}\epsilon_{b_1b_2b_3}, \lb{I3}\earr\eequ
we get, with $\vec w\in\mathbb{Z}^3$, \bequ\barr{ll} G^{(3)}_{\vec
\alpha_1\vec f_1\vec \alpha_2\vec f_2}(x,y)=\!\!
&-\frac1{36}\sum_{\vec w,\vec \beta,\vec h} \langle \epsilon
\psi^3_{\vec \alpha_1\vec f_1}(x)\,\epsilon \bar\psi^3_{\vec
\beta\,\vec h}(p,\vec
w)\rangle^{(0)}\vspace{.15cm}\\&\times\,\langle \epsilon
\psi^3_{\vec \beta\,\vec h}(p+1,\vec w)\,\epsilon \bar\psi^3_{\vec
\alpha_2\vec f_2}(y)\rangle^{(0)}\,,\earr \lb{product}\eequ where
the structure of $\Gamma^{+e^0}$ restricts the spin indices in $\vec
\beta$ to be lower. Hence, the {\em product} structure of Eq.
(\ref{product}) is obtained if we take $\vec \alpha_1$ and $\vec
\alpha_2$ with only lower indices. This formula is also {\em closed}
meaning that the correlation on the l.h.s. of Eq. (\ref{product}) is
the same as the two correlations appearing on the r.h.s. It is
important to stress that, if we start with a gauge invariant average
with generic fields with an odd number of fermions, $\langle
H(x)\bar L(y)\rangle$, we do not need to make an a priori guess of
what the baryon fields are. They emerge as fields to achieve the
product structure, and is consistent with confinement as the baryon
fields are gauge invariant local fields with three quarks.

As explained above, the product property in Eq. (\ref{product})
leads to the existence of particles with isolated dispersion curves,
after eliminating the redundancies due to linear dependencies.

As each $\alpha\,f$ takes four values, there are $64$ possible $\vec
\alpha\,\vec f$'s but, by identifying states which are related by
the totally symmetric property, this number is reduced to $20$. The
list of $12$ representatives out of the $20$ states with their
multiplicities in the intermediate sums of Eq. (\ref{product}) is
the following: $\vec f=(+,+,+),\,(-,-,-)$, $\vec
\alpha=(+,+,+),\,(-,-,-)$, multiplicity $1$; $\vec
f=((+,+,+),\,(-,-,-))$, $\vec \alpha=(+,+,-),\,(-,-,+)$,
multiplicity $3$; $\vec f=(+,+,-)$, $\vec \alpha=(+,+,+),\,(-,-,-)$,
multiplicity $3$; $\vec f=(-,-,+)$, $\vec \alpha=(+,+,+),\,(-,-,-)$,
multiplicity $3$. The other $8$ states are given below.

Eight of the above $12$ states with $\vec f=(+,+,+),\,(-,-,-)$ have
$I=3/2$ and $I_z=3/2,\,-3/2$, respectively, and $s=\pm
3/2,\,\pm1/2$. For the remaining four states, we use the fact (from
properties of the C-G coefficients) that $I=3/2,\,I_z=\pm1/2$ states
formed from tensor products of three isospin $1/2$ states have equal
weights and are symmetric under interchange of any of the individual
isospins. In this way,  these states with $\vec
\alpha=(+,+,+),\,(-,-,-)$ have $I=3/2,\,I_z=\pm 1/2$. Thus, we are
left with $8$  states, with $|I_z|=1/2$ and $|s|=1/2$, given by
(including the respective multiplicities on the right)
$$\barr{l}
1)\,\vec f=(+,-,+),\,\vec \alpha=(+,-,+),\,I_z=\frac12,\,
s=\frac12,\,3;\vspace{.15cm}\\
2)\,\vec f=(-,+,+),\,\vec \alpha=(+,-,+),\,I_z=\frac12,\,
s=\frac12,\,6;\vspace{.15cm}\\
3)\,\vec f=(+,-,+),\,\vec \alpha=(+,-,-),\,I_z=\frac12,\,
s=-\frac12,\,6;\vspace{.15cm}\\
4)\,\vec f=(-,+,+),\,\vec \alpha=(+,-,-),\,I_z=\frac12,\,
s=-\frac12,\,3;\vspace{.15cm}\\
5)\,\vec f=(+,-,-),\,\vec \alpha=(+,-,+),\,I_z=-\frac12,\,
s=\frac12,\,6;\vspace{.15cm}\\
6)\,\vec f=(-,+,-),\,\vec \alpha=(+,-,+),\,I_z=-\frac12,\,
s=\frac12,\,3;\vspace{.15cm}\\
7)\,\vec f=(-,+,-),\,\vec \alpha=(+,+,-),\,I_z=-\frac12,\,
s=-\frac12,\,3;\vspace{.15cm}\\
8)\,\vec f=(+,-,-),\,\vec
\alpha=(-,+,-),\,I_z=-\frac12,\,s=-\frac12,\,6
 \,.\earr$$

We identify linear combinations of these $8$ states which we show to
have $I=3/2$ and $I=1/2$. The proton (neutron) is identified with
the $I=1/2$, $I_z=1/2$ ($-1/2$) state.

Coupling the first two isospins of $1)$ and $2)$ (by antisymmetric
linear combination), to give a zero isospin, than coupling the third
isospin gives the $p_+$ state. Similarly, antisymmetric linear
combination of $3)$ and $4)$, gives $p_-$. $n_+$ ($n_-$) is obtained
by antisymmetric linear combination of $5)$ and $6)$ ($\,7)$ and
$8)\,$). For the $I=3/2$, $I_z=1/2$, $s=1/2$ state with $\vec
\alpha=(+,-,+)$ take the linear combination $1) + 2) + 2)$ and note
that the second $2)$ is not changed by $\alpha_1f_1\leftrightarrow
f_3\alpha_3$. Thus, the resulting state is totally symmetric in any
$\alpha_jf_j\leftrightarrow f_k\alpha_k$; hence it is a $I=3/2$
state. Similarly, the remaining three states are obtained with
$(I,I_z,s)=(3/2,1/2,-1/2)$, $(3/2,-1/2,1/2)$, $(3/2,-1/2,-1/2)$, and
are identified with the linear combinations $3) + 3) + 4)$, $5) + 5)
+ 6)$ and $7) + 8) + 8)$, respectively.

The above $20$ states are unnormalized ones. Using the twenty
normalized states of Section 3.1, and their multiplicities, and
rewriting the intermediate states of Eq. (\ref{product}) in terms of
normalized states, we obtain, for $x^0<y^0$ and $\vec
w\in\mathbb{Z}^3$, \bequ\barr{ll}\!\! \langle B_{II_zs}(x)\bar B_{I
I_zs^\prime}(y)\rangle^{(3)}= &\!\!-\dis\sum_{\vec w,r} \langle
B_{II_zs}(x)\bar B_{I I_zr}(p,\vec
w)\rangle^{(0)}\vspace{.15cm}\\&\!\!\times\,\!\langle
B_{II_zr}(p+1,\vec w)\bar B_{I I_zs^\prime}(y)\rangle^{(0)}.\earr
\lb{product2}\eequ

Similar considerations apply for $x^0>y^0$, and we finally obtain
the product structure for the two-baryon correlation of Eq.
(\ref{GG}) \bequ G^{(3)}_{\ell_1\ell_2}(x,y)= -\sum_{\vec
w,\ell}\,G^{(0)}_{\ell_1\ell}(x,(p,\vec w))
\,G^{(0)}_{\ell\ell_2}((p+1,\vec w),y)\,. \lb{product3}\eequ

Furthermore, for $\kappa=0$, using Wick's theorem in the
$\tilde\psi$ fields, we obtain the normalization property, for any
$x$,
$$ \langle B_{II_zs}(x)\bar B_{II_zs}(x)\rangle^{(0)}=-1\,.$$

To continue, we now use symmetries at the level of correlation
functions to give a more complete treatment of the baryon dispersion
curves than that appearing in Theorem 4 of Ref. \cite{CMP}. Besides
the ordinary time reversal ${\cal T}$, charge conjugation ${\cal C}$
and parity ${\cal P}$, which can be implemented as unitary
(anti-unitary for time reversal) operators acting on the physical
Hilbert space ${\cal H}$, we also have a time {\sl reflection}
symmetry ${\textsc{T}}$ as well as the combination ${\cal C}{\cal
P}\hspace{.1cm}\textsc{T}$.

We summarize these symmetries as follows: \begin{itemize}
\item Time Reflection $\textsc{T}$: The $\tilde \psi$ fields
transform as $\psi_\alpha(x)\rightarrow
A_{\alpha\beta}\psi_\beta(-x^0,\vec x)$,
$\bar\psi_\alpha(x)\rightarrow \bar\psi_\beta(-x^0,\vec x)
B_{\beta\alpha}$ where $A=B=A^{-1}$
$=\left(\barr{cc}0&-iI_2\\iI_2&0\end{array} \right)$,
$f(g_{xy})\rightarrow f(g_{\bar x\bar y})$, with $\bar z=(-z^0,\vec
z)$;
\item Charge Conjugation ${\cal C}$:
$\psi_\alpha(x)\rightarrow \bar\psi_\beta(x) A_{\beta\alpha}$,
$\bar\psi_\alpha(x)\rightarrow B_{\alpha\beta}\psi_\beta(x)$,
$A=-B=B^{-1}=$ $\left(\barr{cc}0&i\sigma^2\\i\sigma^2&0\end{array}
\right)$, $f(g_{xy})\rightarrow f(g^*_{xy})$;
\item Parity ${\cal P}$:
$\psi_\alpha(x)\rightarrow A_{\alpha\beta}\psi_\beta(x^0,-\vec x)$,
$\bar\psi_\alpha(x)\rightarrow \bar\psi_\beta(x^0,-\vec x)
B_{\beta\alpha}$ where $A=B=A^{-1}=\gamma^0$, $f(g_{xy})\rightarrow
f(g_{\bar x\bar y})$, with $\bar z=(z^0,-\vec z)$;
\item Rotation $r_3$ of $\pi/2$ about $e^3$: $\psi_\alpha(x)\rightarrow A_{\alpha\beta}\psi_\beta(x^0,x^2,-x^1,x^3)$,
$\bar\psi_\alpha(x)\rightarrow \bar\psi_\beta(x^0,x^2,-x^1,x^3)
B_{\beta\alpha}$ where $A=B^{-1}={\rm
diag}(e^{-i\pi/4},e^{i\pi/4},e^{-i\pi/4},e^{i\pi/4})$,
$f(g_{xy})\rightarrow f(g_{\bar x\bar y})*$, with $\bar
z=(z^0,z^2,-z^1,z^3\vec z)$;
\item Time Reversal $\cal{T}$: $\psi_\alpha(x)\rightarrow \bar\psi_\beta(-x^0,\vec x)
A_{\beta\alpha}$, $\bar\psi_\alpha(x)\rightarrow
B_{\alpha\beta}\psi_\beta(-x^0,\vec x)$, $A=B=B^{-1}=\gamma^0$,
$f(g_{xy})\rightarrow [f(g_{xy})]^*$.
\end{itemize}
The above symmetries are taken to be order preserving, except for
${\cal C}$ and ${\cal T}$ which are order reversing. For all of
them, except time reversal, the field average equals the transformed
field average; for time reversal the transformed field average is
the complex conjugate of the field average.

Applying the ${\cal P}{\cal C}\hspace{.1cm}\textsc{T}$ symmetry, we
find the following relations, after making a half-integer shift in
the temporal coordinate of the lattice,\bequ\lb{pn} \langle
p_\pm(u)\bar p_\mp(v)\rangle=0\quad ,\quad \langle p_+(u)\bar
p_+(v)\rangle=\langle p_-(u)\bar p_-(v)\rangle\,, \eequ and the same
for the $n$'s.  As shown in Section \ref{sec3}, we also have
$\langle p_\pm(u)\bar n_\mp(v)\rangle=0=\langle n_\pm(u)\bar
p_\mp(v)\rangle$, etc. Setting $\tilde \phi_s(u)\equiv \tilde
B_{\frac 32\frac32s}(u)$, we also have $\langle
\phi_{\frac32}(u)\bar \phi_{-\frac32}(v)\rangle =\langle
\phi_{\frac12}(u)\bar \phi_{-\frac12}(v)\rangle=0$, $\langle
\phi_{\mp\frac32}(u)\bar \phi_{\mp\frac12}(v)\rangle =-\langle
\phi_{\pm\frac12}(u)\bar \phi_{\pm\frac32}(v)\rangle$, $\langle
\phi_{-\frac32}(u)\bar \phi_{\frac12}(v)\rangle =\langle
\phi_{-\frac12}(u)\bar \phi_{\frac32}(v)\rangle$, $\langle
\phi_{\frac32}(u)\bar \phi_{\frac32}(v)\rangle =\langle
\phi_{-\frac32}(u)\bar \phi_{-\frac32}(v)\rangle$, and $\langle
\phi_{\frac12}(u)\bar \phi_{\frac12}(v)\rangle =\langle
\phi_{-\frac12}(u)\bar \phi_{-\frac12}(v)\rangle$ and, using $r_3$,
for $x_r\equiv(x^0,x^2,-x^1,x^3)$, \bequ\lb{r3}\barr{l}\langle
\phi_{\frac32}(u)\bar \phi_{\frac12}(v)\rangle =-i\langle
\phi_{\frac32}(u_r)\bar \phi_{\frac12}(v)\rangle\,,\vspace{.12cm}\\
\langle \phi_{\frac32}(u)\bar \phi_{-\frac12}(v)\rangle =-\langle
\phi_{\frac32}(u_r)\bar \phi_{-\frac12}(v)\rangle\,,\vspace{.12cm}\\
\langle \phi_{\frac32}(u)\bar \phi_{-\frac32}(v)\rangle =i\langle
\phi_{\frac32}(u_r)\bar \phi_{-\frac32}(v)\rangle\,,\vspace{.12cm}\\
\langle \phi_{\frac12}(u)\bar \phi_{-\frac12}(v)\rangle =-i\langle
\phi_{\frac12}(u_r)\bar \phi_{-\frac12}(v)\rangle\,,\vspace{.12cm}\\
\langle \phi_{\frac12}(u)\bar \phi_{-\frac32}(v)\rangle =-\langle
\phi_{\frac12}(u_r)\bar \phi_{-\frac32}(v)\rangle\,,\vspace{.12cm}\\
\langle \phi_{s}(u)\bar \phi_{s}(v)\rangle =\langle
\phi_{s}(u_r)\bar \phi_{s}(v)\rangle\,. \earr\eequ The relations
after Eq. (\ref{pn}) carry over to the two-baryon correlation $G$.

Using Eq. (\ref{pn}), the proton and neutron two-point functions are
diagonal in spin, at all orders in $\kappa$. For spin $3/2$, label
the basis by $3/2$, $-3/2$, $1/2$, and $-1/2$, or, simply, $1$, $2$,
$3$, and $4$, respectively. With this labeling, and remarking that
time reversal ${\cal T}$ parity ${\cal P}$ and the spectral
representation ensures self-adjointness, the two-point function
matrix has the structure
$$\left( \barr{cccc}a&0&c&d\\0&a&\bar d&-\bar c\\\bar c&d&b& 0\\\bar d&-c&0&b\earr\right)\,,$$
with multiplicity two eigenvalues
$$\mu_\pm=\{(b+a)\pm[(b-a)^2+4(|c|^2+|d|^2)]^{1/2}\}/2\,.$$ The
corresponding eigenvectors are $(c,\bar d,\mu_\pm-a,0)$ and
$(\mu_\pm-b,0,\bar c,\bar d)$.  Note that these two linearly
independent vectors have a singular limit when $c,d\rightarrow 0$,
where they become identical. [This limit corresponds to taking
$\tilde\Gamma_{13}\,,\,\tilde\Gamma_{14}\rightarrow 0$ in Eq.
(\ref{lambdapm}) below.]

The same matrix structure is satisfied by its inverse matrix and,
consequently, it also holds for the Fourier transforms $\tilde
G_{ss^\prime}(p^0=i\chi, \vec p)$ and $\tilde
\Gamma_{ss^\prime}(p^0=i\chi, \vec p)$, $\chi\in\mathbb{R}$. Using
the $r_3$ symmetry given above, shows that these matrices are
diagonal for $\vec p=\vec 0$ as well as for $\vec
p=(p^1=0,p^2=0,p^3)$, which is used in Ref. \cite{CMP} to determine
the one-baryon mass spectrum and to show that no mass splitting
occurs up to and including order $\kappa^6$. For $\vec p\not=\vec
0$, recall that the one-baryon dispersion curves $w(\vec p)$ are the
solutions of ${\rm det}\,\tilde\Gamma(p^0=iw(\vec p), \vec p)=0$.
From the above structure, omitting the $\vec p$ dependence, the
determinant factorizes as \bequ\lb{dett}{\rm det}\tilde\Gamma=\left[
\lambda_+\lambda_-\right]^2\,,\eequ which gives two by two identical
dispersion curves associated to the zeroes of
\bequ\lb{lambdapm}\lambda_\pm\equiv \frac
12[\tilde\Gamma_{11}+\tilde\Gamma_{33}]\pm \sqrt{\frac
14[\tilde\Gamma_{11}-\tilde\Gamma_{33}]^2
+|\tilde\Gamma_{13}|^2+|\tilde\Gamma_{14}|^2}\,.\eequ For $\vec
p=\vec 0$, $\lambda_+=\tilde\Gamma_{11}$  and
$\lambda_-=\tilde\Gamma_{33}$, and if they are not equal, mass
splitting occurs.

We now show how to find the solution for $\lambda_+=0$, the
treatment for other solution being similar. Suppressing the $\kappa$
dependence and putting $c_3(\vec p)=c_3\sum_{j=1}^3\,2\,\cos p^j$,
for $c_3=-1/8$ and $c_{30}=-1$ (see Ref. \cite{CMP}), we can
write$$\barr{ll}\tilde\Gamma_{s_1s_2}(\vec p)=&[-1-c_3(\vec
p)\kappa^3-c_{30}\kappa^3(e^{ip^0}+e^{-ip^0})]\vspace{.15cm}\\&\times\,\delta_{s_1s_2}\,+\,\sum^\prime_{n\geq
0,\vec x}\Gamma_{s_1s_2}(n,\vec x)e^{-i\vec p.\vec
x}\vspace{.15cm}\\&\times\,[\delta_{n0}+(1-\delta_{n0})(e^{ip^0n}+e^{-ip^0n})]
\,,\earr$$where $\sum^\prime$ means all terms of order $\kappa^6$ or
higher are maintained, and we have used $\Gamma(x^0,\vec
x)=\Gamma(-x^0,\vec x)$ which follows from $G(x^0,\vec
x)=G(-x^0,\vec x)$ by the spectral representation of Eq.
(\ref{FK1}).

To bring the solution in $p^0$ to the origin as $\kappa\searrow 0$,
as in Ref. \cite{CMP}, we make a nonlinear transformation by
introducing the auxiliary function $H(w,\kappa)\equiv
H(w,\kappa,\vec p)$ and a variable $w$ such that $H(w=-1-c_3(\vec
p)\kappa^3-c_{30}\kappa^3e^{-ip^0},\kappa)=\lambda_+$. $H(w,\kappa)$
is then defined
by\bequ\barr{ll}\lb{2H}H(w,\kappa)=&w+\dfrac{c_{30}\kappa^6}{1+w+c_3(\vec
p)\kappa^3} +\frac12\sum^{\prime\prime}_{n\geq 0,\vec
x}\vspace{.15cm}\\&\times\,(\Gamma_{11}+\Gamma_{33})(n,\vec
x)e^{-i\vec p.\vec
x}\{\delta_{n0}+(1-\delta_{n0})\vspace{.15cm}\\&\times\,(-1)^n[(\frac{1+w+c_3(\vec
p)\kappa^3}{c_{30}\kappa^3})^n+(\frac{c_{30}\kappa^3}{1+w+c_3(\vec
p)\kappa^3})^n]\}\vspace{.15cm}\\&+\sqrt{\zeta_1^2+|\zeta_2|^2+|\zeta_3|^2}
\,,\earr\eequ where
$$\barr{ll}\zeta_1=&\frac12\sum^{\prime\prime}_{n\geq 0,\vec
x}(\Gamma_{33}-\Gamma_{11})(n,\vec x)e^{-i\vec p.\vec
x}\{\delta_{n0}+(1-\delta_{n0})\vspace{.15cm}\\&\times\,(-1)^n[(\frac{1+w+c_3(\vec
p)\kappa^3}{c_{30}\kappa^3})^n+(\frac{c_{30}\kappa^3}{1+w+c_3(\vec
p)\kappa^3})^n]\}\,,\earr$$and similarly for $\zeta_2$ and $\zeta_3$
with the factor $[\Gamma_{33}\!-\!\Gamma_{11}]/2$ replaced by
$\Gamma_{13}$ and $\Gamma_{14}$, respectively. In the above,
$\sum^{\prime\prime}$ means that we only retain terms of order
$\kappa^6$ or higher in the sum, and that we have used the short
distance behavior of $\Gamma_{s_1s_2}$ as given in Theorem 2 of Ref.
\cite{CMP} [the misprinted values of $c_3$ and $c_{30}$ in this
reference are corrected above].

Before we continue the analysis, we give the following short
distance behavior that extends the results of Theorem 2 of Ref.
\cite{CMP} to a higher order in $\kappa$ ($c_6$ here is a
$\kappa$-independent computable constants):
\bequ\lb{highk}G_{s_1s_2}(x)=\left\{\barr{l}s_1=s_2=\frac32\left\{\barr{l}-4c_6\kappa^6,
x=(0,1,1,0);\\ 2c_6\kappa^6,
x=(0,1,0,1),\\\hspace{1,65cm}(0,0,1,1);\earr\right.\vspace{.18cm}\\s_1=s_2=\frac12\left\{\barr{l}4c_6\kappa^6,
x=(0,1,1,0);\\-2c_6\kappa^6,
x=(0,1,0,1),\\\hspace{1.93cm}(0,0,1,1);\earr\right.\\0\,\kappa^6,\,s_1\not=s_2\,.\earr\right.\eequ
We also find the following $\kappa^9$ contributions ($c_9$,
$c_{9,\pm}$ being nonzero constants):
\bequ\lb{highk2}G_{s_1s_2}(x)=\left\{\barr{l}c_9\kappa^9,\,x=(0,1,1,1),\,
s_1=s_2\,,\\c_{9,\pm}\,\kappa^9,\,
x=(0,1,1,1),\,s_1=\frac32,\,s_2=\pm\frac12\,.
\earr\vspace{.18cm}\\\right.\eequ

Note that there are ${\cal O}(\kappa^8)$ contributions to
$\G_{s_1s_2}(x=0)$ arising from two square loops starting and ending
at $x=0$, but with opposite orientations. They give equal diagonal
contributions and there are {\sl no} off-diagonal contributions
using the symmetry properties of Eq. (\ref{r3}).

If any two of the terms in the square root are zero, the square root
in Eq. (\ref{2H}) disappears and, since by Eq. (\ref{2H}) and the
decay properties of $\Gamma$ obtained with the hyperplane method
leads to a joint analyticity in $\kappa$, $w$ and each $\vec p$
component, we can apply the analytic implicit function theorem to
find an analytic $w(\vec p,\kappa)$ such that $H(w(\kappa,\vec
p),\kappa,\vec p)=0$, for small $\kappa$. In fact, the analytic
implicit function $w(\vec p, \kappa)$ becomes explicit and admits
the contour integral representation $$w(\vec
p,\kappa)=\frac1{2\pi\,i}\int_{|w|=r}\,\frac{w\:(\partial H/\partial
w)(w,\kappa,\vec p)}{H(w,\kappa,\vec p)}\,dw\,,$$where the contour
can be taken as a small positively oriented circle (see, for
instance, Ref. \cite{Hille}). This representation is used to obtain
an explicit formula for the Taylor series coefficients in Ref.
\cite{SO2}.

Concerning the proton and the neutron, we know that their dispersion
curves are identical by isospin symmetry. Moreover, as the ${\cal
P}{\cal C}\hspace{.1cm}\textsc{T}$ symmetry [see Eq. (\ref{pn})]
ensures $\tilde\Gamma_{s_1s_2}(\vec p)$ is diagonal (even at $\vec
p\not=\vec0$) and proportional of the identity, the proton
dispersion curves are also identical for their two spin states. The
resulting proton and neutron dispersion curves are given by,
recalling that $p_\ell^2\equiv 2\sum_{i=1}^{3}(1-\cos p^i)$ as in
Eq. (\ref{pl}), \bequ \lb{pndisp}w(\vec p)\equiv w(\vec
p,\kappa)=\left[ -3\ln {\kappa}-3\kappa^3/4+
p_\ell^2\kappa^3/8\right]+r_j(\vec p)\,,\eequ where $r(\vec p)$ is
of  ${\cal O}(\kappa^6)$ and is jointly analytic for each $\vec p$
component,  and $\kappa$. As easily seen, $w(\vec p)$ in Eq.
(\ref{pndisp}) is convex for small $|\vec p|$ and $|\kappa|$, and
increases with $|\vec p|$, as physically expected.

As seen in Eq. (\ref{highk2}), we have non-diagonal contributions
for $G_{s_1s_2}$, $|s_1|\not=|s_2|$. The square root in Eq.
(\ref{2H}) survives. Using $r_3$ symmetry through Eq. (\ref{r3}) and
${\cal P}$, we find the following ${\cal O}(\geq\kappa^6)$
contributions to $\tilde G_{s_1s_2}(\vec p)$:
\bequ\lb{gamatil}\barr{l} \tilde G_{\frac32\frac32}(\vec
p)\propto(-2\cos p^1\cos p^2+\cos p^1\cos
p^3\,,\vspace{.16cm}\\\hspace{4cm}+\cos p^2\cos p^3)\kappa^6\,,\vspace{.16cm}\\
\tilde G_{\frac12\frac12}(\vec p)\propto(2\cos p^1\cos p^2-\cos
p^1\cos
p^3\,,\vspace{.16cm}\\\hspace{4cm}-\cos p^2\cos p^3)\kappa^6\,,\vspace{.16cm}\\
\tilde G_{\frac32\frac12}(\vec p)\propto[(1-i)\sin p^1\cos
p^2\vspace{.16cm}\\\hspace{2.2cm}-(1+i)\cos
p^1\sin p^2]\sin p^3\kappa^9\,,\vspace{.16cm}\\
\tilde G_{\frac32\:-\frac12}(\vec p)\propto\sin p^1\sin p^2\cos
p^3\kappa^9\,,\earr\eequ where we observe that {\sl  all} of these
contributions do vanish at $\vec p=\vec 0$.

We have similar expressions for $\tilde\Gamma_{ij}$. Hence,
$\tilde\Gamma_{\frac32\frac32}(\vec
p)-\tilde\Gamma_{\frac12\frac12}(\vec p)\not= 0$, and we cannot
successfully apply the same method as before, as the $\kappa$
analyticity region shrinks to zero as $\vec p\searrow\vec 0$.
Further analysis is then required.

Using the above results, without using analyticity, we can however
obtain the remaining approximate dispersion curves, but with less
knowledge about their behavior. Particularly, we only have a rough
bound on $r(\vec p)$. To do this job, with $h_2$ denoting the square
root term in Eq. (\ref{2H}), we can write
\bequ\lb{HH}H=w+h_1+h_2\,,\eequ where, uniformly in $\vec p$, we
have $|h_1|\leq c_1\kappa^6$ and $|h_2|\leq c_2\kappa^6$, for some
positive constants $c_{1,2}$, by the above and the bounds of Theorem
2 of Ref. \cite{CMP}. [The $h_2$ bound is due to the $\kappa^6$ term
in $\zeta_1$, which goes to zero as $\vec p\searrow\vec 0$, but we
can possibly have a term of order ${\cal O}(\geq \kappa^{10})$ which
does not go to zero as $\vec p\searrow \vec 0$. Such a term leads to
a mass splitting.]

Writing $H=0$ in Eq. (\ref{HH}) as $$w=-h_1(w,\kappa,\vec
p)-h_2(w,\kappa,\vec p),\;|h_1+h_2|\leq c\kappa^6\,,$$ we plot the
left and right sides as functions of $w$. For $|w|\leq
(1+\alpha)c\kappa^6$, for some $\alpha>0$, we have at least one
solution of $H=0$. Similar considerations apply in working out the
zero determinant solution for $\lambda_-$ in Eq. (\ref{dett}). From
Ref. \cite{CMP}, a Rouch\'e theorem argument shows that there are
exactly four solutions of ${\rm det}\tilde\Gamma=0$, so that there
is exactly one solution for each $\lambda_\pm$, which are presented
in Eq. (\ref{disp}).

We now give an explicit construction of the dispersion curve in the
non-diagonal case. In $H(w,\kappa,\vec p)$, it is convenient to make
another nonlinear variable transformation from $w$ to $u\equiv
w+f_1(w,\kappa)$ and, by the analytic implicit function theorem,
$L(u,w)\equiv w+w+f_1(w,\kappa)-u=0$ yields the analytic function
$w\equiv r(u,\kappa)$ such that $L(r(u,\kappa),u)=0$ for $|w|$,
$|\kappa|$ small.  The condition for the existence of $r$ is
satisfied as $$(\partial L/\partial w)(w,u)=1+(\partial f_1/\partial
w)(w,\kappa)\not= 0\,,$$ for $|w|$, $|u|$, $|\kappa|$ small, since
$|f_1|\leq c_1 \kappa^6$ and $|\partial f_1/\partial w|\leq
c_2\kappa^6$. Furthermore,
$(\zeta_1^2+|\zeta_2|^2+|\zeta_3|^2)(w=r(u,\kappa),\kappa,\vec p)$
is jointly analytic in $u$, $\kappa$, $p^j$ and we write it as
$\sum_n \,a_n\,u^n$.

Thus, we can write the dispersion curve determining equation $H=0$
as the fixed point equation $$u=\left[\sum_n
\,a_n\,u^n\right]^{1/2}\equiv f(u)\,.$$

We obtain the solution ${\underline u}$ by iteration. The
$(\ell-1)$th iteration
gives$$\barr{lll}u&\!\!=\!\!&\left[\sum_{n_1}a_{n_1}\!\left[\sum_{n_2}a_{n_2}\!
\left[\sum_{n_3}a_{n_3}\ldots \!\left[\sum_{n_\ell}a_{n_\ell}
u^{n_\ell}\right]^{\frac{(n_\ell-1)}2}\right.\right.\right.\vspace{.15cm}\\&&
\left.\left.\left.\times
\:\ldots\right]^{\frac{n_2}2}\right]^{\frac{n_1}2}\right]^{1/2}\,.\earr$$

Assuming convergence to the fixed point ${\underline u}$, for the
initial value $u=0$, we take the limit $\ell\rightarrow\infty$ in
the above, with $a_\ell$ replaced by $a_0\,\delta_{\ell0}$, to get
the solution ${\underline u}$.

Additionally, by adapting the subtraction method of Ref. \cite{CMP},
it can be shown that the isolated dispersion curves (the upper gap
property) associated with the two-baryon functions of the above $20$
particles, with asymptotic mass $-3\ln \kappa$, are the only
particles in the full space of states generated by an odd number of
$\psi$ fields, up to the meson-baryon threshold $\approx -5\ln
\kappa$ (the meson asymptotic masses are $\approx -2\ln \kappa$).

To close this Appendix, we point out that for free lattice Fermi
field the matrix structure of the Fourier transform of the two-point
function is the same as in the continuum, i.e.
$p^\mu\gamma^\mu+mI_2$ in the lower subspace.

From this explicit relation, we see that there is {\sl no} spin flip
and that the diagonal elements are the same. Another way to see this
is to use the ${\cal P}{\cal C}{\cal T}$ symmetry which gives
$\langle \psi_3\bar\psi_4\rangle= \langle \psi_4\bar\psi_3\rangle=0$
and $\langle \psi_3\bar\psi_3\rangle= \langle
\psi_4\bar\psi_4\rangle$. Moreover, in the interacting case with a
rotationally invariant interaction, such as in the  ${\rm
U}(N)$-invariant Gross-Neveu model, the same holds using ${\cal
P}{\cal C}{\cal T}$ symmetry.

However, for the three-quark composite local baryon fields, the
${\cal P}{\cal C}{\cal T}$ symmetry, or even the ${\cal P}{\cal
C}\hspace{.1cm}\textsc{T}$ symmetry,  are not enough to conclude
diagonality of the two-point function in the spin, except for
protons and neutrons. As we have seen, it is not diagonal in the
spin $3/2$ case.

In the continuum case, relativistic invariance allows us to pass to
the rest frame, and the two-point function transforms according to
the spatial rotation group ${\rm SU}(2)$. There is no (spin) mass
splitting and there is just one dispersion curve, the relativistic
one.
\appendix{{ \small {\bf APPENDIX C: Coincident point four-baryon
correlation and linear dependence relation}}}
\lb{appc}\setcounter{equation}{0}
\renewcommand{\theequation}{C\arabic{equation}}\vspace{.2cm}\\
Here we obtain the decomposition of $M^{(0)}_0$ in the total spin
basis, and show the pointwise linear dependence relation given in
Section 4.3 [see Eq. (\ref{LD})].

At $\kappa=0$, using the total spin spatial symmetry and
orthogonality properties, the following block decomposition as given
in Eq. (\ref{Mdecompo}) is
$$M^{(0)}(0)=M^{(0)}_1(0)\oplus M^{(0)}_1(0)\oplus
M^{(0)}_1(0)\oplus M^{(0)}_b(0)\,.$$It is convenient to use the
shorthand notation $b_{I_zs}\equiv B_{(I)I_zs}$. $M^{(0)}_b(0)$ is
diagonal by the total spin orthogonality relation and, by the
lowering relations of the z-component of total spin, all of its
elements are equal. The first diagonal element is
$$\barr{ll}\langle {\cal T}^{3/2\,3/2}_{T=3T_z=3}\bar {\cal
T}^{3/2\,3/2}_{T=3T_z=3}\rangle^{(0)}=&4\,\langle
b_{3/2\,3/2}b_{-3/2\,3/2}\vspace{.15cm}\\&\!\times\;\bar
b_{3/2\,3/2}\bar b_{-3/2\,3/2}\rangle^{(0)}=-4\,.\earr$$

The elements of the $2\times 2$ matrix $M^{(0)}_1(0)$, with $S=1$,
$S_z=1$, $T=1$, $T_z=1$, are given by $$ \barr{c}\mu_{11}=2\langle
p_+n_+\bar p_+n_+\rangle^{(0)}\;,\;\mu_{22}=\langle \chi
\bar\chi\rangle^{(0)}\,,\vspace{.15cm}\\
\mu_{21}=\mu_{12}=\langle p_+n_+\bar\chi \bar\chi\rangle^{(0)}\,,
\earr$$where $\chi\equiv$ $\sqrt{3/10}\; b_{3/2\,3/2}$ $
b_{-3/2\,-1/2}$ $-$ $\sqrt{2/5} \;b_{3/2\,1/2}$ $ b_{-3/2\,1/2}$ $+$
$\sqrt{3/10}\;b_{3/2\,-1/2}$ $ b_{-3/2\,3/2}$ $-$ $\sqrt{3/10}\;
b_{1/2\,3/2}$ $ b_{-1/2\,-1/2}$ $+$ $\sqrt{2/5}$ $b_{1/2\,1/2}
b_{-1/2\,1/2}$ $-$ $\sqrt{3/10}\;b_{1/2\,-1/2}$ $b_{-1/2\,3/2}$.

For the sake of checking the calculations, we compute the term of
the four-baryon correlation appearing in $\mu_{ij}$ with $b$ fields
which have the prefactor $1$ since their values are integers. We
denote them by $B^u$ and call them unnormalized fields, and label
the unnormalized products $p^u_+n^u_+$, $b^u_{\frac32\,\frac32}\,
b^u_{\frac{-3}2\,\frac{-1}2}$, $b^u_{\frac32\,\frac12}\,
b^u_{\frac{-3}2\,\frac12}$, $b^u_{\frac32\,\frac{-1}2}\,
b^u_{\frac{-3}2\,\frac32}$, $b^u_{\frac12\,\frac32}\,
b^u_{\frac{-1}2\,\frac{-1}2}$, $b^u_{\frac12\,\frac12}\,
b^u_{\frac{-1}2\,\frac12}$, $b^u_{\frac12\,\frac{-1}2}\,
b^u_{\frac{-1}2\,\frac{3}2}$ appearing above by $1$, $2$, $\ldots$,
$7$, respectively. Here, the corresponding $\kappa=0$ averages are
denoted by $(i,j)$, where the second spot corresponds to the product
of two barred fields. By isospin flip symmetry, we have the
following equalities: $(1,2)=(1,4)$, $(1,5)=(1,7)$, $(2,2)=(4,4)$,
$(2,6)=(4,6)$, $(3,5)=(3,7)$, and $(5,5)=(5,7)$. Thus, all we need
to calculate are the elements listed below with their respective
values:
$$
\barr{lll}(1,1)=-360\:,&(1,2)=144\:,&(1,3)=-96\vspace{.15cm}\\
(1,5)=-144\:,&(2,6)=288\:,&(3,3)=-144\vspace{.15cm}\\
(3,5)=96\:,&(3,6)=240\:,&(4,7)=144\vspace{.15cm}\\
(5,5)=-240\:,&(5,7)=-192\:,&(6,6)=-1296\,. \earr$$

Reinstating the normalization prefactors and substituting in
$\mu_{ij}$ results in the matrix $M^{(0)}_1(0)$ of Eq.
(\ref{Mdec2}).

We now turn to the linear dependence relation of Eq. (\ref{LD}). To
prove that
$${\cal T}_{S=1,S_z} +\frac{\sqrt{5}}2 {\cal T}_{T=1,T_z}=0\,,$$for
$S_z=T_z=-1,0,1$, it is enough to show it holds for $S_z=T_z=1$ and
apply the spin lowering operator on $S_z$ and $T_z$. For
$S_z=T_z=1$, using the definition of the ${\cal T}$ fields in
Section 4.1 [see Eq. (\ref{1/2b})], we write ${\cal T}_{S=1,1}
+\frac{\sqrt{5}}2 {\cal T}_{T=1,1}$ as the sum, with from now on
$b_{I_zs}\equiv B_{(I=3/2)I_zs}$, $s\,=\,\pm 3/2,\,\pm 1/2$,
$$\barr{c}
\sqrt{2}\,p_+n_++\frac{\sqrt{5}}2\left[\sqrt{\frac38}\,
b_{\frac32\frac32}b_{\frac{-3}2\frac{-1}2}
-\sqrt{\frac12}\,b_{\frac32\frac12}b_{\frac{-3}2\frac{1}2}
\right.\vspace{.15cm}\\\left.+
\sqrt{\frac38}\,b_{\frac32\frac{-1}2}b_{\frac{-3}2\frac{3}2}
-\sqrt{\frac38}\,b_{\frac12\frac32}b_{\frac{-1}2\frac{-1}2} +
\sqrt{\frac12}\,b_{\frac12\frac12}b_{\frac{-1}2\frac{1}2}\right.\vspace{.15cm}\\\left.
-\sqrt{\frac38}\,b_{\frac12\frac{-1}2}b_{\frac{-1}2\frac{3}2}\right]\,.\earr
$$
A typical term in the above is labeled by a product of six $\psi$'s,
with indices from one of the following 12 sets of indices: (1)
$1--$; (2) $1-+$; (3) $1+-$; (4) $1++$; (5) $2--$; (6) $2-+$; (7)
$2+-$; (8) $2++$; (9) $3--$; (10) $3-+$; (11) $3+-$; (12) $3++$.

There can be no repeats of any of the twelve indices in the six
factors, by Pauli exclusion. We are left with $21$ linearly
independent products of six $\psi$'s with no repeated indices, and
the coefficients of each of these linearly independent elements sum
to zero.
\appendix{{ \small {\bf APPENDIX D: Decay of the Bethe-Salpeter kernel}}}
\lb{appd}\setcounter{equation}{0}
\renewcommand{\theequation}{D\arabic{equation}}\vspace{.2cm}\\
Using the hyperplane decoupling method, as described in Appendix B
(see also Ref. \cite{CMP} for more details), we obtain improved
temporal decay for the kernel $K$, as compared to that of $M$ and
$M_0$. For now, we use the original and unmodified $M$, and we will
treat the modified $M\equiv M^\prime$ at the end of the Appendix.

The temporal decay follows from a product structure of the sixth
derivatives of the four-point functions $M$ and $M_0$, which in turn
follows from the product structure of the sixth $\kappa_p$
derivative of the basic four-point functions $D$ and $D_0$ defined
in Section 4.2. Recall that $D$, in the individual spin basis is
defined in Eq. (\ref{DD}) and $D_0$ is obtained from $D$ by
erroneously applying Wick's theorem to the baryon fields in $D$. In
a similar way, $M_0$ is obtained from $M$. In what follows, we
suppress the collective index subscripts from $D$ and $D_0$. The
product structure extends from $D$ ($D_0$) to $M$ ($M_0$) as it is
preserved under real orthogonal transformations.

We show that, for $x_1^0=x_2^0\leq   p< x_3^0=x^0_4$, and with $\vec
w$, $\vec w_1$, $\vec w_2\in\mathbb{Z}^3$, \bequ\barr{l}
D^{(6)}(x_1,x_2,x_3,x_4,)= -\frac12 \sum_{\vec w_1\not= \vec
w_2}\vspace{.15cm}\\\hspace{.6cm}\times\,D^{(0)}(x_1,x_2,(p,\vec
w_1),(p,\vec w_2))
\vspace{.15cm}\\\hspace{.6cm}\times\,D^{(0)}((p+1,\vec
w_1),(p+1,\vec w_2),x_3,x_4) \vspace{.15cm}\\\hspace{.6cm}-\frac14
\sum_{\vec w}\,D^{(0)}(x_1,x_2,(p,\vec
w),(p,\vec w))\vspace{.15cm}\\
\hspace{.6cm}\times\,D^{(0)}((p+1,\vec w),(p+1,\vec
w),x_3,x_4)\lb{D6}\,,\earr\eequ and \bequ
\barr{l}D_0^{(6)}(x_1,x_2,x_3,x_4,)= -\frac12 \sum_{\vec w_1, \vec
w_2}\vspace{.15cm}\\\hspace{.6cm}\times\,D_0^{(0)}(x_1,x_2,(p,\vec
w_1),(p,\vec w_2))
\vspace{.15cm}\\\hspace{.6cm}\times\,D_0^{(0)}((p+1,\vec
w_1),(p+1,\vec w_2),x_3,x_4) \lb{D06}\,,\earr\eequ and similarly for
$x_1^0>x_3^0$. We write these results schematically as \bequ\lb{D62}
D^{(6)}= -\frac12 D^{(0)}\circledast D^{(0)}-\frac14
D^{(0)}\circledcirc D^{(0)}\,,\eequ and \bequ\lb{D062} D_0^{(6)}=
-\frac12 D^{(0)}\circ D^{(0)}\,,\eequ using the convolution-like
operations ``$\circledast$", ``$\circledcirc$" and ``$\circ$" given
by Eqs. (\ref{D6}) and (\ref{D06}).

We now show how this structure gives the improved temporal decay for
$K=D_0^{-1} -D^{-1}$. Take $x_1^0\leq p<x_3^0$, for example, and let
$\Lambda$ ($\Lambda_0$) denote the inverse of $D$ ($D_0$). It is
easily seen that $K^{(n)}=0$, $n=0,1,\ldots,5$, which follows from
the fact that the same holds for $D$ and $D_0$. This is seen by
expanding the numerators and using imbalance of fermions. We have,
using Eqs. (\ref{D62}) and (\ref{D062}),
\bequ\lb{K6}\barr{lll}K^{(6)}&=&\Lambda_0^{(0)}D_0^{(6)}
\Lambda_0^{(0)}-\Lambda^{(0)}D^{(6)}\Lambda^{(0)}\vspace{.15cm}\\&=&+
\Lambda_0^{(0)}(-\frac12 D_0^{(0)}\circ D_0^{(0)})
\Lambda_0^{(0)}\vspace{.15cm}\\&&-\Lambda^{(0)}(-\frac12
D^{(0)}\circledast D^{(0)})\Lambda^{(0)}\vspace{.15cm}\\
&&-\Lambda^{(0)}(-\frac14 D^{(0)}\circledcirc
D^{(0)})\Lambda^{(0)}=0
 \,,\earr\eequ for $x^0_1+1<x_3^0$. As this holds
for each hyperplane between $x_1^0$ and $x_3^0$, the above leads to
a $\kappa^6\,\kappa^{7(x_3^0-x_1^0-1)}$ decay, for $x_3^0>x_1^0+1$
and a decay $\kappa^6$, for $x_3^0=x_1^0+1$. The $\kappa^7$ can be
further improved to $\kappa^8$ taking into account fermion imbalance
(or gauge integration) in the calculation of the seventh $\kappa_p$
derivative.

We return to show how Eqs. (\ref{D6}) and (\ref{D06}) are obtained.
For $x_1^0=x_2^0<x_3^0=x_4^0$,
$$\barr{ll}
D_0^{(6)}=&- \langle B_{\ell_1}(x_1)\bar
B_{\ell_3}(x_3)\rangle^{(3)}\, \langle B_{\ell_2}(x_2)\bar
B_{\ell_4}(x_4)\rangle^{(3)}\vspace{.15cm}\\&+\langle
B_{\ell_1}(x_1)\bar B_{\ell_4}(x_4)\rangle^{(3)}\,\langle
B_{\ell_2}(x_2)\bar B_{\ell_3}(x_3)\rangle^{(3)} \,,\earr
$$where the $\ell$'s are collective indices.

Using Eq. (\ref{product2}), and regrouping the terms, Eq.
(\ref{D06}) follows.

To show Eq. (\ref{D6}), in addition to the gauge integral ${\cal
I}_3$ for three coincident bonds of Eq. (\ref{I3}), we need to
compute the gauge integral ${\cal I}_6$ of six coincident bonds
(same orientation!). To compute ${\cal I}_6$, we use the generating
function given in Refs. \cite{Creu2,CMP} and differentiate with
respect to the external sources. It is given by, with $U(g)\equiv g$
and $U^{-1}(g)\equiv g^{-1}$, \bequ\lb{I6}\barr{lll}{\cal
I}_6&=&\dis\int g_{a_1b_1} g_{a_2b_2}g_{a_3b_3}g^{-1}_{a_4b_4}
g^{-1}_{a_5b_5}g^{-1}_{a_6b_6}\,d\mu(g)\vspace{.15cm}\\&=&\dis
\frac{2}{3!4!}\left[
\epsilon_{a_1a_2a_3}\epsilon_{b_1b_2b_3}\epsilon_{a_4a_5a_6}
\epsilon_{b_4b_5b_6}+\right.\vspace{.15cm}\\ &&
\left.\epsilon_{a_1a_2a_4}\epsilon_{b_1b_2b_4}\epsilon_{a_3a_5a_6}
\epsilon_{b_3b_5b_6}+\right.\vspace{.15cm}\\ &&
\left.\epsilon_{a_1a_2a_5}\epsilon_{b_1b_2b_5}\epsilon_{a_3a_4a_6}
\epsilon_{b_3b_4b_6}+\right.\vspace{.15cm}\\ &&
\left.\epsilon_{a_1a_2a_6}\epsilon_{b_1b_2b_6}\epsilon_{a_3a_4a_5}
\epsilon_{b_3b_3b_5}+\right.\vspace{.15cm}\\ &&
\left.\epsilon_{a_1a_3a_4}\epsilon_{b_1b_3b_3}\epsilon_{a_2a_5a_6}
\epsilon_{b_2b_5b_6}+\right.\vspace{.15cm}\\ &&
\left.\epsilon_{a_1a_3a_5}\epsilon_{b_1b_3b_5}\epsilon_{a_2a_4a_6}
\epsilon_{b_2b_4b_6}+\right.\vspace{.15cm}\\ &&
\left.\epsilon_{a_1a_3a_6}\epsilon_{b_1b_3b_6}\epsilon_{a_2a_4a_5}
\epsilon_{b_2b_4b_5}+\right.\vspace{.15cm}\\ &&
\left.\epsilon_{a_1a_4a_5}\epsilon_{b_1b_4b_5}\epsilon_{a_2a_3a_6}
\epsilon_{b_2b_3b_6}+\right.\vspace{.15cm}\\ &&
\left.\epsilon_{a_1a_4a_6}\epsilon_{b_1b_4b_6}\epsilon_{a_2a_3a_5}
\epsilon_{b_2b_3b_5}+\right.\vspace{.15cm}\\ &&
\left.\epsilon_{a_1a_5a_6}\epsilon_{b_1b_5b_6}\epsilon_{a_2a_3a_4}
\epsilon_{b_2b_3b_4}\right] \,.\earr\eequ By performing appropriate
(index) contractions in ${\cal I}_6$, we verify that ${\cal I}_4$,
${\cal I}_3$ and ${\cal I}_2$ are obtained, where (see Refs.
\cite{Creu2,CMP,2flavor2meson}) \bequ {\cal I}_2=\int g_{a_1b_1}
g^{-1}_{a_2b_2}\,d\mu(g)= \frac 13\,
\delta_{a_1b_2}\,\delta_{a_2b_1}\,,\lb{I2}\eequ ${\cal I}_3$ is
given in Eq. (\ref{I3}) and \bequ \barr{lll}{\cal I}_4&=&\int
g_{a_1b_1} g_{a_2b_2}^{-1} g_{a_3b_3}
g_{a_4b_4}^{-1}\,d\mu(g)\vspace{.14cm}\\&=& \frac {1}{8}\,
[\delta_{a_1b_2}\delta_{a_3b_4}\delta_{b_1a_2}\,\delta_{b_3a_4} +
(a_2\rightleftarrows a_4, b_2\rightleftarrows b_4)]
\vspace{.14cm}\\&& - \frac {1}{24}\,
[\delta_{a_1b_2}\delta_{a_3b_4}\delta_{b_1a_4}\,\delta_{b_3a_2}+
(a_2\rightleftarrows a_4, b_2\rightleftarrows b_4)]
\lb{I4}\,.\end{array}\eequ

Eq. (\ref{I6}) can be written in a compact notation as
$$
{\cal I}_6=\frac{2}{3!4!}\,\sum_{{\cal P}}\,E_R\,E_S\,,
$$
where ${\cal P}$ is the set of the ten disjoint partitions of
$T\equiv\{1,2,3,4,5,6\}=R\cup S$, $|R|=3=|S|$, and for
$R=\{i,j,k\}$, $E_R=\epsilon_{a_ia_ja_k}\,\epsilon_{b_ib_jb_k}$,
where the elements in $E_R$ and $E_S$ are written in increasing
order.

$D^{(6)}$ is obtained by writing $D=N/Z$ and expanding the
exponential in the numerator of $N$. In this way,
$D^{(6)}=N^{(6)}/Z^{(0)}$, where $N^{(6)}=N^{(6)}_n+N^{(6)}_o$.
$N^{(6)}_n$ arises from the terms where there are two nonoverlapping
bonds, indexed by $w_1$ and $w_2$, each bond with three positively
oriented gauge fields. After performing the gauge integrals using
the formula for ${\cal I}_3$ of Eq. (\ref{I3}), we get
$$\barr{lll}\!
{N_n^{(6)}}\!\!\!\!&=\!\!\!&-
\frac1{2\times6^4}\,\epsilon_{a_1a_2a_3}\epsilon_{a_4a_5a_6}\epsilon_{b_1b_2b_3}
\epsilon_{b_4b_5b_6}\,\sum_{\vec w_1\not=\vec
w_2,\{\alpha_i\},\{f_i\}}\vspace{.15cm}\\\!\!&\!\!&\times\, \langle
B_{\ell_1}(x_1)
B_{\ell_2}(x_2)(\bar\psi_{a_1\alpha_1f_1}\bar\psi_{a_2\alpha_2f_2}
\bar\psi_{a_3\alpha_3f_3})(p,\vec
w_1)\vspace{.15cm}\\\!\!&\!\!&\times\,
(\bar\psi_{a_4\alpha_4f_4}\bar\psi_{a_5\alpha_5f_5}
\bar\psi_{a_6\alpha_6f_6})(p,\vec
w_2)\rangle^{(0)}\vspace{.15cm}\\\!&\!&\times\, \langle
(\psi_{b_1\alpha_1f_1}\psi_{b_2\alpha_2f_2}
\psi_{b_3\alpha_3f_3})(p+1,\vec
w_1)\vspace{.15cm}\\\!\!&\!\!&\times\,
\psi_{a_4\alpha_4f_4}\psi_{a_5\alpha_5f_5}
\psi_{a_6\alpha_6f_6})(p+1,\vec w_2)
\vspace{.15cm}\\\!\!&\!\!&\times\,\bar B_{\ell_3}(x_3)\bar
B_{\ell_4}(x_4)\rangle^{(0)} \,,\earr
$$
where the numerical factor comes from $(-1/2)^6\!\times\!(1/6!)\!
 \times\! 20\!\times \!(-2)^6\!\times\!(1/2)\!\times\!(1/6)^2$; 20 is a combinatorial factor,
$(-2)^6$ comes from the six $\Gamma^{e^0}$'s and $(1/6)^2$ coming
from the two integrals ${\cal I}_3$. We observe that the sums over
the $\alpha_i$'s are restricted to lower components $\alpha=3,4$ due
to the structure of $\Gamma^{e^0}$.

Resumming over the intermediate states, after taking into account
multiplicities and normalizations of the baryon fields, gives the
first term of Eq. (\ref{D6}).

The contribution  to $N^{6}_o$, for which there are overlapping
bonds, after performing the ${\cal I}_6$ gauge integral, is
$$\barr{l}
-\frac1{4\times6^4\times 10}\,\sum_{\vec w,\{\alpha_i\},\{f_i\}}\,[
\epsilon_{a_1a_2a_3}\epsilon_{b_1b_2b_3}\epsilon_{a_4a_5a_6}
\epsilon_{b_4b_5b_6}\vspace{.15cm}\\+ \text {\:nine\: remaining\:
terms}]\:\langle B_{\ell_1}(x_1) B_{\ell_2}(x_2)\vspace{.15cm}\\
\times\,(\bar\psi_{a_1\alpha_1f_1}\ldots
\bar\psi_{a_6\alpha_6f_6})(p,\vec w)\rangle^{(0)} \vspace{.15cm}\\
\times\,\langle (\psi_{b_1\alpha_1f_1}\ldots
\psi_{b_6\alpha_6f_6})(p+1,\vec w)\bar B_{\ell_3}(x_3)\bar
B_{\ell_4}(x_4)\rangle^{(0)}\,,
 \earr$$where the numerical factor comes from $(-1/2)^6 \times(1/6!)
\times (-2)^6\times[2/(3!\times 4!)]$; the term in square brackets
is the contribution of the ${\cal I}_6$ integral. Again, resumming
over the intermediate states gives the second term of Eq.
(\ref{D6}). In both cases, a factor $6^{-4}$ is absorbed in the
normalization process.

By considering hyperplanes in each spatial direction, and additional
complex parameters associated with them, spatial decay of the B-S
kernel can also be obtained as in Ref. \cite{SO}.

We now describe how the product structure of Eqs. (\ref{D62}) and
(\ref{D062}) extend to $M^{(6)}$ and $M_0^{(6)}$. First, we take
isospin C-G linear combinations of the external fields present in
$D^{(6)}$, expressed in the individual spin and isospin basis, to
get total isospin zero states. Then, again using the inverse C-G
coefficients, we express the intermediate states in terms of states
of fixed total isospin $I$ and its $z$-components $I_z$. At this
point, the isospin orthogonality properties are used to get only
zero total isospin intermediate states. In this way, we arrive at
the following product structure formula for $M^{(6)}$
$$
M^{(6)}=-\frac12 M^{(0)}\circledast M^{(0)}-\frac14
M^{(0)}\circledcirc M^{(0)}\,,
$$
where we remember that we are still in the individual spin basis. We
can pass to the total isospin zero, total spin basis by taking spin
C-G linear combinations. The product structure is maintained as it
is invariant under real orthogonal transformations. A similar
argument applies to $M_0^{(6)}$.

We are now ready to explain and motivate the adoption of a
$h$-modified $M\equiv M^\prime$ as introduced in Section 4.3. For
this, we consider the following leading contributions to the
unmodified $K(x_1,x_2,x_3,x_4)$ and the corresponding contributions
to the generalized potential $\hat K(\vec\xi, \vec \eta, k^0)$: {\em
i)} the $\kappa^0$ order, $K^{(0)}$ at coincident points, and
corresponding space-range zero, energy independent constant
potential $\hat K(\vec 0, \vec 0, k^0)$; {\em ii)} the $\kappa^6$
contribution with time-range one and space-range zero
$K^{(6)}(0,0,e^0,e^0)$, with energy-dependent zero space-range
potential  $\hat K(0,0,k^0=i(2\bar m-\epsilon))$ of order $\kappa^2
e^{-\epsilon}$ ; and, {\em iii)} the $\kappa^2$ quasimeson exchange
$\kappa^2$ contribution of space-range one $K^{(2)}(0,0,e^j,e^j)$,
$j=1,2,3$, with space-range one energy-independent potential $\hat
K(0,e^j,k^0)$ of order $\kappa^2$.

So, the resulting potential is seen to be energy-dependent. We make
our modification to $M$ and then $K$ so that only an
energy-independent potential arises which is more in keeping with
our usual intuition for understanding the occurrence of bound
states.

As the $k^0$-dependent part of the effective potential is associated
with a temporal distance-one $\kappa^6$ contribution to $K$, we
modify $M$ so that this contribution is smaller. This happens
because $M$ is modified so that $M^{(6)}$ has the same product
structure as $M_0^{(6)}$. By doing so, the decay of the modified $K$
is improved for temporal distance one. In contrast to the unmodified
$M$, for our modified $M\equiv M^\prime$, we get a
$\kappa^{8|x_3^0-x_1^0|}$ decay for any distance $|x_3^0-x_1^0|\geq
1$. This comes about as the modified $M^{(6)}$ also has a product
structure so that the modified $K(x_1,x_2,x_3,x_4)$ in Eq.
(\ref{modifBS}) has a zero sixth derivative $K^{(6)}=0$, for
$x_1^0=x_2^0<x_3^0=x_4^0$, since the last two terms on the r.h.s. of
Eq. (\ref{K6}) have now the same coefficient as the first term.

There is still a possibility of having a $\kappa^8$ contribution to
the modified $K(x_1,x_2,x_3,x_4)$ of temporal distance one leading
to an energy-dependent potential of order $\kappa^2$, which would be
of the same order of the quasi-meson exchange potential. These
$\kappa^8$ contributions arise from $M$ and $M_0$ with temporal
distance one, but vanish due to imbalance of fermion components.

With this modification, the space-range zero potential is zero and
the dominant contribution is the quasimeson space-range one exchange
potential.
\appendix{{ \small {\bf APPENDIX E: Determination of the space
range-one exchange potential}}} \lb{appe}\setcounter{equation}{0}
\renewcommand{\theequation}{E\arabic{equation}}\vspace{.2cm}\\
In this Appendix, we determine the quasi-meson exchange potential
$20\times 20$ matrix of Section 4.3 in the individual spin basis,
i.e. $v^{(2)}_{s_1s_2s_3s_4}$, $v^{(2)}_{s_1s_2t_3t_4}$,
$v^{(2)}_{t_1t_2s_3s_4}$, and $v^{(2)}_{t_1t_2t_3t_4}$. We give
reduced formulas which we use for computation, and present the
derivation explicitly for $v^{(2)}_{s_1s_2s_3s_4}$ given in Eq.
(\ref{v}), the elements of which, in the total spin basis, are given
below Eq. (\ref{K22}). The other elements are obtained in an
analogous manner.

In Eq. (\ref{v}), consider the $\kappa=0$ averages expressed as a
sum of all contractions using Wick's theorem. If the explicit
$\bar\psi\psi$ contract in the first average then $\alpha_1=\beta_2$
and
$\Gamma^{e^1}_{\alpha_1\beta_1}\Gamma^{-e^1}_{\alpha_2\alpha_1}=0$;
the same for the second average. Thus, only contractions involving
one explicit internal $\tilde \psi$ and one quark field of $\tilde
B$'s contribute. Also, recalling that the $\tilde\psi$'s on the
$\tilde B$'s only have lower spin indices, this forces the spin
index of the explicit $\tilde\psi$'s to be also lower in order not
to get a zero contribution. Next, noting that in the subspace of
lower components the spin matrix $\Gamma^{\pm e^1}$ reduces to minus
the identity, and after using the $\kappa=0$  isospin flip symmetry
in the second factor, we can write \bequ\lb{vvvv2}\barr{ll}\!\!
v^{(2)}_{s_1,s_2,s_3,s_4}\!=\!\!&\!\!\displaystyle\sum_{I_1,I_3}\,
c^{\small \frac 12\,\frac12}_{I_1\,-I_1} c^{\small \frac
12\,\frac12}_{I_3\,-I_3} \,\times\vspace{.14cm}\\&\!\!\langle
B_{{\small \frac12}I_1s_1}\bar B_{{\small \frac12}I_3s_3}
:\bar\psi_{a\alpha f_1} \psi_{a\beta
f_2}:\rangle^{(0)}\,\times\vspace{.14cm}\\&\!\!\langle B_{{\small
\frac12}I_1s_2}\bar B_{{\small \frac12}I_3s_4} :\bar\psi_{b\beta
\,-f_2} \psi_{b\alpha \,-f_1}:\rangle^{(0)}\,,\earr\eequ with all
fields at a same site and where the Wick order $:\::$ forbids
contraction between the explicit internal $\tilde\psi$'s. To
implement the exclusion of this contraction, we note that up to a
global sign and omitting the detailed field indices and summations
we have $\langle B\bar B\psi \bar\psi \rangle^{(0)}\thicksim \langle
\psi_1\psi_2\psi_3\bar \psi_5\bar \psi_6\bar \psi_7 \psi_4\bar\psi_8
\rangle^{(0)}\thicksim \langle \psi_1\psi_2\psi_3\psi_4\bar
\psi_5\bar \psi_6\bar \psi_7 \bar\psi_8 \rangle^{(0)}={\rm det}
[\delta_{ij}]$, $i=1,...,4$ and $j=5,...,8$, such that the Wick
order in $:\psi_4\bar\psi_8:$ is equivalent to replacing
$\delta_{78}$ by zero in the previous fermionic Gram determinant
${\rm det}[\delta_{ij}]$.

To continue, we use symmetries and isospin and spin restrictions to
reduce the number of $v^{(2)}$'s to be calculated. Fixing a matrix
element in $v^{(2)}$, we consider the terms in the sums labeled by
$(I_1I_3)$, $(I_1J_3)$, $(J_1I_3)$ or $(J_1J_3)$. Isospin flip
symmetry gives equality of the $(J_1J_3)$ and $(-J_1\,-J_3)$ terms,
etc..., where we  take into account the properties of the C-G
coefficients. Isospin conservation forces the restrictions
$I_1+f_2=I_3+f_1\Rightarrow |I_1-I_3|\leq 1$,
$I_1+f_2=J_3+f_1\Rightarrow |I_1-J_3|\leq 1$,
$J_1+f_2=J_3+f_1\Rightarrow |J_1-J_3|\leq 1$. In addition, we have
spin sum restrictions $s_1-s_3=s_4-s_2=\alpha-\beta$, $|s_1-s_3|\leq
1$; $s_1-t_3=t_4-s_2=\alpha-\beta$, $|s_1-t_3|\leq 1$;
$t_1-t_3=t_4-t_2=\alpha-\beta$, $|t_1-t_3|\leq 1$. Also, for
$\kappa=0$, conjugation (denoted by ${\cal C}_0$) is a symmetry.
Using ${\cal C}_0$ gives the equality of the $(I_1I_3)$ and
$(I_3I_1)$ ($(J_1J_3)$ and $(J_3J_1)$) product of summed
expectations if $s_1=s_3$ and $s_2=s_4$  ($t_1=t_3$ and $t_2=t_4$),
such as diagonal elements. In this way, we show that many of the
terms are zero. We can also, after applying ${\cal C}_0$ to both
factors, change $f_1\leftrightarrow -f_1$, $f_2\leftrightarrow -f_2$
to get equality of the $(I_1I_3)$ and $(I_3I_1)$ terms if $s_1=s_4$
and $s_2=s_3$; similarly for the $J$'s. We can also impose the
prohibited contraction condition to show that certain elements are
zero.

In addition to the fact that the $20\times 20$ potential matrix is
symmetric there are other symmetries. By making the change of
labeling $\alpha\leftrightarrow \beta$, $f_1\leftrightarrow -f_2$,
$f_2\leftrightarrow -f_1$, the following symmetry properties are
obtained
$$\barr{c}v^{(2)}_{s_1s_2s_3s_4}=v^{(2)}_{s_2s_1s_4s_3}\;\;,\;\;
v^{(2)}_{s_1s_2t_3t_4}=v^{(2)}_{s_2s_1t_4t_3}\,,\vspace{.17cm}\\
v^{(2)}_{t_1t_2t_3t_4}=v^{(2)}_{t_2t_1t_4t_3}\,.\earr$$

Also, the spin flip change $\alpha\leftrightarrow -\alpha$,
$\beta\leftrightarrow -\beta$ gives the equality of the $v^{(2)}$'s
for $s_k\rightarrow -s_k$ $t_j\rightarrow -t_j$, for all $j$'s and
$k$'s.

Thus, there is symmetry in the upper left $4\times 4$ (lower right
$16\times 16$) matrix about the secondary diagonal of the submatrix.

Using each symmetry alone or coupling them explains the repeated
values occurring in the matrix with the following two exceptions:
the equality of the $(6\,9)$ and $(8\,11)$ elements and the equality
between $(11\,11)$ and $(11\,14)$. By spin flip at $\kappa=0$, we
see that $(1\,1)=(4\,4)$, $(2\,2)=(3\,3)$, $(1\,7)=(4\,18)$,
$(1\,10)=(4\,15)$, $(1\,13)=(4\,12)$, $(2\,8)=(3\,17)$,
$(2\,11)=(3\,14)$, $(2\,14)=(3\,11)$; by spin flip at $\kappa=0$ and
$1\leftrightarrow 2$, $3\leftrightarrow 4$ exchanges
$(1\,13)=(4\,18)$, $(7\,7)=(12\,12)$; by $1\leftrightarrow 2$,
$3\leftrightarrow 4$ exchanges $(6\,6)=(9\,9)$. By applying ${\cal
C}_0$ to both average factors, we obtain $(7\,10)=(10\,13)$. Thus,
we only have to calculate e.g. the elements $(1\,1)$, $(2\,2)$,
$(2\,3)$, $(1\,7)$, $(1\,10)$, $(1\,13)$, $(2\,8)$, $(2\,11)$,
$(2\,14)$, $(5\,5)$, $(6\,6)$, $(6\,9)$, $(7\,7)$, $(7\,10)$,
$(8\,8)$, $(8\,11)$, $(10\,10)$, $(11\,11)$ and $(11\,14)$.

The values of the above elements are given below Eq. (\ref{K22}). As
a check in calculating the elements of Eq. (\ref{v}), etc, it is
useful to give the decomposition of the total value before
performing the sums in the equivalent expression of Eq.
(\ref{vvvv2}). We now give the nonzero values for the partial terms
$(I_1I_3)$ or $(I_1J_3)$ or $(J_1J_3)$ for each matrix element; the
values for $(-I_1\,-I_3)$ or $(-I_1\,-J_3)$ or $(-J_1\,-J_3)$ are
the same. The sum of the values for each relevant partial matrix
element given below multiplied by a factor $2$ is the value of the
matrix element. The values, given here with precision of two digits,
are as follows. For the matrix element $(1\,1)$, which corresponds
to $s_1=s_2=1/2$, $s_3=s_4=1/2$, in Eq. (\ref{vvvv2}), we need the
contributions $(I_1I_3)=(\frac12\, \frac 12)$, which is $0.78$, and
$(I_1I_3)=(\frac12\, -\frac 12)$, which is $-0.94$. Denoting this
contribution as $(1\,1)$: $(\frac12\, \frac 12)$ $0.78$, $(\frac12\,
-\frac 12)$ $-0.94$, and using this notation for the other elements
we find $(2\,2)$: $(\frac12\, \frac 12)$  $1.22$, $(\frac12\, -\frac
12)$ $0.44$; $(2\,3)$: $(\frac12\, \frac 12)$ $-0.44$, $(\frac12\,
-\frac 12)$ $-1.39$; $(1\,7)$: $(\frac12\, \frac 32)$  $-0.41$,
$(\frac12\, \frac 12)$  $-0.27$, $(\frac12\, -\frac 12)$ $-0.14$;
$(1\,10)$: $(\frac12\, \frac 32)$ $0.47$, $(\frac12\, \frac 12)$
$0.31$, $(\frac12\, -\frac 12)$ $0.16$; $(1\,13)$: $(\frac12\, \frac
32)$ $-0.41$, $(\frac12\, \frac 12)$ $-0.27$,  $(\frac12\, -\frac
12)$ $-0.14$; $(2\,8)$: $(\frac12\, \frac 32)$  $0.71$, $(\frac12\,
\frac 12)$ $0.47$, $(\frac12\, -\frac 12)$ $0.24$; $(2\,11)$:
$(\frac12\, \frac 32)$ $-0.47$, $(\frac12\, \frac 12)$ $-0.31$,
$(\frac12\, -\frac 12)$ $-0.16$; $(2\,14)$: $(\frac12\, \frac 32)$
$0.24$, $(\frac12\, \frac 12)$ $0.16$, $(\frac12\, -\frac 12)$
$0.08$; $(5\,5)$: $(\frac32\, \frac 12)$ $-0.75$, $(\frac12\, \frac
32)$ $-0.75$, $(\frac12\, \frac 12)$ $1.00$, $(\frac12\, -\frac 12)$
$-1.00$; $(6\,6)$: $(\frac32\, \frac 12)$ $-0.50$, $(\frac12\, \frac
32)$ $-0.50$, $(\frac12\, \frac 12)$ $0.67$, $(\frac12\, -\frac 12)$
$-0.67$; $(6\,9)$: $(\frac32\, \frac 12)$  $-0.25$, $(\frac12\,
\frac 32)$ $-0.25$, $(\frac12\, \frac 12)$ $0.33$, $(\frac12\,
-\frac 12)$ $-0.33$; $(7\,7)$: $(\frac32\, \frac 12)$  $-0.25$,
$(\frac12\, \frac 32)$ $-0.25$, $(\frac12\, \frac 12)$ $0.33$,
$(\frac12\, -\frac 12)$ $-0.33$; $(7\,10)$: $(\frac32\, \frac 12)$
$-0.29$, $(\frac12\, \frac 32)$ $-0.29$, $(\frac12\, \frac 12)$
$0.38$, $(\frac12\, -\frac 12)$ $-0.38$; $(8\,8)$: all zero;
$(8\,11)$: $(\frac32\, \frac 12)$ $-0.25$, $(\frac12\, \frac 32)$
$-0.25$, $(\frac12\, \frac 12)$ $0.33$, $(\frac12\, -\frac 12)$
$-0.33$; $(10\,10)$: $(\frac32\, \frac 12)$ $-0.42$, $(\frac12\,
\frac 32)$ $-0.42$, $(\frac12\, \frac 12)$ $0.56$, $(\frac12\,
-\frac 12)$ $-0.56$; $(11\,11)$: $(\frac32\, \frac 12)$ $-0.33$,
$(\frac12\, \frac 32)$ $-0.33$, $(\frac12\, \frac 12)$ $0.44$,
$(\frac12\, -\frac 12)$ $-0.44$; $(11\,14)$: $(\frac32\, \frac 12)$
$-0.33$, $(\frac12\, \frac 32)$ $-0.33$, $(\frac12\, \frac 12)$
$0.44$, $(\frac12\, -\frac 12)$ $-0.44$.

The values given below Eq. (\ref{K22}) are the exact values for
$v^{(2)}$ obtaining by summing up the exact values for the above
partial decomposition values and multiplying by two.
\appendix{{ \small {\bf APPENDIX F: Uncorrelated four-baryon
function and spectral representation}}}
\lb{appf}\setcounter{equation}{0}
\renewcommand{\theequation}{F\arabic{equation}}\vspace{.2cm}\\
Here we show how to obtain a spectral representation for $\hat
M_{0,\L}(x_1,x_2,x_3,x_4)$, and its total spin version; the
components of which are obtained by summing over the appropriate C-G
combinations of the $\hat F_{0,\L}(x_1,x_2,x_3,x_4)$. Our starting
point is Eq. (\ref{m0}): $$
F_{0,1234}=-G_{13}G_{24}+G_{14}G_{23}\,,$$where we use a shorthand
notation for the collective indices $(\ell,x)$.

In lattice relative coordinates and in the equal time representation
\bequ\lb{F0}\barr{ll} F_{0,1234}(\vec \xi,\vec
\eta,\tau)=&-G_{13}(\tau+\vec \xi)\,G_{24}(\tau +\vec
\eta)\vspace{.15cm}\\&+G_{14}(\tau+\vec \xi+\vec
\eta)\,G_{23}(\tau)\,, \earr\eequ and setting $\vec k=\vec 0$ in the
Fourier transform of $\tau$ only, with dual variable $k$, our goal
is to obtain a spectral representation for $$\hat F_{0,1234}(\vec
\xi,\vec \eta,k^0)=\int\,F_{0,1234}(\vec \xi,\vec
\eta,\tau)\,e^{-ik^0\tau^0}d\tau\,.$$ For $x^0\not= 0$, and using
the spectral representation for $G_{ij}(x)$ of Eq. (\ref{FK1}), we
obtain \bequ \lb{FKK}\barr{lcl}
 G_{ij}(x)=-\dis\int_{-1}^1\,\dis
 \int_{{\mathbb T}^3}\,(\la^0)^{|x^0|-1}
 e^{i\vec p.\vec x}  d_{\la^0}\alpha_{\vec p,ij}(\la^0)d\vec p\,,
 \earr\eequ
where, by parity, $\alpha_{\vec p,ij} =\alpha_{-\vec p,ij}$.

Separating the $\tau^0=0$ contribution and substituting the spectral
representation for the $G's$ into Eq. (\ref{F0}), after performing
the integral over $\tau$, we get \bequ\lb{FFF}\barr{ll}\! \hat
F_{0,1234}(\vec \xi,&\!\!\vec \eta,k^0)\!=\!(2\pi)^{-3}\!
\int_{{\mathbb{T}}^3} \left[ -\tilde G_{13}(\vec p) \tilde
G_{24}(\vec p)\,\Xi_+(\vec p)
\right.\vspace{.15cm}\\&\!\left.+\tilde G_{14}(\vec p) \tilde
G_{23}(\vec p)\,\Xi_-(\vec p)\right] d\vec p \vspace{.15cm}\\&\!
+(2\pi)^{3}\int_{-1}^1\int_{-1}^1 \int_{{\mathbb{T}}^3}
f(k^0,\la^0\la^{\prime\, 0})\vspace{.15cm}\\&\!\times\!\left[
-d_{\la^0}\alpha_{\vec p,13}(\la^0)\,d_{\la^{\prime 0}}\alpha_{\vec
p,24}(\la^{\prime\, 0})\,\Xi_+(\vec
p)\right.\vspace{.15cm}\\&\!\left.+ d_{\la^0}\alpha_{\vec
p,14}(\la^0)\,d_{\la^{\prime 0}}\alpha_{\vec p,23}(\la^{\prime\,
0})\,\,\Xi_-(\vec p) \right]\,d\vec p\;,\earr \eequ where we recall
that $\tilde G (\vec p)=\sum_{\vec x}\,e^{-i\vec p.\vec
x}G(x^0=0,\vec x)$, and where we have $$\Xi_{\pm}(\vec
p)\equiv\Xi_{\pm}(\vec p,\vec\xi,\vec \eta)= \cos \vec p.\vec
\xi\cos \vec p.\vec \eta\pm \sin \vec p.\vec \xi\sin \vec
p.\vec\eta\,.$$

To proceed in our deduction of $\hat M_0$, we make the approximation
that $\tilde G$ is diagonal in spin,  and independent of the isospin
$I$ and spin (which holds up to and including ${\cal O}(\kappa^5)$,
see Ref. \cite{CMP}), and we denote it simply by $\tilde G(\vec p)$,
with no index; and the same for the measure $d_{\la^0}\alpha_{\vec
p}$.

Under this hypothesis, carrying out the sum over the isospin C-G
coefficients we obtain, for the $11$ element of $\hat M_0$ in the
individual spin basis \bequ\lb{FFF2}\barr{ll}\!\! \hat
M_{0,11,s_1s_2s_3s_4}\!\!\!&(\vec \xi,\vec
\eta,k^0)=\vspace{.15cm}\\&\!(2\pi)^{-3}\! \int_{{\mathbb{T}}^3}
[\tilde G(\vec p)]^2 \left[
-\delta_{s_1s_3}\delta_{s_2s_4}\,\Xi_+(\vec
p)\right.\vspace{.15cm}\\&\left.\!-\delta_{s_1s_4}\delta_{s_2s_3}\,\Xi_-(\vec
p)\right] d\vec p \vspace{.15cm}\\&\!
+(2\pi)^{3}\int_{-1}^1\int_{-1}^1 \int_{{\mathbb{T}}^3}
f(k^0,\la^0\la^{\prime\, 0})\,\times\vspace{.15cm}\\&\left[
-\delta_{s_1s_3}\delta_{s_2s_4}\,\Xi_+(\vec
p)-\delta_{s_1s_4}\delta_{s_2s_3}\,\Xi_-(\vec p)
\right]\vspace{.15cm}\\&\!\times\,d_{\la^0}\alpha_{\vec
p}(\la^0)\,d_{\la^{\prime 0}}\alpha_{\vec p}(\la^{\prime\, 0})d\vec
p\;.\earr \eequ

The $22$ element of $\hat M_0$ is obtained in the same way, by
replacing $s_j$ by $t_j$ in the above. The off-diagonal elements are
zero.

Finally, we obtain the total spin basis representation for $\hat
M_0$ by carrying out the sums over the spin C-G coefficients and
using the explicit values and symmetry properties of the C-G
coefficients, we obtain the expression given in Eq. (\ref{m00}).


\end{document}